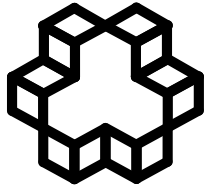

**1925**
**K. N. Toosi University of Technology**
**Faculty of Electrical Engineering**

# Tunable Graphene-based Pulse Compressor for Terahertz Application

**A Thesis Submitted in Partial Fulfillment of the Requirements for the Degree**

**of Doctor of Philosophy in Electrical Engineering**

By:

**Seyed Mohammadreza Razavizadeh**

Supervisor:
**Prof. Ramezanali Sadeghzadeh**

Advisor:
**Dr. Zahra Ghattan-kashani**

**September, 2020**

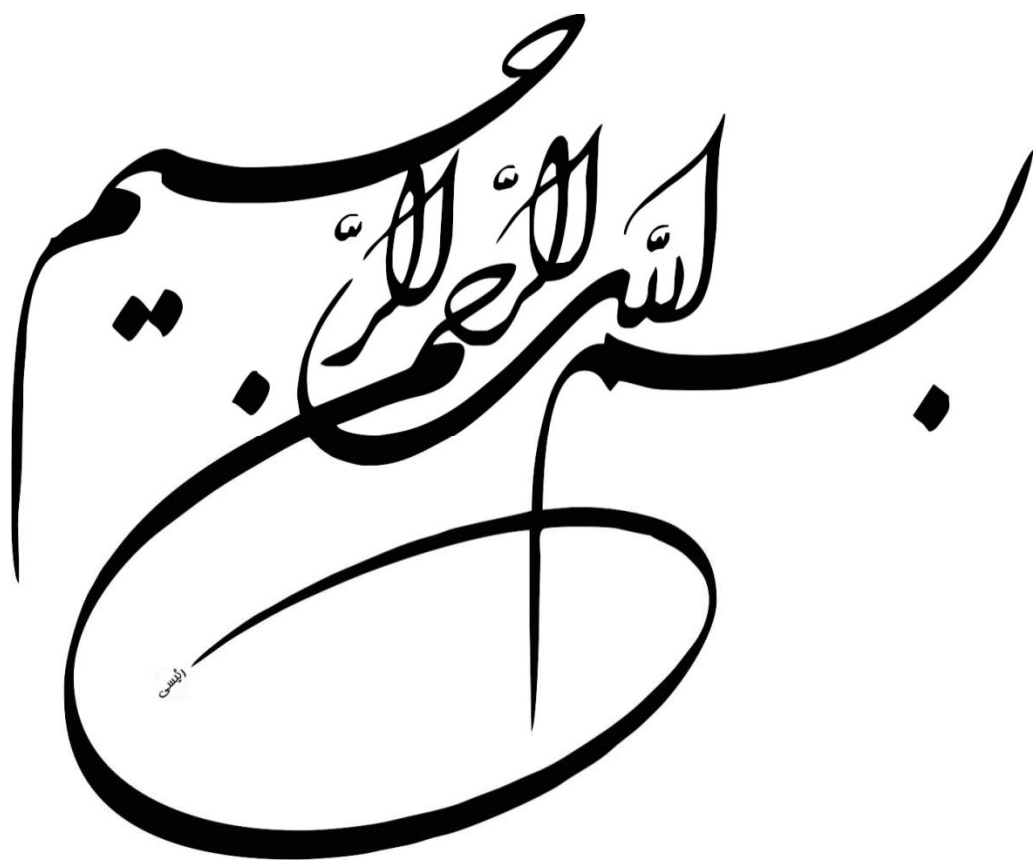

## *Dedication*

I would like to dedicate this thesis to little prince of Imam *Hussain* (As), Baby *Ali Asghar* (As)…

"*Who was scarified in way of his grandfather Prophet Muhammed*"

and to:

my beloved wife, mother and the precious memories of my father who was my biggest supporter in my life and in choosing telecommunication engineering as my major like himself, and to follow my dreams. Alas, no words can express my love, feelings for him. May God Bless Him.

# تائیدیه هیات داوران جلسه دفاع از رساله دکتری

نام دانشکده: دانشکده مهندسی برق

نام دانشجو: سیدمحمدرضا رضوی زاده

عنوان رساله: طراحی و تحلیل سامانه فشرده کننده پالس تراهرتزی کنترل پذیر مبتنی بر ساختار نوار گرافنی مارپیچی

تاریخ دفاع: ۱۳۹۹/۶/۳۱

رشته: مهندسی برق

گرایش: مخابرات میدان

| امضا | دانشگاه | مرتبه دانشگاهی | نام و نام خانوادگی | سمت | ردیف |
|---|---|---|---|---|---|
| | خواجه نصیرالدین طوسی | استاد | دکتر رمضانعلی صادق زاده | استاد راهنما | ۱ |
| | خواجه نصیرالدین طوسی | استادیار | دکتر زهرا قطان کاشانی | استاد مشاور | ۲ |
| | علم و صنعت ایران | استاد | دکتر همایون عریضی | استاد مدعو خارجی | ۳ |
| | تهران | استادیار | دکتر محمد نشاط | استاد مدعو خارجی | ۴ |
| | خواجه نصیرالدین طوسی | استاد | دکتر محمدصادق ابریشمان | استاد مدعو داخلی | ۵ |
| | خواجه نصیرالدین طوسی | استادیار | دکتر سید آرش احمدی | استاد مدعو داخلی | ۶ |

<div dir="rtl">

**تائیدیه صحت و اصالت نتایج**

بسمه تعالی

اینجانب سیدمحمدرضا رضوی زاده به شماره دانشجویی ۹۳۰۰۵۷۶ دانشجوی رشته مهندسی برق گرایش مخابرات میدان مقطع تحصیلی دکتری تخصصی تایید می‌نمایم که کلیه نتایج این رساله حاصل کار اینجانب و بدون هرگونه دخل و تصرف است و موارد نسخه‌برداری شده از آثار دیگران را با ذکر کامل مشخصات منبع ذکر کرده‌ام. در صورت اثبات خلاف مندرجات فوق، به تشخیص دانشگاه مطابق با ضوابط و مقررات حاکم (قانون حمایت از حقوق مولفان و مصنفان و قانون ترجمه و تکثیر کتب و نشریات و آثار صوتی، ضوابط و مقررات آموزشی، پژوهشی و انضباطی ...) با اینجانب رفتار خواهد شد و حق هرگونه اعتراض درخصوص احقاق حقوق مکتسب و تشخیص و تعیین تخلف و مجازات را از خویش سلب می‌نمایم. در ضمن، مسئولیت هرگونه پاسخگویی به اشخاص اعم از حقیقی و حقوقی و مراجع ذی‌صلاح، اعم از اداری و قضایی، به عهده اینجانب خواهد بود و دانشگاه هیچ مسئولیتی در این خصوص نخواهد داشت.

نام و نام خانوادگی: سیدمحمدرضا رضوی زاده

تاریخ: ۳۱ / ۰۶ / ۱۳۹۹

امضا:

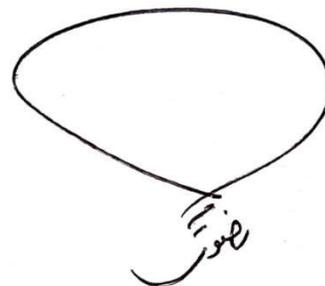

</div>

# مجوز بهره‌برداری از پایان‌نامه

بهره‌برداری از این پایان‌نامه در چهارچوب مقررات کتابخانه و با توجه به محدودیتی که توسط استاد راهنما به شرح زیر تعیین می‌شود بلامانع است:

☐ بهره‌برداری ازین پایان‌نامه برای همگان بلامانع است.

☑ بهره‌برداری ازین پایان‌نامه با اخذ مجوز از استاد راهنما، بلامانع است.

☐ بهره‌برداری ازین پایان‌نامه تا تاریخ ................................... ممنوع است.

نام استاد راهنما: رمضانعلی صادق زاده

تاریخ:   ۳۱ / ۰۶ / ۱۳۹۹

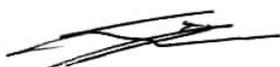

امضا:

# ACKNOWLEDGEMENTS

I would like to thank my wife first, as without her love, care, and constant support I wouldn't be here. I am greatly thankful to my advisers, Professor Ramezanali Sadeghzadeh and Dr. Zahra Ghattan, for their insight, guidance, and patience during the course of this research project. I also thank them for the encouragement, helpful advice, and for teaching me all of the terahertz technology that I know today. In addition, I would like to express sincere appreciation to Professor Miguel Navarro-Cia for his key comments and tips regarding our publications.

It was a great pleasure to my former graduate teachers, who were around at the time when I started: Professor Mohammad Sadegh Abrishamian and Dr. Arash Ahmadi  for sharing their invaluable experimental tips, and their guidance. I would also like to thank Dr. Oriezi, , and Dr. Neshat for being on my defense committee. I am thankful to Mrs Eeieni, and the staff at the K. N. Toosi University for their assistance in all the general matters.

I would like to extend many heart-felt thanks to my family, to my parents, and especially to my wife who was the primary motivational factor in achieving the Doctor of Philosophy degree.

# ABSTRACT

Terahertz short pulses have been extensively employed in many frontier modern applications, where the transmitted signals are usually frequency-chirped or phase-modulated to increase the time-bandwidth product, leading to an increased propagation length. Thus, a key component for communication systems, the pulse-shaper, remains elusive outside the lab. Frequently, temporally short pulses are realized with linear dispersion or nonlinear dispersion compression using external dispersive elements that have large chromatic dispersion. This function might be performed by a variety of photonic components such as gratings, prisms, chirped-mirrors and gas-filled hollow-core fibers. Nevertheless, they lack tunability, which is crucial for communication applications, and maybe difficult to integrate with commercial systems. A promising solution may rely on low-loss dispersion-tunable waveguides, wherein graphene, with excellent tunability at THz frequencies is exploited. Such a waveguide can be realized with helically-corrugated circular waveguides which have been successfully used for "gyro" applications, whose helical corrugation is a hybrid metal-graphene ribbon that allows electrical tunability. In this thesis, the possibility of the linear tunable compression of chirped pulses in the positive group velocity dispersion (GVD) region of a dielectric-lined circular waveguide loaded with a helical graphene ribbon has been discussed. We will show that the proposed structure introduces a good tunability of the compression factor via the graphene electrostatic bias. In addition, this thesis also presents that the waveguide dispersion, as the first component of the chromatic dispersion compared to the second component, i.e. material dispersion that introduced by graphene, plays the main role which is caused by the shape the graphene-ribbon helix. It is demonstrated that there is an optimal compression waveguide length over which THz chirped pulses reach the maximum compression. It is shown that by applying an electrostatic controlling gate voltage ($V_g$) of 0 and 30 V on the helical graphene ribbon, the temporal input pulses of width 8 and 12 ps, propagating through two different lengths (700 μm and 1700 μm), can be tuned by 5.9% and 8%, respectively, in the frequency range of 2.15–2.28 THz. Another outstanding achievement of this research was to provide a comprehensive system model by incorporating the full-wave time-domain simulations and the numerical transfer function estimation approach. The use of the system transfer function to analyze the structure is preferable to the full-wave simulation because of saving the execution time.

# TABLE OF CONTENTS

# LIST OF FIGURES

xi

# LIST OF TABELS

# LIST OF ABBREVIATIONS

**Acronyms / Abbreviations**

| | |
|---|---|
| A | absorption |
| Au | Gold |
| CVD | chemical vapor deposition |
| D | dispersion coefficient |
| DL | dielectric lined |
| FDTD | Finite Difference Time Domain |
| FEM | finite element method |
| FIT | finite integral technique |
| FWHM | Full Width at Half Maximum |
| GaAs | Gallium Arsenide |
| GD | group delay |
| GV | group velocity |
| GVD | group velocity dispersion |
| Gyro | gyrotron |
| HCW | hollow-core waveguide |
| HDPE | high density polyethylene |
| HE | hybrid electric mode |
| HIS | high impedance surface |
| IR | infrared |
| LFM | linear frequency modulation |
| LTI | linear time invariant |
| PEC | perfect electric conductor |
| PMMA | poly methyl meth acrylate |
| QCL | quantum-cascade lasers |
| MW | microwave |
| RF | radio frequency |
| Si | silicon |
| SiC | silicon carbide |
| SiO2 | silicon dioxide |
| SLG | single layer grapgene |
| SWS | slow wave structure |
| R | reflection |
| T | transmission |
| T | temprature |
| TE | transverse electric |
| TIIS | terahertz interferometric imaging system |
| TWTA | traveling wave tube amplifier |
| THz | terahertz |
| THz-TDS | THz Time Domain Spectroscopy |

# LIST OF PHYSICAL SYMBOLS

| Symbol | Meaning | Typical SI units |
|---|---|---|
| $\varepsilon$ | Electric permittivity | F/m |
| $e$ | Electron energy ($1.602176634 \times 10^{-19}$) | eV |
| $\alpha$ | Absorption coefficient | m$^{-1}$ |
| $\beta$ | Axial phase constant | Rad/m |
| $\delta$ | Skin depth | M |
| $\omega$ | Angular frequency | Rad/s |
| $\pi$ | Pi number | - |
| $\sigma$ | Conductivity | S/m |
| $h$ | Plank constant ($6.62607015 \times 10^{-34}$) | J.s |
| $\hbar$ | Modified Plank constant ($h/\pi$) | J·s |
| $k_B$ | Boltzmann constant ($1.380\,649 \times 10^{-23}$) | JK$^{-1}$ |
| $\mu$ | Magnetic permeability | N·A$^{-2}$ |
| $\Gamma$ | Wave reflection coefficient | 1/s |
| $\gamma$ | phenomenological electron scattering rate | 1/s |
| $E_F$ | Fermi-level | eV |
| $\mu_C$ | Chemical potential | eV |
| $\mu$ | Chirp factor | Rad·s$^{-2}$ |
| $\tau$ | Relaxation time | S |
| $v_F$ | Fermi velocity | m/s |
| $\omega_p$ | Plasma angular frequency | Rad/s |
| $\omega_0$ | Resonance angular frequency | Rad/s |

# Chapter 1

# Introduction

This chapter provides the general background for specifics of pulse compression techniques, and graphene-based terahertz devices, motivating the work presented in the majority of this thesis.

## 1.1   Temporary Pulse Compression and Stretching Development

Pulse shaping, compression, and stretching is an important area of research in anywhere part of the electromagnetic spectrum of frequencies from microwave up to optical frequencies, with important applications for high-resolution imaging, linear accelerators, radar, and non-linear testing. Principles and methods of pulse compression are quite different, depending on its application [1]. Here we survey some of the major approaches that can be common between microwave and optical frequencies.

Numerous uses of THz radiation have been explored, including trace gas detection, medical diagnosis, security screening, and defect analysis in complex materials. Many THz applications rely on the use of broadband pulses for time-domain analysis and spectroscopic applications [2].

In recent times, short pulses have received much attention owing to their applications in various areas such as optical sampling systems, time-resolved spectroscopy, ultrafast physical processes. Although several techniques are being used for generating short pulses [3].

Pulse shaping, compression, and stretching is an important area of research in anywhere part of the electromagnetic spectrum of frequencies from microwave up to optical frequencies, with important applications for high-resolution imaging, linear accelerators, radar, and non-linear testing. Principles and methods of pulse compression are quite different depending on their application. Here we survey some of the major approaches that can be common between microwave and optical frequencies.

Numerous uses of THz radiation have been explored, including trace gas detection, medical diagnosis, security screening, and defect analysis in complex materials. Many THz applications rely on the use of broadband pulses for time-domain analysis and spectroscopic applications.

In recent times, short pulses have received much attention owing to their applications in various areas such as optical sampling systems, time-resolved spectroscopy, ultrafast physical processes. Although several techniques are being used for generating short pulses [3].

In radar applications, for good system performance, a transmitted waveform is desired that has 1) wide bandwidth for high range resolution and 2) long duration for the high-velocity resolution and high transmitted energy. In a pulse-compression system, a long pulse of duration T and bandwidth B (product of T and B greater than one) is transmitted. The ration of the duration of the long pulse to that of a short pulse is an important system parameter called the compression ratio [4].

Pulse compression ratio (CR) in radar is the ratio of the range resolution of an unmodulated pulse length T to that of the modulated pulse of the same length and bandwidth B $CR = \frac{c T/2}{c/2B} = BT$

The quantity BT is called the "time-bandwidth product" or "BT product" of the modulated pulse.

### 1.1.1 Microwave Pulse Compression Approaches

In modern radar systems, for increasing the radar range resolution and detection distance, where the transmitted signals are usually frequency-chirped or phase-modulated to increase the time-bandwidth product (TBWP), the microwave pulse compression has been extensively employed [5].

In a dispersive medium with the anomalous-group velocity dispersion (GVD), the group velocity of any wave propagating through it is dependent, descending on the frequency of the wave. Therefore, if a microwave pulse is produced in which the wave is swept from a frequency with a low group velocity to a frequency with a high group velocity, the tail of the pulse will move to overtake the front of the pulse, resulting in pulse shortening and a corresponding growth in amplitude if the losses are sufficiently small. One example of performing of this approach is using the dispersive delay line (DDL). The Bo Xiang, was proposed through his PhD thesis [6] an integrated DDL to provide dispersion to the chirped signal, which operates from 11 GHz to 15 GHz and is capable of compressing a 2 ns pulse into a 1 ns pulse.

The use of a metal waveguide with a helically corrugated wall, as a dispersive medium, was proposed as a sole way to generate a multi-megawatts short pulse in ultra-high range radar applications. This produces an ideal dispersion i.e. has a large change in group velocity with frequency in the operating region, which is a required condition for an efficient high-intensity microwave pulse compression. The input down-chirped pulse had a frequency sweep from 9.6 to 9.2 GHz, with an initial pulse duration of 67 ns and 5.7 kW power at the fundamental $TE_{11}$ operating mode of the waveguide. Experimental optimization resulted in a compressed pulse of 2.8 ns duration and 68 kW peak power, giving a peak power compression ratio of 12 at the compressor length of 2.08 m. The major advantages of this method, in such metal waveguide-based, are high power capability and convenient sizes [7]. Such a device is known as the passive pulse compressor.

The helically corrugated waveguide [8], was constructed from copper and had threefold right-handed corrugations (see Fig1-1), which coupled the right-handed circularly polarized $TE_{2,1}$ near-cutoff mode with the left-handed circularly polarized $TE_{1,1}$ for from cutoff mode. The resultant group velocity was calculated using the measured data from the VNA, and a conventional x-band TWT was used to amplify the radiation generated by the solid-state oscillators and mixers, up to a power between 5 and 8 kW, and the losses of the compressor were measured as a function of frequency (using the SNA), see Fig 1-2. The optimum frequency modulation of the quasi-rectangular input pulse (Fig. 3) was found from the analysis of the measured helical waveguide dispersion.

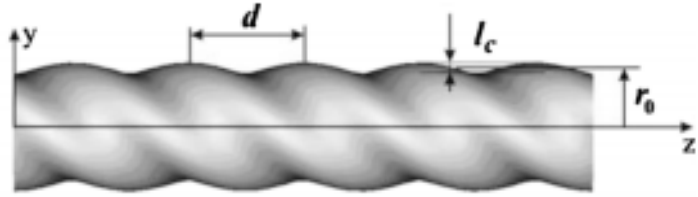

**FIG. 1-1.** Helically corrugated waveguide the surface equation in cylindrical (J. Appl. Phys. <u>108, 054908 (2010)</u>); doi: 10.1063/1.3482024).

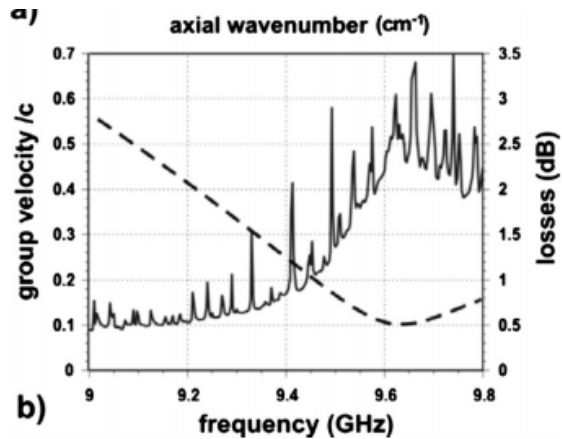

**FIG. 1-2.** Measured characteristics of helically corrugated waveguide used in the compression experiment: group velocity of the operating mode dashed line and losses solid line for the whole compressor (J. Appl. Phys. <u>108, 054908 (2010)</u>); doi: 10.1063/1.3482024).

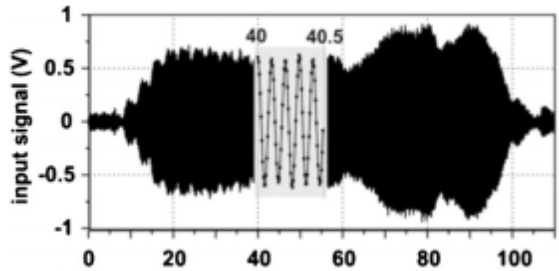

**FIG. 1-3.** Input pulse details for pulse compression experiment (J. Appl. Phys. <u>108, 054908 (2010)</u>); doi: 10.1063/1.3482024).

The sweep started from the frequency corresponding to the minimum value of the group velocity (approximately 10% the speed of light at 9.61 GHz) and extended down to the frequency of higher group velocity (approximately 55% the speed of light at 9.10GHz, in such a manner, that the inverse group velocity was linearly decreasing function of time over an input pulse whose duration was determined by the length of the waveguide.

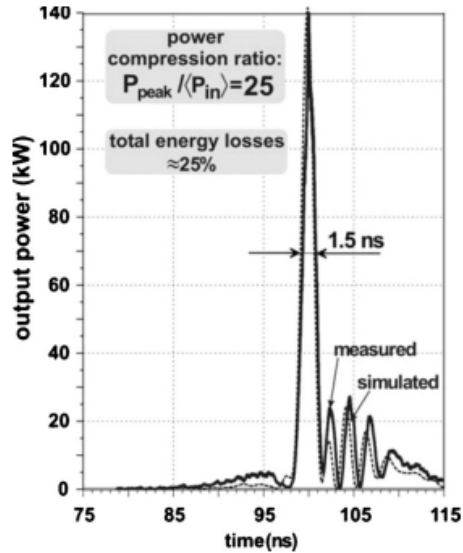

The maximum available (for this compressor) power compression ratio of 25 was achieved, with a 5.6 kW (averaged over the pulse), 80ns duration microwave pulse, compressed to a 1.5ns duration pulse with a peak power of 140 kW (Fig.4).

Approximately 50% of the input energy is compressed to the main body of the compressed pulse (assuming the main body as a Gaussian-shaped waveform with 1.5 ns full width half maximum), and approximately 25% of the energy is retained in the train of lower amplitude secondary pulse.

## 1.1.2 Optical Pulse Compression Approaches

Several techniques are used for pulse shaping and pulse compression based on infrared radiation into the visible spectral range by linear and non-linear optics methods [9]. This section gives a brief overview of a few linear and non-linear techniques used for pulse compression. A pulse linearly compressed when the input pulses are chirped, and their duration reduced by removing this chirp. De-chirping can be performed by passing the pulses through an optical element with a suitable amount of chromatic dispersion.

Methods of linear dispersion compression are then used to compress such pulses with optical elements having negative chromatic dispersion. A variety of optical components may perform this function, both discrete (diffraction or Bragg grating, prisms, chirped mirrors, etc.) and fiber-based [10]. Chirped mirror, pair of diffraction grating [11], fiber Bragg grating, and prism pair [12], are the most common linearly dispersive optical elements [13]. Pulse compression in fiber was first investigated by Mollenauer et al. [3]. Manimgalai and et al., have identified the design of photonic crystal fibers (PCFs) capable of providing high-quality compressed pulses at 1550nm. They design the PCF to get a maximum possible dispersion, which helps realize a compact compressor as the corresponding dispersion length becomes short. Figure 1-5 shows the 3D picture of the proposed Tapered-PCF.

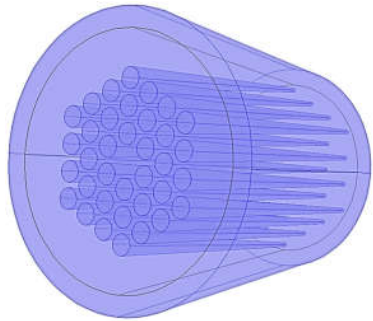

**FIG. 1-5 3-D picture of the proposed single-mode TPCF**.

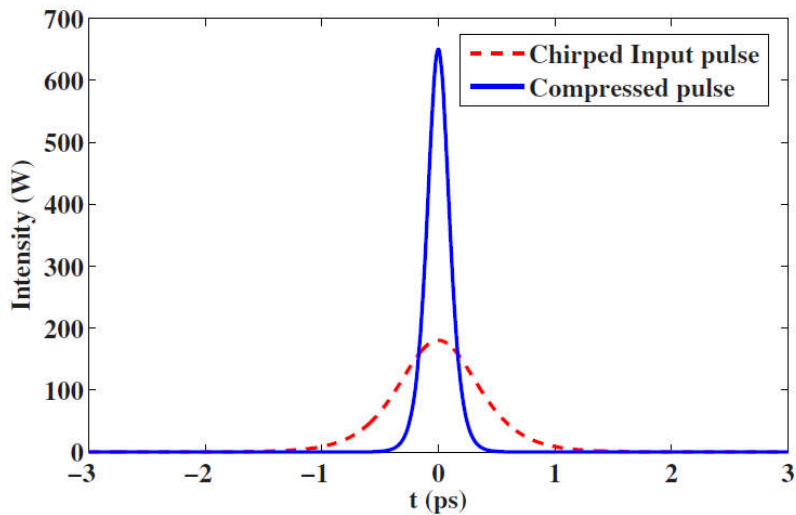

**FIG. 1-6 The input and output pulse of TPCF at 1550 nm.**

For both increasing dispersion and nonlinearity profiles, which are obtained by arbitrarily vary the design parameters, including air-hole diameter and pitch. Fig. 1-6 shows the compression of the chirped self-similar pulse at 1550 nm [3].

In a non-linear pulse compressor, first of all, the optical bandwidth typically with a non-linear interaction such as self-phase modulation is increased. Spectral broadening of the optical pulses can be performed via self-phase modulation using noble-gas-filled hollow fibers. Subsequent compression with chirped mirrors shortens the pulses by more than a factor 10.[14].

Generally, non-linear pulse compression depends on the interplay between the self-phase modulation and group-velocity dispersion (GVD) [3]. In most cases, this results in shortened chirped pulses in duration in compared with the original input pulses. Moreover, the pulse duration can be sharply reduced by linear compression, which removes or at least decreases the chirp. The single-mode laser can have a very narrow optical frequency spectrum. However, short pulse generation requires a broad optical bandwidth and hence multiple longitudinal mode operation. Methods for forcing a great number of modes to oscillate to obtain broad bandwidths and characteristics of a laser, which we assume already have the multimode operation. To generate ultrashort optical pulses, a mode-locked laser is needed via incorporation either active or non-linear pulse-forming elements (modulators) into the laser cavity. Passive mode-locking refers to situations in which the pulse forms its own modulation through nonlinearities; both amplitude and phase nonlinearities can be important. Active mode-locking is achieved with the aid of an externally driven intracavity loss modulator. During the early 1990s, passively mode-locked dye lasers were largely supplanted by passively mode-locked solid-state lasers. Today, such lasers produce femtosecond pulses at a variety of wavelengths, with pulses down to roughly 5fs [15] and with average powers up to several watts. This solid-state laser mode-locking is usually achieved using artificial saturable absorbers based on the non-linear refractive index, also known as the optical Kerr effect. The non-linear refractive index is usually written

$$n = n_0 + n_2 I(t) \tag{1.1}$$

Where *I(t)* is the pulse intensity.

The optical Kerr effect is allowed for gases, liquids, and solids. Furthermore, the non-linear refractive index also leads to new effects due to self-phase modulation (SPM). For a medium with $n_2 > 0$, SPM gives rise to lower frequencies (red shift) on the front edge of the pulse and higher frequencies (blue shift) on the trailing edge. This variation of instantaneous frequency as a function of time is called a chirp, in this case e.g. an up-chirp, means that this instantaneous frequency raises with time, while a down-chirp means that the instantaneous frequency decreases with time. [16].

Quasi-phase-matching (QPM) is a technique for compensating phase-velocity mismatch through periodic inversion of the non-linear coefficient. Chirped QPM structures have been used for the demonstration of non-linear crystals with engineered phase responses suitable for pulse compression and dispersion management [16]. Additionally, longitudinally-patterned QPM devices have enabled the demonstration of pulse compression during the second harmonic generation (SHG) [17]. The hollow-core fiber filled with a noble gas, can spectrally broaden high-energy input pulses by non-linear interaction with a noble gas of adjustable gas pressure inside a hollow fiber [18]. Terahertz (THz) spectroscopy is gaining increasing interest thanks to its several potential applications. THz pulse shaping has been addressed in literature by manipulating the generation with photoconductive antennas and periodically poled Lithium Niobate. In a recent report, Sato et al., demonstrated strong control on the generated THz pulse using shaping the pulse generation. Pulse shaping has also been achieved through linear filtering of a freely propagating THz in masks and waveguide [19]. M. Shalaby and et al. [20], presented a tunable THz pulse shaping technique operating in the time domain, capable of tailoring the temporal and spectral wave contents. Their technique operates via non-linear excitation of free carriers in semiconductors using the optical pump-THz probe technique [21].

## 1.2 Fixed and Tunable Graphene-based THz Devices

### 1.2.1 Graphene

Graphene was discovered in 2004 by Prof Andre Geim and Prof Kostya Novoselov, who were awarded the 2010 Nobel Prize in physics for their pioneering research on graphene at the University of Manchester. Graphene is a one-atom-thick layer of carbon atoms arranged in a hexagonal lattice. It has attracted remarkable attention in the field of fundamental science and applied research because of its distinctive properties such as high electrical conductivity, high thermal conductivity, good mechanical properties, huge surface area, excellent chemical properties etc. Another fascinating property of graphene is the possibility of controlling its electrical conductivity by an external electrostatic or magnetostatic field. This property undoubtedly opens up a new route to design a large range of tunable electronic and optical devices. Nowadays, many tunable graphene-based devices in terahertz science and technology have already been successfully fabricated and characterized, including absorber, phase-shifter, modulator and etc.

### 1.2.2 EM characterization of graphene

Figure 1-7 depicts infinite graphene lying in the xy-plane at the interface between two different mediums generally characterized by $\mu_1$ ,$\varepsilon_1$ for z≥0 and $\mu_2$ ,$\varepsilon_2$ for z<0 , respectively. From an electromagnetic point of view, graphene can be modeled in two regimes of the linear and non-linear regime. The linear model of graphene is characterized based on the Kubo formula. In the linear or small-signal model, the graphene sheet is modeled as an infinitesimally-thin metal like, by a 2D tensor surface conductivity of $\sigma = \sigma_x a_x + \sigma_y a_y$. This conductivity is a multi-variable dependent function as below:

$$\hat{\sigma}(\omega(rad/s), \mu_C(E_o), B_o, T) = \sigma_x \hat{a}_x + \sigma_y \hat{a}_y \qquad (1.2)$$

where ω is radian frequency, $\mu_C$ is the chemical potential [or Fermi level which can be controlled by an applied electrostatic bias field $E_0 = E_0 a_z$, or by doping, γ is a phenomenological electron scattering scattering rate that is assumed to be independent of energy, T is temperature, and $B_0$ is an applied magnetostatic bias field. Generally, based on bias applied on a graphene sheet, three cases of Eq.(1.2) be considered. Spatial dispersion, neither electrostatic nor magnetostatic bias ($E_0=B_0=0$), is applied on the graphene. Electrostatic bias, no magnetostatic bias, nor spatial dispersion ($E_0 \neq 0$, $B_0=0$). In this case, the conductivity matrix is diagonal:

$$\begin{cases} \sigma_{xx} = \sigma_{yy} = \sigma_d(\mu_C(E_0)) \\ \quad\quad \sigma_{xy} = \sigma_{yx} = 0 \end{cases} \tag{1.3}$$

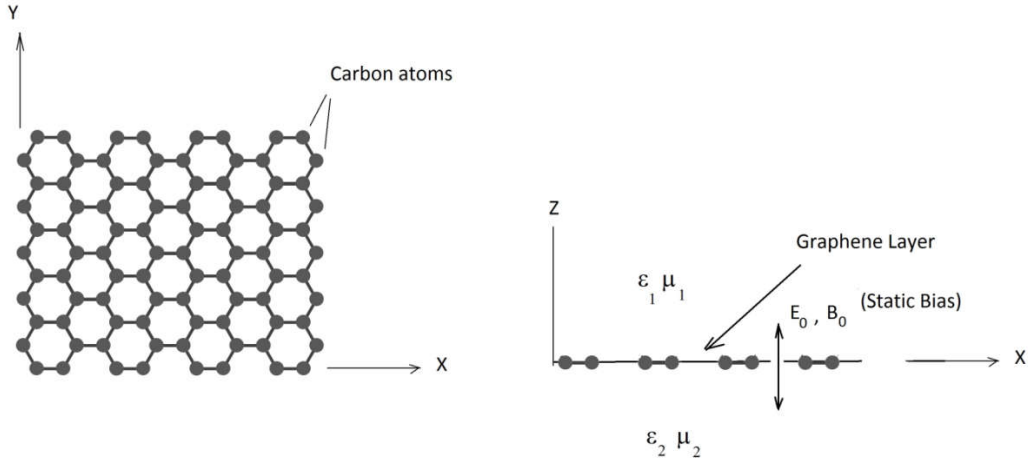

**FIG. 1-7** (a) The 2D model of graphene lattice structure, and (b) external perpendicular electric and magnetic fields applied on it as the controlling static bias.

When the graphene sheet is applied by a magnetostatic external field ($B_0 \neq 0$), and possibly electrostatic bias ($E_0 \neq 0$), this case refers to the local Hall Effect regime. In chapter 4 we describe the modeling of graphene in detail.

### 1.2.3  Graphene based circuit/components

The tunability of the graphene has attracted tremendous research interests in the realization of tunable terahertz-graphene-based devices include plasmonic filter [22], waveguide [23] antenna related [24-26], metamaterial [27] and modulators [28] in recent years. Figure 1-8 shows a new beam reconfigurable antenna is proposed for THz application based on switchable graphene high impedance surface (HIS) [24]. The graphene-based HIS has a switchable reflection characteristic, due to the voltage-controlled surface conductivity of graphene.

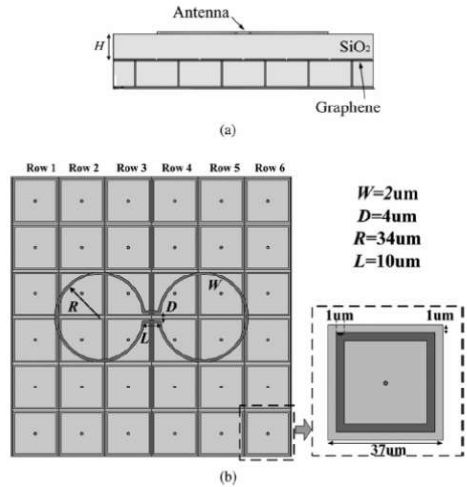

**FIG.1-8** Schematic of the antenna with HIS: (a) cross-sectional view and (b) top view (Fig.3 of [24]).

Based on this unique property, a THz antenna with the frequency of interest around 0.860THz is designed over the HIS. The proposed antenna over the switchable graphene-based HIS has a relatively stable efficiency for different cases and the beam reconfiguration is achieved with a deflection range of ±30 deg.

A graphene-based reflect-array was proposed in [26] to realize a DC-controlled reconfigurable-beam antenna array (Fig. 1-9). The controlling DC voltage applied between the graphene patch elements and the back metal electrode, the polycrystalline silicon layer allows tuning the patch reactance, and thus the reflection phase of each unit cell of the graphene-based reflectarray.

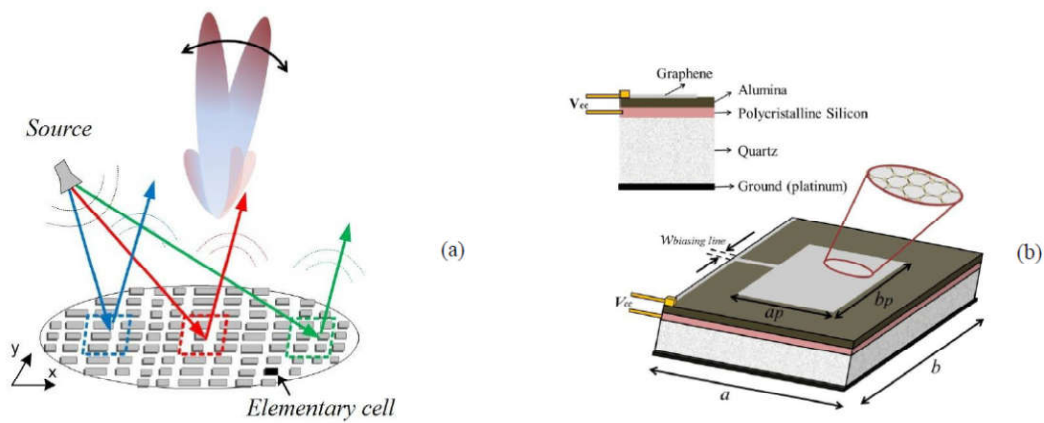

**FIG. 1-9** (a) General reflectarray concept, and (b) Graphene reflective cell.

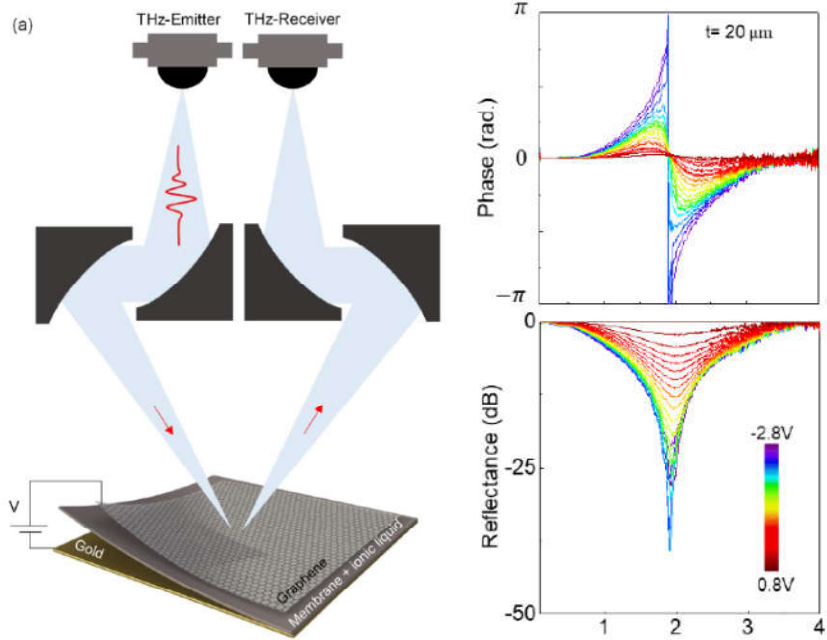

**FIG. 1-10** (a) Experimental setup used for the terahertz time-domain reflection measurements. (b) Variation of phase (radians) and intensity (dB) of the THz reflection of the devices with different DC-controlling voltages.

An efficient terahertz phase and amplitude modulation using electrically tunable graphene devices is presented in [28]. The device structure consists of electrolyte-gated graphene placed at quarter wavelength distance from a reflecting metallic surface. As shown in Fig. 1-10, in this geometry graphene operates as a tunable impedance surface, which yields an electrically controlled reflection phase. Terahertz time-domain reflection spectroscopy reveals the voltage-controlled phase modulation of π and the reflection modulation of 50 dB.

## 1.2.4 Motivation for Tunable Terahertz Pulse compression

Terahertz frequency range (0.1-10 THz) has received considerable attention in recent decades, due to its huge bandwidth and broad applications. High-field terahertz radiation is important to probe for a variety of fundamental scientific investigations and a wide range of applications.

The [29] summarized the two decades of advances of THz imaging, and how it is worth noting the role of THz time-domain spectroscopy (THz-TDS) as a versatile technique for medical and industrial imaging.

Terahertz radar has recently attracted great interest from both the academic and the industrial societies due to its applications of high center frequency, as well as large bandwidth, which possesses high penetration, and anti-stealth capabilities to support high-resolution imaging, under cloth detection, and so on [30]. A number of terahertz short pulse generation systems have recently been established based on an optical approach, which typically needs expensive and bulky setup and high energy laser pulse [31]. In digital data transmission, pulse broadening is a limitation on the maximum achievable data rate due to inter symbol interference [32]. Terahertz time-domain spectroscopy (THz-TDS) is a spectroscopic technique in which the properties of matter are probed with short pulses of terahertz radiation. The high-energy THz short pulse generation is vital for application in non-linear THz optics and spectroscopy.

Currently, intense THz radiation pulses exceeding tens of micro-Joules can be obtained from large accelerator facilities such as linear accelerators, synchrotrons, and free-electron lasers. However, due to the high cost of building and operating those facilities and limited access, there is a present and growing demand for high-energy, compact THz sources at a tabletop scale.

Generally, in order to measure an event in the time domain, a shorter one has been needed. If a shorter reference pulse is available, then it can be used to measure the unknown pulse, such method is named as time-domain interferometry. The terahertz interferometric imaging systems (TIIS) is used to examine THz imaging of biological tissue using terahertz pulses [33-34].

These applications provide strong motivation to advance the state of the art terahertz graphene-based technology for DC-controlled pulse compression.

In this thesis, we explore for the first time numerically possibilities and limitations of linear compression of chirped pulses in the positive dispersion region of a helical graphene ribbon-loaded hollow-core waveguide. Interest in this type of temporal pulse compression stems from this fact that a graphene-based linear compressor complies with the concept of all waveguide configurations and may be simply integrated with a THz chirped pulse generator.

## 1.2.5  Thesis Outline

This thesis starts with an overview of the pulse compression development in the microwave and optical frequencies, using the graphene in tunable optical and terahertz devices and a brief study of the dielectric and conductivity modeling of graphene in optics and terahertz region are given. In Chapter 2, a literature review in passive terahertz pulse compression has been presented.

Chapters 3 introduce important concepts and fundamental theory in pulse compression based on dispersion controlling techniques. In Chapter 4, we discuss how the tunable chromatic dispersive properties of a helical graphene ribbon can be exploited in a compact waveguide component to generate a DC-controlled compressed pulse. Validation of numerical analyses have been provided in Chapter 5 based on introducing a fixed terahertz pulse compressor with replacing the helical graphene ribbon with gold one, with two different numerical methods of Finite integral techniques (FIT) and Finite elements method (FEM). Moreover, a system identification transfer function for the proposed fixed terahertz pulse compressor is presented in Chapter 5. A comparative study on how the material of helical ribbon can affect the pulse compressor performance between non-biased graphene, Perfect electric conductor (PEC) and gold, as illustrated in Chapter 6. Chapter 7 concludes this thesis, with a summary of work done and contributions, along with possible future work related to this thesis.

# Chapter 2

# Literature Review

## 2.1 Terahertz Pulse Compressors

Nowadays, broadband THz pulses have found a wide range of applications, from imaging to communications. Although in high-speed broadband systems, dispersion is one of the crucial factors in the performance degradation of microwave and optical transmission media [35], it could be a beneficial characteristic in many specific applications. Nonlinear optical phenomena such as electro-optic effect and optical rectification are attractive options for the short pulse generation and detection of terahertz radiation due to the dispersion management [2, 35-37].

With the proliferation of advanced solid-state laser technology, methods based on nonlinear optical frequency conversion of high power infrared or far-infrared laser radiation laser have gained ground. Certainly, they have produced the best optical-to-terahertz conversion efficiencies and peak electric fields to date [2, 36-39].Besides, laser-driven approaches offer precise synchronization possibilities, which are instrumental for spectroscopic experiments. The significant problem of such intense laser-based THz sources is bulky set-up and necessary to cryogenic temperature.In contrast to the optical and near-infrared (NIR) domain where ultrashort pulse generation can be readily achieved in devices such as mode-locked semiconductor diodes and vertical external cavity surface emitting lasers, to generate ultrashort pulses, the following components are required i) a gain medium within a laser cavity; ii) a mode-locking mechanism such as the fast modulation of the losses or gain at the cavity round-trip and iii) dispersion compensation. Nevertheless, a table-top, compact and tunable source of THz short pulses is most interested by researchers for different applications ranging from spectroscopy to imaging. In Ref. [40], a monolithic on-chip compensation scheme is realized for a mode-locked QCL, permitting THz pulses to be considerably shortened from 16ps to 4ps.

In this thesis, it was demonstrated that the tunable chromatic dispersive properties of a helical graphene ribbon could be exploited in a compact waveguide component to generate a DC-controlled compressed pulse. To the best of our knowledge, there are a few published works on engineered dispersion devices made possible passively temporal THz pulse shortening. Moreover, no other studies have examined the combination of graphene and pulse-compression techniques in the THz regime. It is helpful, however, to review some relevant literature to quantitatively justify our work with different passive pulse compression using dispersion in THz-gap and both lower and upper-frequency spectrums, microwave and optics, respectively.

## 2.1.1 Approach(THz): Temporal delay/advancement when a pulse passes through the aperture

Short electromagnetic pulses experience significant spectral and temporal deformation when they are diffracted on subwavelength apertures. Temporal delay/advancement is one of the effects that occur when a pulse passes through the aperture. Mitrofanov and et al. [41], were demonstrated that the intrinsic negative chirp of terahertz pulses is the origin of the temporal advancement in the limit that the aperture is much smaller than the wavelength of the pulse. The advancement is shown to disappear for unchipped terahertz pulses. Based on this phenomenon, they proved that a circular aperture in a gold-film GaAs could be considered as a passive THz pulse compressor.

The chirped pulses are naturally generated using 100 fs pulses from the Ti:sapphire laser, which excite a biased photoconductive antenna fabricated on the low-temperature grown GaAs. The same pulses are used to gate the detecting antenna. The spectrum of the detected pulse peaks at 0.5 THz with a bandwidth of 0.57 THz and the pulse duration is 2.0 ps. After transmission through a 10 µm circular aperture, the center of the chirped pulse shifts forward by 0.3 ps, as shown in Fig. 1. The trailing edge of the pulse is suppressed, which results in pulse compression from 2.0 ps to 1.4 ps with FWHM of 0.07 THz. In such a passive pulse compression system, the bandwidth of the input pulse increases to 0.64 THz.

One of the significant drawbacks of this structure is the extremely low output to input peak intensity ratio of $1.4 \times 10^{-4}$, as shown in <span style="color:red">Fig. 2-1</span>.

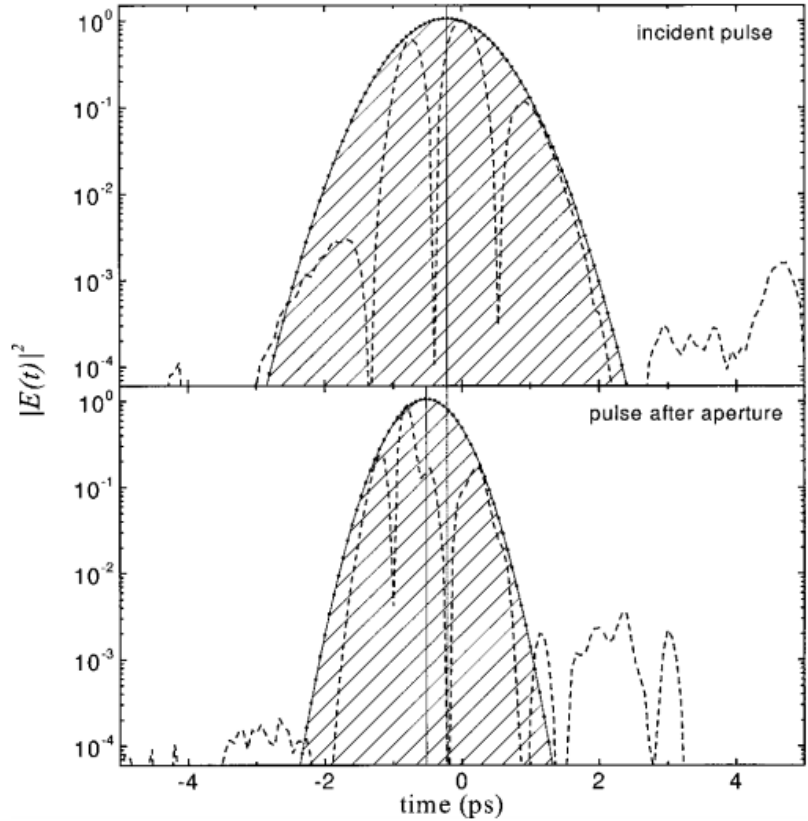

**<span style="color:red">FIG. 2-1</span>.Temporal deformation of the intrinsically chirped THz pulse due to transmission through the 10 mm aperture. The dashed lines indicate squared, normalized electric-field wave functions, and the shaded area shows the Gaussian envelopes of the pulses. Ratio of the peak intensity of the transmitted pulse to that of the incident pulse is 1.4×10⁻⁴ (<span style="color:red">Fig. 3</span> from [<span style="color:red">41</span>]).**

### 2.1.2 Approach (Microwave):

In a series of publications [<span style="color:red">1</span>, <span style="color:red">42</span>], the well-known phenomenon of passive pulse compression reduction in duration accompanied by an increase in amplitude, widely used in the microwave. The method is based on the generation of a frequency modulated FM pulse by a relativistic backward-wave oscillator (RBWO) following its compression due to propagation through a dispersive media (DM) in the form of a hollow metallic waveguide with helical corrugation of the inner surface. In [42], Bratman and et al, present the pulse compression experiment in which multigigawatt peak power of X-band radiation has been achieved. The experimental setup comprises two key elements Fig. 2-2: the source of FM radiation, an RBWO, and the DM a helical-waveguide (HW) compressor.

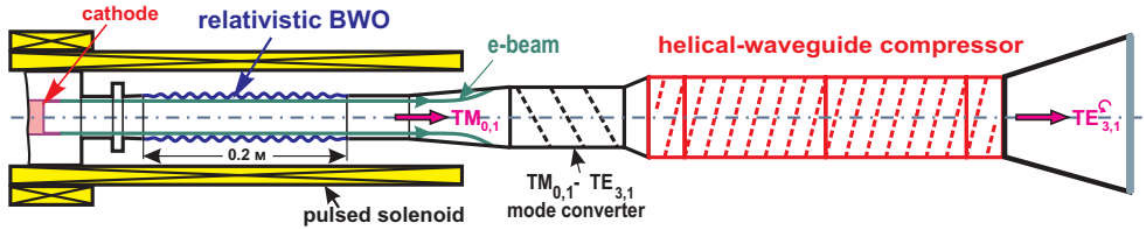

**FIG. 2-2. Scheme of the experimental setup [42].**

The most favorable frequency region for the pulse compression is a part of the dispersion characteristic where the operating wave normalized group velocity ($V_{gr}/c$) has a large negative gradient as a function of frequency (where $c$ denotes the speed of light in free space) which, for the HW under consideration, is the region from 9.4 to 9.95 GHz Fig. 2-3(a).

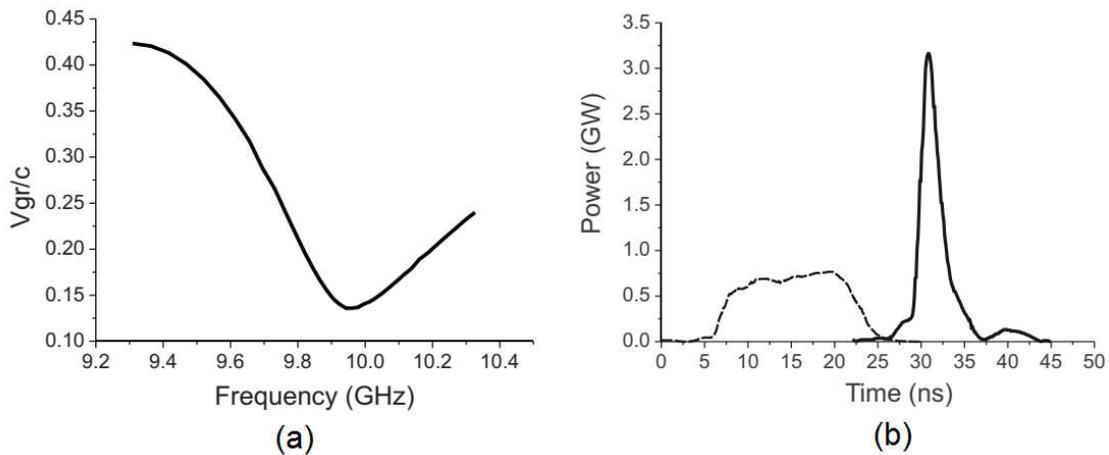

**FIG. 2-3. (a) Group velocity normalized to the speed of light of the helical waveguide operating wave(Fig. 2 from [18]), and (b) the resultant compressed pulse solid curve shown in contrast with the relativistic backward-wave oscillator output pulse dashed curve (Fig. 4 from [42]).**

Therefore, to be effectively compressed, an input pulse should have a negative frequency sweep within the mentioned frequency interval. As a result, the maximum peak power of the compressed pulse was measured to be as high as 3.2 GW while its Full width at half maximum (FWHM) amounted to 2.2 ns due to ~15 ns frequency chirped input pulse ( see Fig. 2-3(b)).

### 2.1.3 Approach (Optics-region):

In [43], Liu and et al., show how group velocity dispersion and enhanced nonlinear phase modulations at the band edges of sub-mm thick, transparent but highly dispersive cholesteric liquid crystals (CLC) can work in concert to compress picoseconds - the femtosecond laser pulses (see Fig. 2-4). This is an example ofdirect passive compression of ultra-fast laser pulses using a dispersive and nonlinear thick (550 μm) cholesteric liquid crystal sample, at optical wavelength. After passing the laser pulses across this material, a broadening of 1.8 THz around the center frequency of 387 THz (777 nm) has been achieved.

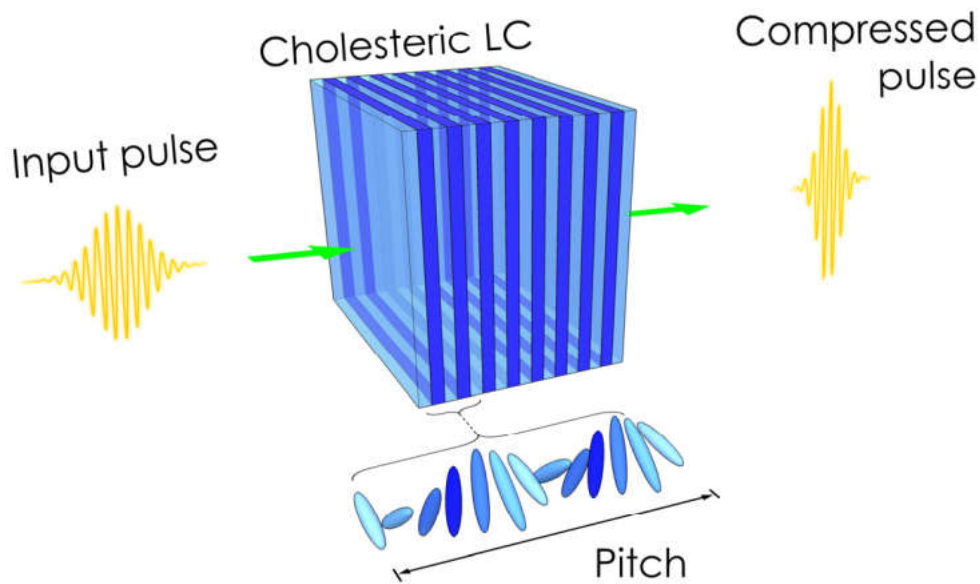

**FIG. 2-4.** Schematic depiction of a cholesteric liquid crystal cell (CLC) for femtoseconds laser pulse compression or stretching, caused by nonlinear phase modulation and dispersion at the band edge [43].

To design a terahertz pulse compression structure with simply electronically tunable pulse shaping, it would be found that graphene with the excellent dispersion characteristics tunability at THz frequencies [44-46] could be one of the best candidates. Furthermore, using a dielectric-lined hollow-core waveguide where loaded with a helical graphene ribbon based on large negative gradient profile of the normalized group velocity ($V_{gr}$/c) due to the waveguide and material dispersions introduced by helical geometry and graphene, respectively, (as the main requirement to realize a passive pulse compression aforementioned in microwave approach) was motivated us for getting the main idea to design a tunable graphene-based device in this dissertation.

In the next chapter we review the fundamental concepts and basic theory of dispersion and linear passive pulse compression.

# Chapter3

# Important Concepts and Basic Theory

As noted in Chapter 1, control of dispersion is key to obtaining short pulse in both RF and optical pulse shaping systems, including high power LFM microwave pulse compressor and femtosecond pulse generation. One of the most frequently used ways to generate ultra-short temporal pulses is linear dispersion compression by means of external dispersion elements [47]. Before discussing using dispersion control devices in a pulse compression application, it is useful briefly to review dispersion fundamentals and concepts.

1
2
## 3.1   Chromatic Dispersion

Dispersion and attenuation are the key propagation properties determining the ultimate performance of any transmission medium used in telecommunication applications. It is worthwhile reviewing the fundamental mechanisms giving rise to pulse dispersion in waveguides and fiber systems. Pulse spreading or pulse dispersion results from both waveguide and material properties of a fiber. Dispersive effects are most important for optical and terahertz fibers. We may note that shorter pulses are affected more strongly by dispersion than longer ones. Longer pulses have a narrower spectrum and because of the Fourier-transform relation: $\Delta f \cdot \Delta T$=constant, where $\Delta f$ is the spectral and $\Delta T$ is the transform-limited temporal width.

The important dispersion mechanisms are material dispersion, waveguide dispersion, nonlinear effects, and polarization mode dispersion, the two most important are material and waveguide dispersion when taken together, are referred to as chromatic dispersion.

## 3.1.1   Material Dispersion: Modeling of Dispersive Material in Terahertz

The dispersion in any single-mode waveguides originates from material dispersion and waveguide dispersion. Material dispersion arises from the frequency-dependent variation of the intrinsic material electrical and magnetic properties, whereas the waveguide dispersion part is due to the frequency-dependent power distribution in the non-homogenous cross-section of the waveguide.

In optical dielectric-based waveguide such as single-mode (SM) and multi-mode (MM) fiber, all-dielectric materials are dispersive, this means that the refractive index varies with wavelength, i.e., n=n($\lambda$). There are several ways to measure dispersion in transparent materials. A simple measure is the Abbe number, $V_D$. The Abbe number is obtained by measuring the index at several key wavelengths, or

$$V_D = \frac{n_D - 1}{n_F - n_c} \tag{3-1}$$

where $n_F$, $n_D$, and $n_c$, are the refractive indices at three standard visible wavelengths : $\lambda_F$=486.1 nm (blue), $\lambda_D$=589.2 nm (yellow), and $\lambda_c$=656.3 nm (red). High dispersive materials have low $V_D$, and materials with low dispersion introduce large $V_D$. Another measure of dispersion is the derivative dn/d$\lambda$, which may be most easily obtained from dispersion data like dispersion coefficients vs. wavelength, once a curve fits the data has been made to determine *n* as a function of wavelength. It is quite common to fit the dispersion data using a Sellmeier power-law fit. The Sellmeier equation is of the form

$$n^2 = 1 + \sum_i \frac{A_i \lambda^2}{\lambda^2 - \lambda_{0i}^2} \tag{3-2}$$

where $A_i$ and $\lambda_{oi}$ are the Sellmeier constants, $\lambda_{oi}$ is a vacuum wavelength related to natural vibrational frequencies $\lambda_o/v_{oi}$=c. The Sellmeier constants for many of the standard IR materials are given in the Handbook of Optical Constants Series by Palik [48].

The Drude model can be used to describe the dielectric function of a conducting medium as

$$\tilde{\varepsilon}(\omega) = \varepsilon_\infty - \frac{\omega_p^2}{\omega(\omega - j\Gamma)} \tag{3-3}$$

$\Gamma$ is the momentum relaxation time of the material and the plasma frequency, $\omega_p$, is defined as

$$\omega_p = \frac{Ne^2}{\varepsilon_0 m_*} \tag{3-4}$$

with m* being the effective mass.

### 3.1.2 Waveguide Dispersion

The waveguide dispersion affected by waveguide dimensions and geometry as well as the magnitude of material properties ($\varepsilon_r$ and $\mu_r$) is a strong function of frequency. This phenomenon can be observed in several of the examples presented in literature, such as ridged waveguide [49], photonic band-gap structure [50], corrugated wall waveguide [42], and grating surfaces [51].

## 3.2   Dispersion Analysis

Propagation phase constant when the pulse propagates through an unbounded medium with a real part of the complex refractive index of *n(ω)*, is commonly written in terms of its Taylor series expansion, around the center frequency $\omega_0$, as follows

$$\beta(\omega) = n(\omega)\frac{\omega}{c_0} = \beta(\omega_0) + \frac{\partial\beta}{\partial\omega}(\omega - \omega_0) + \frac{1}{2}\frac{\partial^2\beta}{\partial\omega^2}(\omega - \omega_0)^2 + \frac{1}{6}\frac{\partial^3\beta}{\partial\omega^3}(\omega - \omega_0)^3 + \cdots \text{(3-5)}$$

where $\beta^n = \frac{\partial^n\beta}{\partial\omega^n}$ , and all the derivatives are evaluated at $\omega = \omega_0$. If only the dielectric constant data are available, the *n(ω)* can be replaced with $\sqrt{\varepsilon_l(\omega)}$, where $\varepsilon_l(\omega)$ is the real part of complex electric permittivity.

### 3.2.1 Group Velocity

In a transmission system, the carrier travels at the phase velocity, whereas the envelope and also EM energy travel at the group velocity. Consider an input pulse $e_{in}(t)$ with spectrum $E_{in}(\omega)$. After passing through a dispersive system, the output pulse as follows

$$e_{in}(t) = \frac{1}{2\pi}\int d\omega \ E_{in}(\omega)e^{j\omega t}e^{j\beta(\omega)L}$$

(3-6)

Now if we introduce the typical notation for envelope function, where the positive frequency part of $E_{in}(\omega)$ is replaced by $A(\widetilde{\omega})$ and $\widetilde{\omega} = \omega - \omega_0$ , we can rewrite the output pulse as follows

$$e_{out}(t) = Re\{e^{j(\omega_0 t - \beta_0 L)}a_{out}(t)\} \qquad (3\text{-}7)$$

where

$$a_{out}(t) = \frac{1}{2\pi}\int d\omega \ A(\widetilde{\omega})e^{j[\omega t - (\beta_1 \widetilde{\omega} + (\beta_2/2)\widetilde{\omega}^2 + \dots)L]} \qquad (3\text{-}8)$$

Thus $e_{out}(t)$ is the product of a carrier term $e^{j(\omega_0 t - \beta_0 L)}$ and the envelope function $a_{out}(t)$ . The carrier term propagates at the phase velocity

$$v_p = \frac{\omega_0}{\beta_0} \qquad (3\text{-}9)$$

and is unaffected by the variation of $\beta$ with $\omega$. The envelope function $a_{out}(t)$ , however, it is affected by the form of $\beta(\omega)$. One significant case occurs when it is a linear function of $\omega$.

$$\beta(\omega) = \beta_0 + \beta_1 \widetilde{\omega} \qquad (3\text{-}10)$$

In this case, we find that

$$a_{out}(t) = a_{in}(t - \beta_1 L) \qquad (3\text{-}11)$$

Thus, the output pulse is an undistorted replica of the input pulse, which travels with a velocity $\beta_1^{-1}$. This velocity is called the group velocity $v_g$, where

$$v_g = \beta_1^{-1} = \left(\frac{\partial \beta}{\partial \omega}\Big|_{\omega=\omega_0}\right)^{-1} \qquad (3\text{-}12)$$

We now proceed to give a useful formula for the group velocity, specifically the material dispersion, encountered during propagation in a bulk medium. Starting with

$$\beta(\omega) = \frac{\omega n(\omega)}{c} = \frac{\omega \sqrt{\varepsilon_r(\omega)}}{c} \tag{3-13}$$

We find that

$$V_g = \frac{1}{d\beta/d\omega} = \frac{c}{n + \omega\left(\frac{dn}{d\omega}\right)} \tag{3-14}$$

It turns out that it is also useful to express vg in terms real part of complex permittivity of $\varepsilon_r(\omega)$

$$V_g = \frac{c}{\sqrt{\varepsilon_r} + \omega\left(\frac{d\sqrt{\varepsilon_r}}{d\omega}\right)} \tag{3-15}$$

### 3.2.2 Group Delay

We can also look at the concept of group delay (τ) using simple Fourier transform identities. The Fourier transform of a delayed pulse $a(t - \tau)$ is given by $A(\omega)e^{-j\omega\tau}$, so $\tau = -\frac{d\varphi}{dt} = -\frac{d(\beta L)}{dt}$, using this equation and eq.(4-8), we find that $\tau = \beta_1 L = L/V_g$, which isconsistent with our discussion above, where $L$ is an electromagnetic wave is traveling distance. As in all full-wave, electromagnetic solvers, the scattering parameters results of [S] are available after frequency domain calculations, we can simply compute group delay (GD) parameters via phase angle of insertion loss of S$_{21}$ (φ$_{21}$) using Eq. (3-16).

$$GD(sec.) = -\frac{1}{360}\frac{\partial\varphi_{21}(deg.)}{\partial f(Hz)} \tag{3-16}$$

It is noteworthy that the other required dispersion parameters such as group velocity (GV) and axial phase constant ($\beta$) can be calculated in postprocessing using above mentioned relation $\tau$ and $V_g$ incorporation with Eq. (3-16).

### 3.2.3 Group Velocity Dispersion

When $\beta(\omega)$ is not a linear function of $\omega$ [i.e., Eq.(3-10) does not hold], $a_{out}(t)$ is changed compared to $a_{in}(t)$. The part of $\beta(\omega)$ that is not linear in $\omega$ is called dispersion. The $\partial^2\beta/\partial\omega^2$ term contributes a quadratic spectral phase variation, which leads to a linear variation in delay with frequency; this imparts a linear chirp to the output pulse. The $\partial^3\beta/\partial\omega^3$ term contributes a cubic spectral phase leading to a quadratic variation in delay with frequency. This results in an asymmetric pulse distortion with an oscillatory or ringing on the tail part of the pulse.

Group velocity dispersion (GVD) is important in any fast and ultrashort optical systems when the path lengths are relatively large and the pulses are very short. In the case of nonzero GVD, different frequencies have slightly different traveling times, and this can be an important pulse broadening mechanism. Now we turn our attention to the quadratic dispersion term. In this case we can mathematically treat the dispersion by considering the frequency dependence of the propagation constant $\beta$, based on a second-order Taylor series expansion, as follow

$$\beta(\omega) = \beta(\omega_0) + \beta_1(\omega - \omega_0) + 0.5\beta_2(\omega - \omega_0) \tag{3-17}$$

For a particular frequency component, the time required to propagate through a length $L$ of a dispersive medium is $L/v_g(\omega)$. The propagation time relative to that corresponding to the frequency $\omega_0$ is given by

$$\tau(\omega) - \tau(\omega_0) = \Delta\tau(\omega) = \frac{\partial \tau}{\partial \omega_0}(\omega - \omega_0)$$
$$= \frac{\partial(v_g^{-1})}{\partial \omega}(\omega - \omega_0)L = \beta_2(\omega - \omega_0)L \tag{3-18}$$

Using Eq.(3.18), we obtain

$$\Delta\tau(\omega) = \left(2\frac{dn}{d\omega} + \omega\frac{d^2n}{d\omega^2}\right)\frac{(\omega - \omega_0)}{c}L \tag{3-19}$$

In some applications, notably fiber optics, it is customary to express $V_g$ and $\Delta\tau$ in terms of the wavelength.

Here we use

$$\lambda = \frac{2\pi c}{\omega} \ and \ \frac{d\lambda}{d\omega} = -\frac{2\pi c}{\omega^2} = \frac{-\lambda^2}{2\pi c} \tag{3-20}$$

The results using the chain rule are

$$v_g = \frac{c}{n - \lambda(dn/d\lambda)} \tag{3-21}$$

$$\Delta\tau(\lambda) = \frac{\partial(v_g^{-1})}{\partial\lambda}(\lambda - \lambda_0)L = \frac{-2\pi c\beta_2\Delta\lambda L}{\lambda^2} \tag{3-22}$$

Fiber dispersion is usually described in terms of a dispersion parameter D with units ps nm$^{-1}$ km$^{-1}$, defined by

$$\Delta\tau(\lambda) = \left(\frac{-2\pi c\beta_2}{\lambda^2}\right)\Delta\lambda L = D\Delta\lambda L \tag{3-23}$$

where

$$D = \frac{\partial(v_g^{-1})}{\partial\lambda} = \frac{-2\pi c\beta_2}{\lambda^2} = \frac{-\lambda d^2 n}{cd\lambda^2} \tag{3-24}$$

## 3.3   Normal and Anomalous Dispersion

In the classical optics terminology, normal and anomalous dispersion appears to refer specifically to the sign of $dn/d\lambda$. However, in the ultrafast optics community, one is usually more interested in temporal dispersion, which in the case of material dispersion is governed by $d^2n/d\lambda^2$. It is therefore customary to utilize the terms normal and anomalous material dispersion to mean $d^2n/d\lambda^2 > 0$ or $d^2n/d\lambda^2 < 0$, respectively. However, these can lead to confusion since they are used differently by different communities. Within ultrafast optics, these terms usually refer to $\partial\tau/\partial\omega$, whereas in fiber optics, they usually refer to D, which is proportional to $\partial\tau/\partial\lambda$ [15].

Refer to Eq.(3-18), generally, dispersion can be interpreted based on GVD. Thus, for $\beta_2 > 0$ (called normal dispersion), higher frequencies (shorter wavelengths) travel more slowly and are displaced toward the trailing edge of the pulse. This leads to an up-chirped. For $\beta_2 < 0$ (anomalous dispersion), higher frequencies move faster and are displaced to the leading edge of the pulse, leading to down-chirp.

## 3.4  Dispersion-based Pulse Compressor

Although in high-speed broadband systems, dispersion is usually considered as a negative effect, it could be a beneficial characteristic in many specific applications such as pulse shaping, including pulse stretcher or compressor. The use of waveguide operating close its cutoff frequency, as an approximation to a dispersive line, is one of the historical approaches in a passive generation a short pulse in high-resolution microwave radar systems. LFM pulse compression techniques based on using waveguide operating close its cutoff frequency, as an approximation to a dispersive line has been demonstrated in early radar systems. In such compression systems, the dispersive line being the vital component. R. A. Biomley and et al.[52], more than fifty years ago, showed how a 91.5 meter length short-circuited rectangular copper waveguide No. 11A, can be successfully used to realize a dispersive line for S-band pulse compression applications. For a pulse compressor based on conventional waveguide the operating frequencies should haven close (few percents above) to the cutoff [53], where the wave dispersion is sufficiently large, but one problem in close to the cutoff frequency is the higher attenuation losses. Phelps and et al. [54], show that with enhancing waveguide dispersion component of chromatic dispersion via creation a helical corrugation on a circular waveguide wallpermits the realization of the necessary dispersion for pulse compression without having to operate close to cutoff which makes it possible to use such a compressor at the output of a high-power amplifier.The most favorable region for pulse compression in the helically corrugated waveguide is the part of the dispersion characteristic, where the group velocity has a negative gradient as a function of frequency (see Fig.2-3a). Correspondingly, the input pulse should be a linear frequency-modulated (chirped), which has a negative slope(down-chirped) in contrast to the case when a smooth waveguide is used. In this case, all frequency components can be reached simultaneously at the end of the waveguide to make a short or compressed pulse.

### 3.4.1 Chirped Linear FM Pulse

The basic idea of pulse compression is based on Linear Frequency Modulation (LFM) which was invented by R. Dicke in 1945 for RADAR applications. The algorithm for LFM pulse compression involves mainly two steps which are: generation of LFM waveform and the management of the traveling time delay of the head and the tail of the square-shaped LFM pulse using a dispersive element.

An LFM signal (Fig. 3-1.) is a frequency modulated waveform in which carrier frequency is linearly swept over a specific period of time.

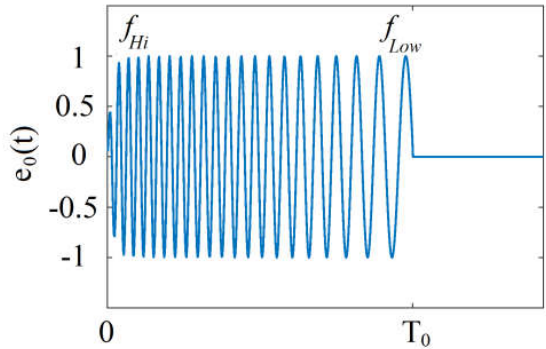

**FIG.3-1 – A time domain illustration of a down-chirped sinusoidal linear FM pulse.**

To investigate the pulse-compression performance of a passive linear dispersion-based pulse compressor with the group velocity, which has a negative gradient as a function of the frequency component, it is assumed that the input pulse is a chirped sine function, as stated in Eq. (3-25),

$$E = \begin{cases} E_0 \sin(2\pi f_{Hi} t - \mu t^2), & if \ t \le T_0 \\ 0, & other \end{cases} \qquad (3\text{-}25)$$

where $f_{Hi}$ is the upper-band frequency of the negative dispersion region, $T_0$ is the input pulse width, $E_0$ is the electric field amplitude, and $\mu$ is the chirp factor. The chirp factor for a uniform input pulse spectrum in the positive GVD region from $f_{Lo}$ to $f_{Hi}$, respectively, can be calculated using Eq. (3-26):

$$\mu = \frac{\pi(f_{Hi} - f_{Lo})}{T_0} \qquad (3\text{-}26)$$

A pulse compressor uses a system with a proper time delay or equivalently group velocity profile in terms of frequency (Fig.3-2a) to delay one end of the pulse relative to the other one and consequently, to produce a temporal narrower with the greater peak amplitude. This phenomenon can be simply presented by two lower and higher frequency edge components of the LFM waveform (Fig. 3-2b). This system could be a matched filter or a dispersion managed waveguide.

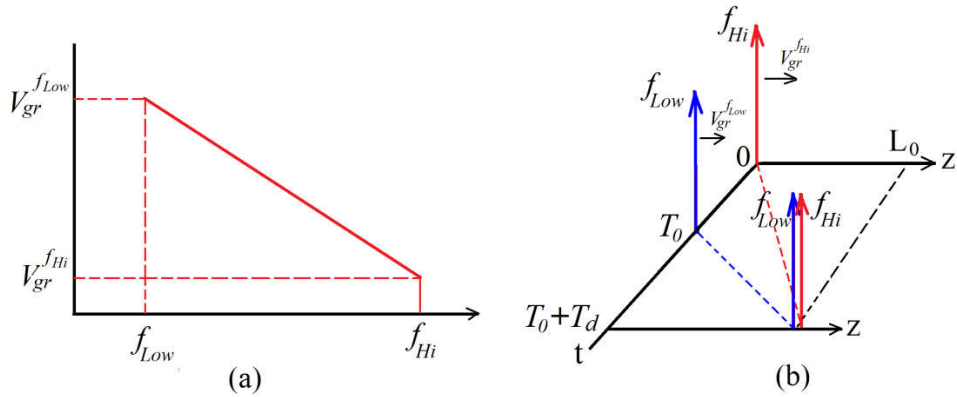

(a)                    (b)

**FIG. 3-2(a) group velocity profile with negative gradient variation in terms of frequency of a dispersion managed device and (b) a representation of how a down-chirped LFM waveform with duration of $T_0$ can produce a grater peak amplitude after traveling an optimal length of a the dispersive waveguide.**

Based on the normalized group velocity profile, shown in Fig.3-2(a), and the instantaneous frequency, which can be obtained from the time derivative of the phase argument of Eq. (3-25) proves that the tail of the chirped pulses with low-frequency components and faster group velocities will move to overtake the front high-frequency components of the pulses, resulting in temporal pulse shortening with growth in amplitude.

A unique optimal compression length can be easily achieved based on a significant value of the chirped LFM pulse duration ($T_0$) and the group velocity profile as follows:

$$T_0 = T_d(f_{Hi}) - T_d(f_{Lo}) \tag{3-27}$$

$$L_0 = V_{gr}(f_{Hi})T_d(f_{Hi}) = V_{gr}(f_{Lo})T_d(f_{Lo}) \tag{3-28}$$

By using the relation of $T_d(f_{Hi})$ given in (1), the optimal length for maximum compression is computed in terms of the LFM input pulse duration (T0) and the boarder values of group velocity values as follows:

$$T_d(f_{Lo}) = \frac{T_0}{\frac{v_{gr}(f_{Lo})}{v_{gr}(f_{Hi})}-1} \text{and } (3\text{-}28) \ \Rightarrow \ L_0 = \frac{v_{gr}(f_{Lo})}{\frac{v_{gr}(f_{Lo})}{v_{gr}(f_{Hi})}-1}T_0 \tag{3-29}$$

## 3.5   Broadening Factor

One of the important parameters in the pulse compression applications where the compressed pulse in the time domain is not symmetric is the broadening factor (*F*), which is defined as the ratio of full-width at half-maximum (FWHM) of the compressed pulse to the FWHM of the input pulse. The percentage of the spectral broadening change as a figure of merit of a tunable pulse compressor can be defined. This parameter is also known as the modulation depth defined by ($\Delta F /F_0$)×100%, where $\Delta F$ is the changes of *F* due to tuning factor, and $F_0$ is the average of the minimum and maximum obtainable *F* [55].

## 3.6   Attenuation

Absorption is caused by three different mechanisms: Impurities in material, intrinsic absorption, and radiation defects. The attenuation of an electromagnetic wave power flow through a lossy medium is followed an exponential form of *exp*(-$\alpha$L) where $\alpha$ is called the absorption coefficient, and L is the propagation distance. In general, the absorption coefficient ($\alpha$) can be related to the imaginary parts of the effective refractive index by $2k_0 Im(n_{e\,f\,f})$, where is effective refraction index of the medium or equivalent medium of a radiating system such as fiber or waveguide. In an equivalent radiating system, the conservation of energy leads to the statement that the sum of the transmission (*T*), reflection (*R*) ,and absorption (*A*) of electromagnetic power flow is equal to unity, i.e., *T* + *R* + *A* = 1 wherein the absence of nonlinear effect (i.e. Raman effect, etc), this equation is held for any angular frequency of $\omega$ as $T(\omega) + R(\omega) + A(\omega) = 1$ [56]. As the $T(\omega)$ and $R(\omega)$ are related to the square of the scattering parameters of $|S_{21}|$ and $|S_{11}|$, respectively, the $A(\omega)$ easily can be calculated.

## 3.7  LFM Generation

The generation of a high quality linear FM pulses for RADAR applications are commonly with frequency range centered at microwave and millimeter-wave bands from gigahertz to tens of gigahertz. In these frequency ranges the most common method to achieve LFM signals is using voltage controlled oscillator (VCO) circuits.

Recently, practical generations of LFM waveforms in sub-millimeter and terahertz bands have been reported based on some optical techniques such as optical interferometer and optoelectronic oscillator (OEO).

# Chapter 4

# Tunable Terahertz Pulse Compressor Design Based on Helical Graphene Ribbon

Novel ideas for tunable THz pulse compression using a graphene-based waveguide are proposed here. In this research, we explore for the first time numerically possibilities and limitations of linear compression of negatively chirped pulses in the positive group velocity dispersion (GVD) region of a dielectric-lined circular waveguide loaded with a helical graphene ribbon. We show that the proposed structure introduces a good tunability of the compression factor via the graphene electrostatic bias.

This chapter is dedicated to explaining the design details and results for terahertz's direct pulse compression in a dielectric-lined circular waveguide with tunability based on a strong both waveguide and material dispersions introduced by helical graphene-loading. A theoretical investigation backed by transient numerical analysis has led to a thorough understanding of the passive pulse compression principles and constraints.

3
## 4.1 Model Description and Design Principles

Microwave, infrared, and optical regions have mature technologies as a result of extensive development. However, in the THz band, i.e., from 0.3 to 3 THz, components, system architectures and experimental techniques are still emerging. In the quest to develop the THz band, THz waveguides have become an active research area [60]. Among the microwave and optical waveguide technologies, the use of optical fibers for THz application has attracted considerable attention in recent years [61]. However, conventional optical fibers such as silica glass ones are not suitable for guiding THz wave owing to their high transmission loss mainly coming from intrinsic absorption responses of core and clad dielectrics, scattering loss and low free-space THz coupling efficiency. Hollow-core waveguides (HCWs) due to the inherent advantage of their air-core have received a great deal of attention for THz applications. As mentioned in [62] at the IR region, HCWs are categorized based on the refractive index of the inner dielectric coating material into two groups, which are leaky guides if the refractive index is greater than one and the attenuated total reflectance(ATR) guides otherwise.

Recently, researchers have proposed, fabricated, and tested a variety of single-mode low-loss waveguides entitled "dielectric-lined hollow metallic waveguides (DL-HMWs)."DL-HMWs become a strong candidate for the effective single-mode terahertz propagation with possible transmission loss below 1dB/m [61], [63-65]. Since the goal of this context is to design a single-mode waveguide having low-loss with tunable dispersion, the DL-HMW is an appropriate candidate.

Another important element of the structure is the helical ribbon. Helix generally used in slow-wave structures (SWS) in high power microwave sources such as traveling wave tube (TWT) or Gyro which should have two crucial characteristics of low-loss and controlled dispersion in a wide operating bandwidth. In recent years, several reports can be found in the literature demonstrating the techniques to shape the dispersion characteristics of a helix RF circuit with acceptable electronic efficiency across the entire operating frequency band [66].

### 4.1.1 Model Description

Figure 1 shows the 3D geometry and side view of the proposed graphene ribbon loaded waveguide: a graphene ribbon with thickness $t_{gr}$, ribbon width $w$, and pitch size $p$ is helically wrapped and inserted inside a metalized dielectric tube of thickness $t_d$. By carefully choosing the geometric parameters, dielectric material, and DC-controlling bias, an optimum condition for an arbitrary dispersion profile, large mode confinement, minimum mismatch, and propagation loss with maximum tunability can be achieved. The length of the waveguide section loaded with the helical graphene ribbon is $L_0$. As shown in the side-view illustration of the $yz$-plane, the two ends of the waveguide are tapered sections to ensure the lowest possible return loss when interfacing the THz pulse compressor with a standard smooth circular waveguide.

These identical sections are two simple conical circular waveguides of length $L_2$ that gradually connect the helix to the ends of $L_1$ long circular waveguides. Next section provides initial concepts of design methodology principle of the proposed graphene-based tunable pulse compressor.

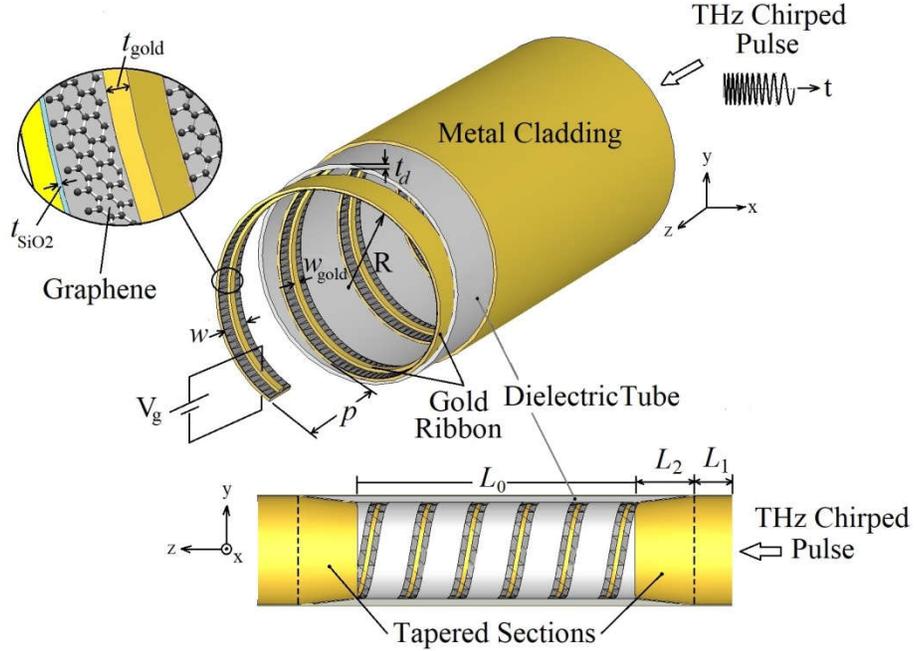

**FIG. 4-1.** Schematic-diagram, 3D geometry, and side view of an internally dielectric-coated hollow-core waveguide loaded by a helical graphene ribbon (inset: the graphene ribbon is grown on a thin SiO$_2$ layer and sandwiched between two gold ribbons to facilitate the DC controlling of the device).

### 4.1.2  Design Methodology Principle

As the first design parameters, we start with calculating the cutoff frequency of fundamental transverse electric field of TE$_{11}$ modes in a circular waveguide with a radius of $R$, filled with air as follows:

$$f_{c,TE_{11}} = \frac{c}{2\pi} \frac{1.841}{R} \tag{4-1a}$$

$$f_{c,TE_{11}}(THz) = \frac{87.901}{R(\mu m)} \tag{4-1b}$$

When a waveguide is filled partially with dielectric with relative permittivity larger than air ($\varepsilon_r > 1$), one can expect that the cutoff-frequency shifts to lower. Another factor that can be affected by the cutoff-frequency is internal loading waveguide with a helically wrapped conductive ribbon, which shifts up the cutoff-frequency inversely proportional to helix inner radius related to our proposed device. It should be noted that for a circular waveguide loaded with a dielectric-coated helically wrapped graphene ribbon, which is a material with Re[$\varepsilon_r$]<0 and Im[$\varepsilon_r$] >0, we cannot expect analytically to find an exact estimation of the cutoff-frequency. Hence, based on the initial value of *R*, which is calculated based on Eq.(4-1) for operating bandwidth around our target center frequency of 2 THz, the structure design numerically is optimized using a finite integral technique (FIT) [67]. We previously stated in Section 3.4 that one problem in close to the cutoff frequency is the higher attenuation losses, an air-filled circular waveguide with *R*=100 μm has a cutoff frequency of about 0.9 THz, which is far from the center frequency of 2 THz for minimum insertion loss.

As shown in Fig.1, adding a dielectric coating (stated as a dielectric tube with a thickness of $t_d$) on the inner surface of the metallic cylindrical hollow THz waveguide has two major benefits. Firstly, this dielectric film coating allows the propagation of the fundamental mode $TE_{11}$ for which the tangential electric field component on the metal boundary surface can be significantly reduced and minimizes the ohmic losses due to the reduced penetration of the electric field into the waveguide walls [64]. Secondly, this inner dielectric layer substantially is required as a tuning parameter to adjust wave-material interaction with respect to the relative distance between helical graphene-ribbon and metal waveguide walls or metal-cladding.

Miyagi and Kowakami [68] efforts in theoretically loss analysis for an oversized DL-HCW composed of a single-dielectric layer which carries fundamental hybrid modes of $HE_{11}$, the minimum loss occurs when the thickness $t_d$ is given by a multiplication factor of the central free-space wavelength as

$$t_d = \left(\frac{1}{2\pi} \cdot (n_d^2 - 1)^{-1/2} \cdot tan^{-1}\left[n_d/(n_d^2 - 1)^{1/4}\right]\right) \cdot \lambda_o \qquad (4\text{-}2)$$

where $n_d$ is the real part of the refractive index of the dielectric film. A dielectric coating layer much smaller than this estimated thickness may cause a condition for only traveling the TE modes, which is considered for our normal-sized DL-HCW [69]. As we state later, non-absorptive polymeric dielectrics recently used as coating layer in the THz DL-HCW waveguides have refractive index $n_d$ in range of (1.3-1.6), using Eq.(4-2) result in $t_d$ around 30 μm, that for only TE mode $t_d \ll$ 30 μm is recommended.

The design procedure of helix as slow-wave structure (SWS) in TWTA application is based on the desired phase velocity ($V_p \cong V_e$), where $V_p$ and $V_e$ are phase velocity and electron-beam velocities in TWT, respectively, and has been performed in terms of the radial propagation constant ($k_r$) of the traveling EM wave along the helix axis. There are many configurations of helix-SWS commonly used in TWT applications with or without metal shields as well as dielectric coatings. The important structural parameters in the design of all helix-SWS configurations are the inner helix radius ($a$), the metal shield inner radius ($a_{sh}$), the tape-width ($w$) for the tape-helix (or the diameter of a wire wounded helix), and the pitch size ($p$). The $k_r r$ is known as the most important parameter, where $r$ is cylindrical radial coordinate which is in $0<r<a_{sh}$. It is due to the fact that the variation of the axial electric field inside the helix, in the radial direction, is a function of $I_o(k_r r)$, where $I_o$ is the modified Bessel function of the first kind and zero-order. In order to minimize the field variations across the bore area of a helix, it is desirable to keep $k_r a$ above 1.5 [70].It is worth mentioning that the strip width over the pitch size ratio ($w/p$) and the inner radius of the helix are the most influencing parameters to control the overall dispersion of the helix. With the help of the numerical results mentioned in [71], it has been clarified that, in TWTA applications, the helix tape width ($w$) should be chosen in the $0.4<w/p<0.8$ range. This reduces the transmission loss for the fundamental wave by minimizing the coupling between the fundamental and harmonic space modes. Finally, as we use helix in our structure as a main dispersion controlling element, the inequality $w/p<0.8$ can be used as a design margin.

As in THz frequencies, most materials, including metals and dielectrics, can show a frequency-dependent or dispersive behavior, the next section is devoted to discussing all material considerations that we use in the proposed device.

## 4.2 Definition of Material

### 4.2.1 Graphene

The demand for tunable or reconfigurable components at terahertz and optical frequencies increased during the last years. Tunable THz devices for controllable systems such as tunable filters, modulators, or phase shifter are becoming more and more important. At mm-wave frequencies, piezoelectric materials and actuators have been proposed for the dynamic reconfiguration of phase-shifting surfaces. At lower mm-wave frequencies, micro-electromechanical systems (MEMs) component when is integrated to each element of the periodic structure, such as high impedance surface (HIS) or metamaterials, can be realized tunable phase shift. The actuators exhibit a displacement under a voltage bias, which is translated to a dynamic control of the reflection phase [72]. But for shorter wavelengths, including Far-IR, Mid-IR, Near-IR, and optical regions, the aforementioned techniques are not feasible. At this range of frequencies, a promising solution may rely on using graphene-based devices. Moreover, excellent tunability of graphene in terahertz [44–46] motivated us to use the graphene as the main element of the proposed to realize a tunable terahertz pulse compressor.

As this work deals with a passive compact pulse compression can be expected that the graphene relay in the linear regime, thus the graphene sheet can be modeled as an ultra-thin, nonlocal two-sided surface characterized by a surface conductivity tensor $\hat{\sigma}(\omega, E_F(E_0), \tau, T, B_0)$ based on Kubo formalism [73] where $\omega$ is the angular frequency, $E_F$ is Fermi energy (or $\mu_C$ the electrochemical potential which can be controlled by a normal applied electrostatic bias field $E_0$, or by doping), $\tau$ is the carrier relaxation time, $T$ is temperature, and $B_0$ is a normal applied magnetostatic bias field.

For a graphene sheet (in xy-plane) under only an electrostatic field bias, the conductivity is a scalar:

$$[\sigma] = \begin{bmatrix} \sigma_g & 0 \\ 0 & \sigma_g \end{bmatrix} \qquad \text{(4-3)}$$

graphene's scalar conductivity ($\sigma_g$) is described with the combination of two components of interband and intraband as follows:

$$\sigma_g = \sigma_g^{inter} + \sigma_g^{intra} \quad \text{(4-4)}$$

In the THz spectral region where the photon energy $\hbar\omega \ll E_F$, the interband part ($\sigma_g^{inter}$) is negligible compared to the intraband part ($\sigma_g^{intra}$).

**Drude-like Model**

Accordingly, in the THz regime and in case of no magnetostatic bias, graphene could be well introduced by the Drude-like surface conductivity:

$$\sigma_g^{intra}(\omega) = \frac{2k_B Te^2}{\pi\hbar^2} \ln\left(2\cosh\frac{E_F}{2k_B T}\right)\frac{i}{\omega + j\tau^{-1}} \quad \text{(4-5)}$$

Where $k_B$ is the Boltzmann constant, and $\hbar$ is the modified Plank constant ($h/\pi$). The relaxation time at the band of interest (<10 THz), can be considered $\tau = \mu\hbar\sqrt{n\pi}/(q_e v_F) \cong 10^{-13}s$ [71], where $v_F$ is the Fermi velocity ($\cong 10^6$ m/s), $q_e$ is electron charge and $n$ is the carrier density.

**Effective permittivity of Graphene**

For the numerical simulation monolayer-graphene is usually represented as a layer of material of a very small thickness $\Delta$ with two in-plane($\varepsilon_{eff,t}$) and normal-plane($\varepsilon_{eff,n}$) effective permittivities as following [73, 75]:

$$\varepsilon_{eff,t}(\omega) = 1 + j\frac{\sigma_g^{intra}(\omega)}{\varepsilon_0 \omega\Delta} \quad , \varepsilon_{eff,n}(\omega) = 1 \quad \text{(4-6)}$$

The effective permittivity of a multilayer-graphene, which is frequently observed in the chemical vapor deposition (CVD) synthesis, can also be calculated using Eq. (4-6) with a good precision [76-77].

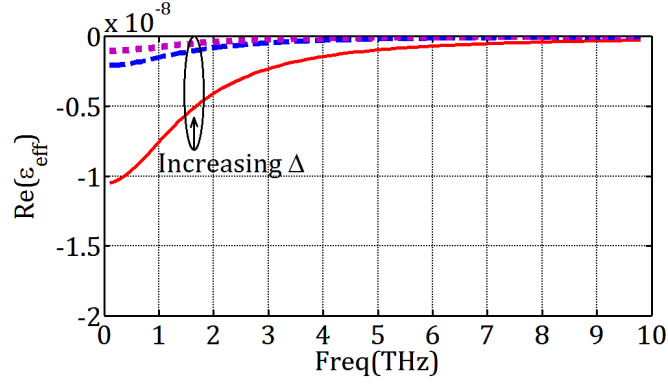

**(a)**

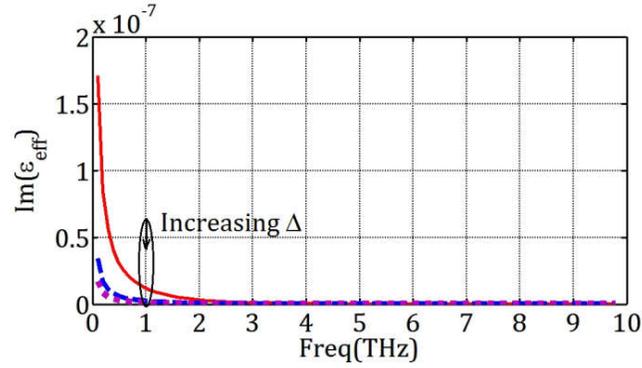

**(b)**

**FIG. 4-2.** The THz spectral of real and imaginary parts of the effective permittivity of an infinite graphene sheet for T=300K, $E_F$=0eV, and different values of Δ=[1, 10, 50]nm.

Fig.4-2 shows the real and the imaginary parts of the effective permittivity of an infinite graphene sheet in the spectral range of interest (0.1-10 THz). The effective permittivity of graphene for T=300K and $E_F$ =0eV and the different values of *Δ* is shown in Fig.4-2. For a single-layer graphene (SLG) the thickness was reported between 0.4 and 1.7 nm [78]. As shown in Fig.4-2 (a) and (b) with increasing the thickness of graphene, the variation rate of both real and imaginary parts of the $\varepsilon_{eff}$ drastically decreases. The electrochemical dependence of the equivalent effective dielectric constant of single layer graphene (SLG) is illustrated in Fig. 4-3.

The most important conclusion of these curves can be the best frequency limits for maximum tunability, which might be from 1.8 THz up to 5 THz, which the real part of graphene permittivity has maximum variation respect to $E_F$. In contrast, it has a good guard-band far from region that the imaginary part of graphene permittivity has relatively large variation respect to $E_F$ that it might cause a strong amplitude distortion on the traveling electromagnetic field waves.

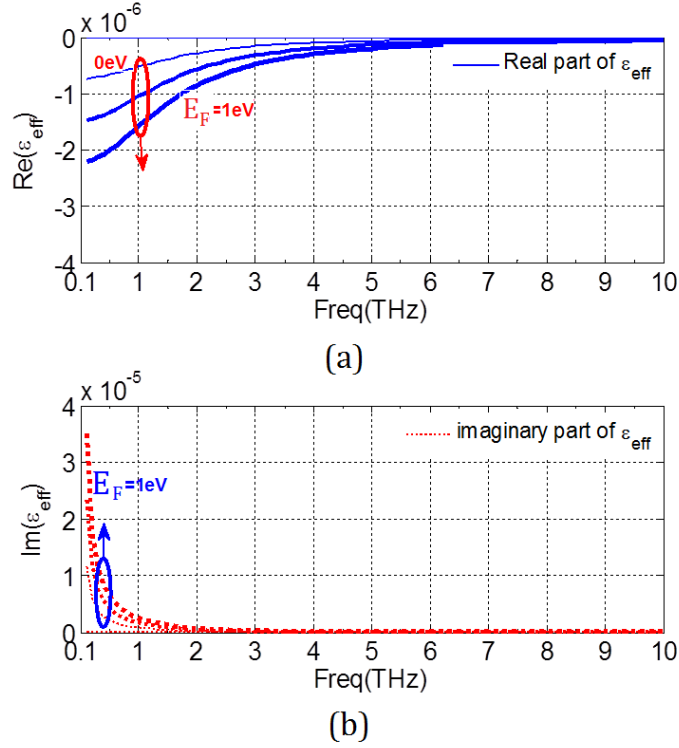

**FIG. 4-3** The THz spectral of (a) real and (b) imaginary parts of the effective permittivity of an infinite graphene sheet for T=300K,Δ =1 nm, and different values of $E_F$= [0.0-1.0] eV.

**The Effect of Substrate**

Silicon dioxide is a desirable choice for dc-controlled graphene-based terahertz device. The graphene ribbon lies on a thin $SiO_2$ layer with thickness $t_{SiO2}$. An internal helical narrow gold film ($w_{gold}$) and the fully covered back gold films facilitate the required DC-controlling bias of the graphene ribbon. The Fermi level $E_F$ can be controlled over a wide range by applying a transverse electric field using a gate voltage $V_g$.

An approximate expression to relate $E_F$ and $V_g$ is given by $V_g = E_F^2 q_e t_{SiO2}/\pi\varepsilon_0\varepsilon_{SiO2}(\hbar v_F)^2$ [22], where $q_e$ is the electron charge, $v_F$ is the Fermi velocity, and $\hbar$ is the reduced Plank's constant. Generally, the Fermi level can be tuned within the range from -1.0 to 1.0 eV, but due to negligible variation of the group velocity for $E_F$ greater than 0.5 eV, such a value of $E_F$ is considered the maximum. The typical values of $V_g$ that ensure safe bias electrostatic fields on the order of several volts per nanometer applied on the graphene are in the range of 0–100 V. Figure 4-4 shows how $V_g$ changes with $E_F$ for different $t_{SiO2}$.Therefore, for a maximum safe bias of about 30 V, the thickness of $t_{SiO2}$ has been chosen as 150 nm.

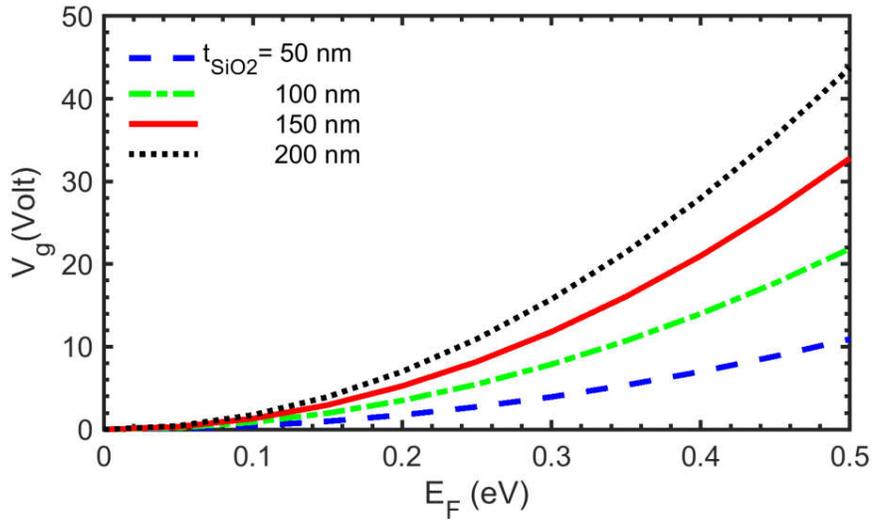

**FIG. 4-4.** DC controlling voltage $V_g$ versus Fermi level $E_F$ in terms of various SiO2 thickness.

### 4.2.2 HDPE

High-density polyethylene (HDPE) is used for the dielectric tube given its good THz transparency ($n = 1.524$ and $\alpha = 0.2$ cm$^{-1}$ at 1 THz [79]).

### 4.2.3 Nobel Metals: Gold

As shown in Fig. 1, the dielectric tube is completely covered by a gold film layer as a metal cladding. Gold is modeled with the Drude parameters as follows: $\varepsilon_\infty = 9.1$, $\omega_p = 1.38 \times 10^{16}$ rad/s, and $\gamma = 1.075 \times 10^{14}$ Hz [80].

### 4.2.4 Silicon Dioxide

From a technological perspective, the effect of a substrate and the impurities therein cannot be ignored for graphene-based, microwave optical, and terahertz device applications. Sule and et al. [81] demonstrate that the substrate has a significant influence on carrier transport at low frequencies (< 1 THz), while its influence becomes less pronounced at higher frequencies. In this work, the graphene ribbon lies on a thin SiO$_2$ layer with $\varepsilon_r = 3.9$ and tan $\delta = 0.0056$ [82].

## 4.3 Results and Discussion

The optimized dimensions of the proposed tunable graphene-based waveguide for optimal pulse compression performance and maximum tunability are listed in Table 4-1.

**Table 4-1.** Optimal dimensions of the proposed graphene-based pulse compressor.

| $R$ | $P$ | $w$ | $w_{gold}$ | $t_{gold}$ |
|-----|-----|-----|------------|------------|
| 100 μm | 126 μm | 42 μm | 30 μm | 1μm |
| $t_{SiO2}$ | $t_d$ | $t_{gr}$ | $L_1$ | $L_2$ |
| 150 nm | 7.5 μm | 100 nm | 40 μm | 100 μm |

### 4.3.1 DC-Controlled Dispersion Mechanism

The inner dielectric layer of the circular waveguide and the helical graphene ribbon provides the main functionality in the proposed graphene-based pulse compressor. The propagation characteristics of an electromagnetic structure can be studied with two common approaches: one is directly focused on circuit theory using the equivalent circuit model, and the other is based on electromagnetic mode analysis. The circuit model of the helical waveguide, as shown in Fig. 4-5 can be expressed by one inductance per unit length and two radial and inter-turn capacitances per unit length [83, 84]. Thus, a strong dispersion due to helix-shield capacitive coupling voltages and helix equivalent inductive currents can be introduced. This effect mainly depends on the dielectric tube parameters ($t_d$, $\varepsilon_d$) for the former, and the helix characteristics ($\sigma$, $w$, $p$) for the latter, where $\sigma$ is the conductivity of helix-tape that in our proposed structure is an equivalent dispersive conductivity resulted from hybrid graphene-ribbon and DC-bias facilities of gold-ribbon as shown in Fig. 4-1.The dependence of the propagation constant in optical and RF waveguides, including shape, size, and inner-coating material layer, has been explored and shown in articles [85-87]. In principle, the mode propagation constant can be obtained from the dispersion equation solved from electromagnetic field solutions. The numerical analysis for tackling these problems is reportedly available in the literature [86, 87].

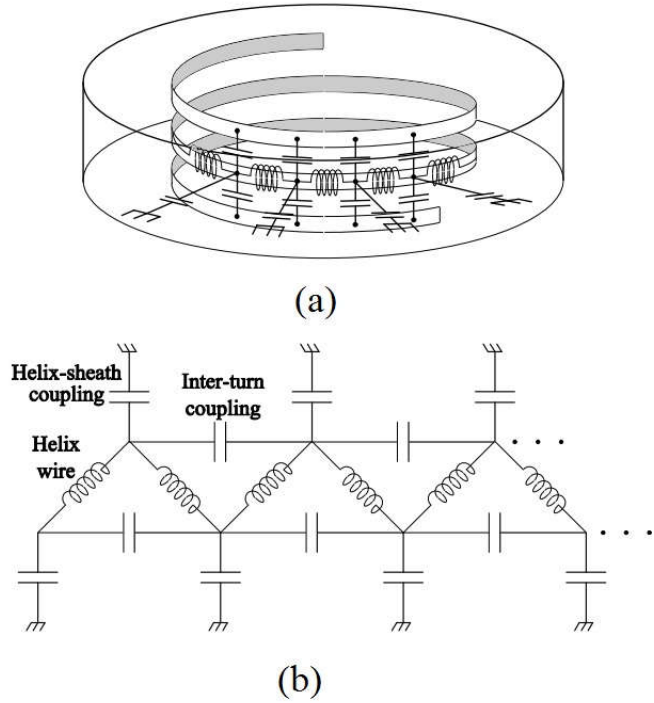

(a)

(b)

**FIG. 4-5.** The equivalent circuit modeling of a cylindrical metallic tube which contains a metallic helix from [83].

In a simple structure such as an internally coated circular waveguide with a lossy layer, the approximated analytical solutions of the characteristic equation in studying the physical features of the propagation constants have been presented in Ref. [86]. In this case, Luo et al. show how the variation of the lossy layer thickness ($t$) affects the propagation constants for all five modes (TE$_{11}$, TM$_{01}$, TE$_{21}$, TM$_{11}$, and TE$_{01}$); the effect in the propagation constants is great when $t/\delta$ <1.5 (where $\delta$ is the skin depth of electromagnetic waves in the good conductor), while for $t/\delta$ >2.5 the propagation constants become stable (see Fig.4-6).

Rao [87], using the boundary value method, has obtained the dispersive equation of a circular waveguide that is coated on the inside with a metamaterial and compared with a dielectric coated one. He shows that the variation of the normalized phase constant ($\beta/k_0$) with frequency for dielectric ($\varepsilon_r$>0, $\mu_r$>0) and metamaterial coating ($\varepsilon_r$<0, $\mu_r$<0) is exactly opposite. Thus, for a circular waveguide loaded with a dielectric-coated helically wrapped graphene ribbon, which is a material with Re[$\varepsilon_r$] <0 and Im[$\varepsilon_r$] >0, we can expect strong dispersion suitable for passive pulse compression. The influence of the Fermi level on the propagation properties depicted in Figs. 4-5 and 4-6.

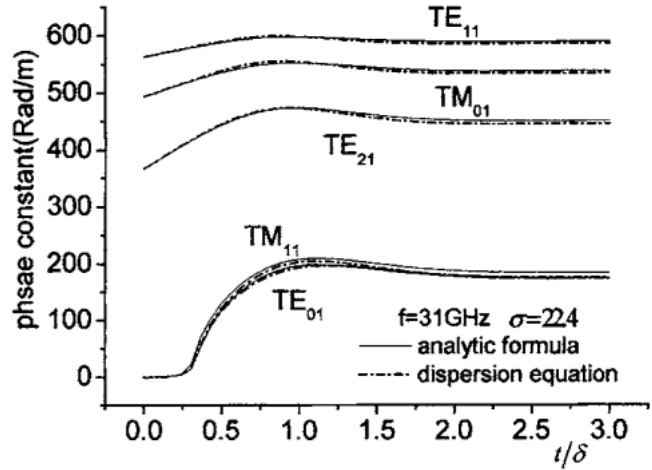

**FIG. 4-6. Variation of phase constant for five modes TE$_{11}$, TM$_{01}$, TE$_{21}$,TM$_{11}$, and TE$_{01}$ vs relative thickness *t*/δ at =22.4 S/m and *f*=31 GHz, Fig.6b of [86].**

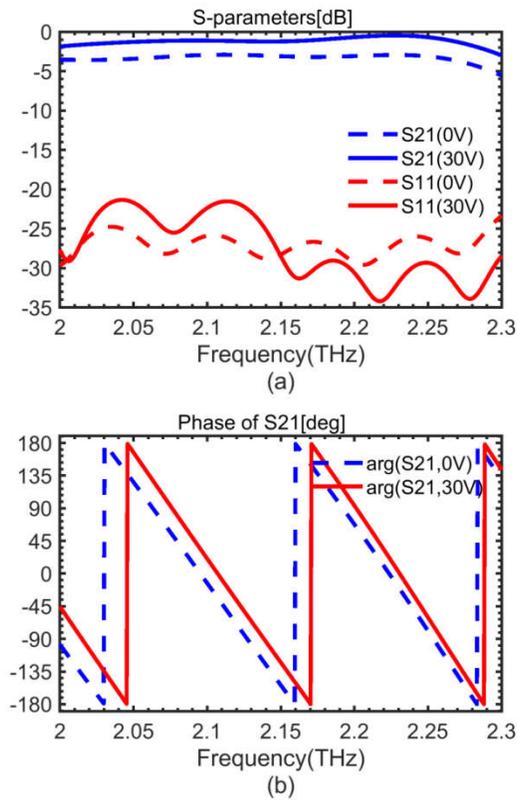

**FIG. 4-7. (a) Magnitude and (b) phase of the S-parameters of the proposed device biased at *Vg* D 0 and 30 V.**

The pulse-shaper component has a good matching and low insertion loss (an average of -1.4 dB) in a bandwidth >200 GHz around the center frequency of 2.15 THz, and it is independent of the graphene bias. The spectral phase response of the structure shows a tunable phase change in terms of the applied DC voltage [Fig.4-7(b)].The GVD and normalized group velocity of the component is illustrated in Figs. 4-8(a) and (b), respectively, where the Fermi level of the graphene ribbon changes in the range of 0.0 eV (nonbiased) and 0.5 eV ($V_g$= 30 V). For Fermi level $E_F$= 0.5 eV, the waveguide simultaneously shows negative and positive GVD responses at 2.06–2.15 THz and 2.15–2.28 THz, respectively. This phenomenon can be explained by the fact that when the effective permittivity of the graphene ribbon changes with the increase of the Fermi level, the aforementioned equivalent capacitances, and inductances of the structure can manipulate the traveling wave in a dispersive manner.

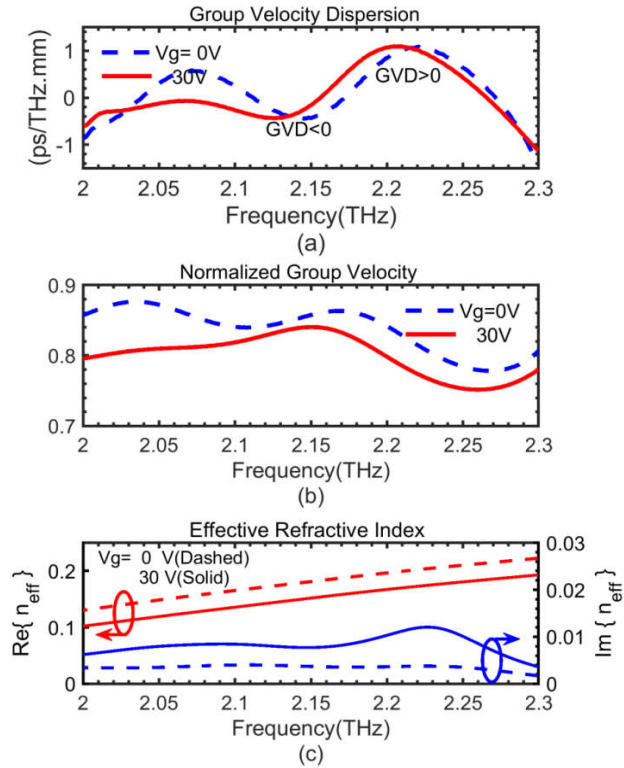

**FIG. 4-8.** (a) Group velocity dispersion (GVD), (b) normalized group velocity, and (c) real and imaginary parts of the effective refractive index of the proposed device biased at $V_g$= 0 and 30 V.

Second, as the overall wavefront inside a metallic hollow waveguide has been formed by successive reflections from the inner walls of the waveguide, one can expect the presence of a very high chromatic dispersive structure such as a helical graphene ribbon to be able to manipulate the phase of the resultant propagation mode.

For completeness, Fig. 4-8(c) Re[$n_{eff}$]= $\beta/k_0$, where $k_0$ is the free-space wave number, $\beta$ is the wave propagation phase constant, and Im[$n_{eff}$]. The former encapsulates the group velocity and group velocity dispersion, whereas the latter accounts for the loss. Notice that Im[$n_{eff}$]is easily computed via the absorption coefficient $\alpha = 2k_0$Im[$n_{eff}$] [56], which is estimated from the scattering parameters |$S_{21}$| and |$S_{11}$| shown in Fig. 4-7.

## 4.3.2  Pulse Compression Performance

To investigate the pulse-compression performance of the component, it is assumed that the input pulse is a $y$ polarized chirped sine function, as stated in Eq. (3-25), where $f_0$ is the upper-band frequency of the negative dispersion region, $T_0$ is the input pulse width, $E_0$ is the electric field amplitude, and $\mu$ is the chirp factor. The chirp factor for a uniform input pulse spectrum in the positive GVD region from $f_{Lo}$ to $f_{Hi}$, 2.15 to 2.28 THz, respectively, can be calculated using Eq. (3-26):

$$E = \begin{cases} E_0 \sin(2\pi f_{Hi}t - \mu t^2), & if \ t \leq T_0 \\ 0, & other \end{cases} \qquad (3-25)$$

$$\mu = \frac{2\pi(f_{Hi}-f_{Lo})}{T_0} \qquad (3-26)$$

The instantaneous frequency can be obtained from the time derivative of the phase argument of Eq. (3-25) proves that the tail of the chirped pulses with low-frequency components and faster group velocities move to overtake the front high-frequency components of the pulses, resulting in temporal pulse shortening with growth in amplitude. To validate the performance of the device and highlight the presence of a unique optimal length for each arbitrarily selected pulse width, two optimal compression lengths corresponding to the two chosen pulse widths of 8 and 12 ps have been numerically obtained as720 and 1700 μm, respectively. The optimization goals in our design are maximum compression ratio and minimal residual pulse pedestal, which ensures a high-quality compression. Figures 4-9 and 4-10 show the temporal waveform evolution of the envelope of the device for these two optimal lengths, 720 and1700 μm, for the two bias states, zero-bias, and 30 V.

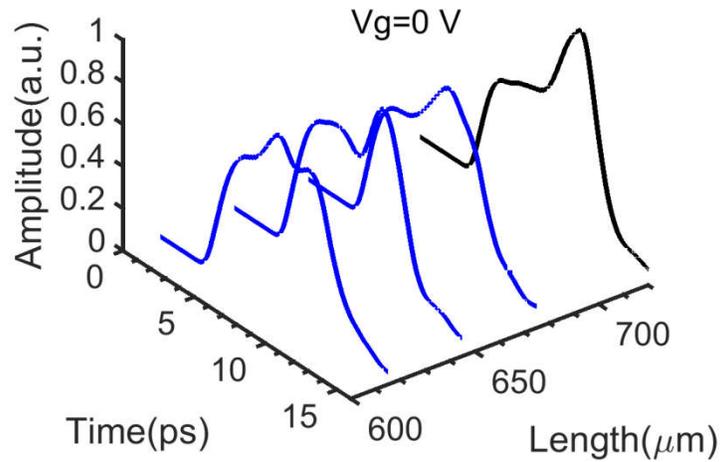

(a)

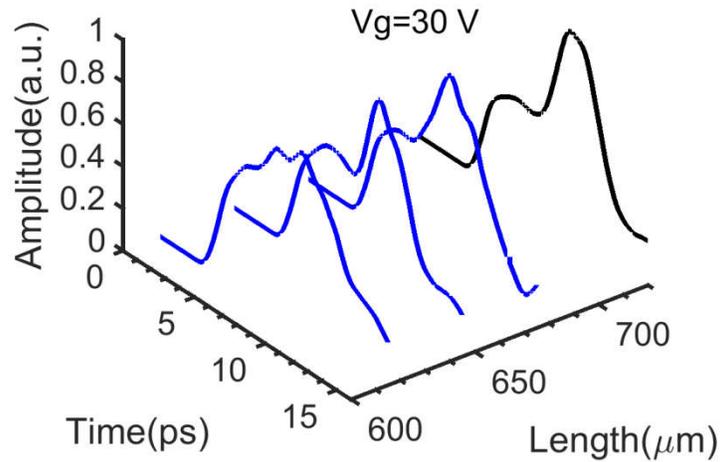

(b)

**FIG. 4-9. Waveform evolution of the proposed device with the optimal length of 720 μm, corresponding to input pulse duration of 8 ps, based on two border values of controlling DC voltage (a) 0 V and (b) 30 V.**

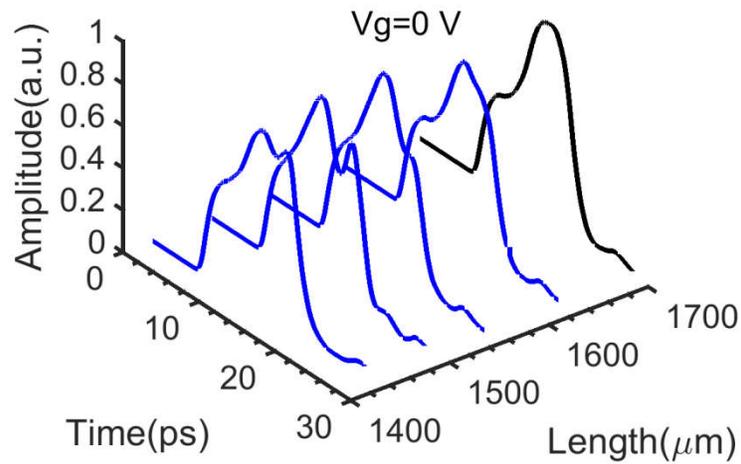

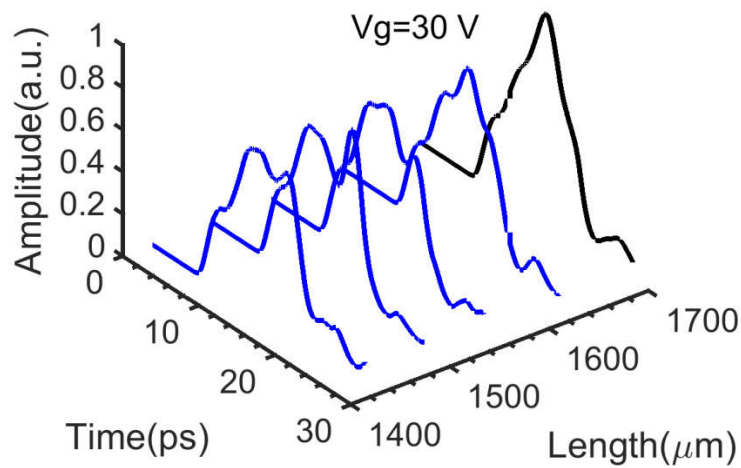

**FIG. 4-10**. Waveform evolution of the proposed device with the optimal length of 1700 μm, corresponding to input pulse duration of 12 ps, based on two border values of controlling DC voltage (a) 0 V and (b) 30 V.

## Proof of the TE₁₁ mode of the proposed structure

As shown in Figure 4-11, it can be seen that the final output temporal pulse axial electric field component ($E_z$) of the structure with increasing the length of the structure (or equivalently periods number) is sharply reduced.

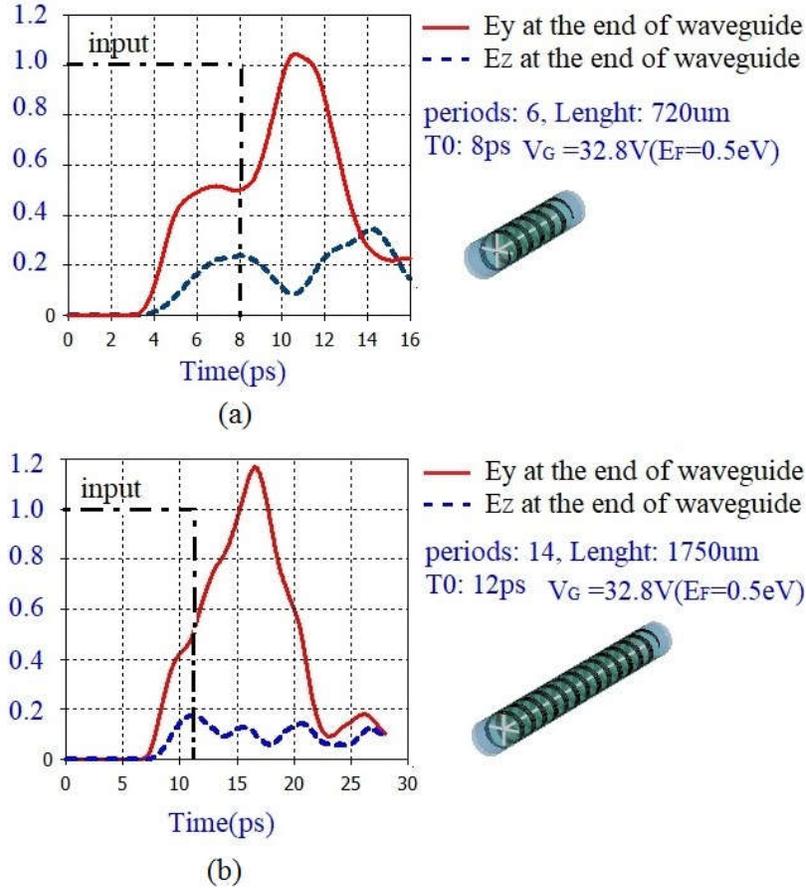

(a)

(b)

**FIG. 4-11.** Envelope of the compressed output pulses monitored transverse and axial electric field components of $E_y$ and $E_z$, respectively, for two model lengths of (a) 720µm and (b) 1700 µm.

A comparison between transverse ($E_y$) and axial ($E_z$) electric field components of the proposed structures in two different lengths when the helical Graphene ribbon gating voltage is 32.8V (or 0.5eV).As the second reason, which confirmed that we have a likely pure-TE$_{11}$ mode due to use a regular size circular waveguide, while for HE$_{11}$ mode, we should have an over-sized waveguide.

The electric field distributions of $E_y$ and $E_z$ illustrated in Fig.4-12 at 2.15 THz on the output aperture of the waveguide, based on these field patterns, it could be seen that the maximum bore field spot devoted to the $E_y$ component, which is fundamental of TE$_{11}$ mode.

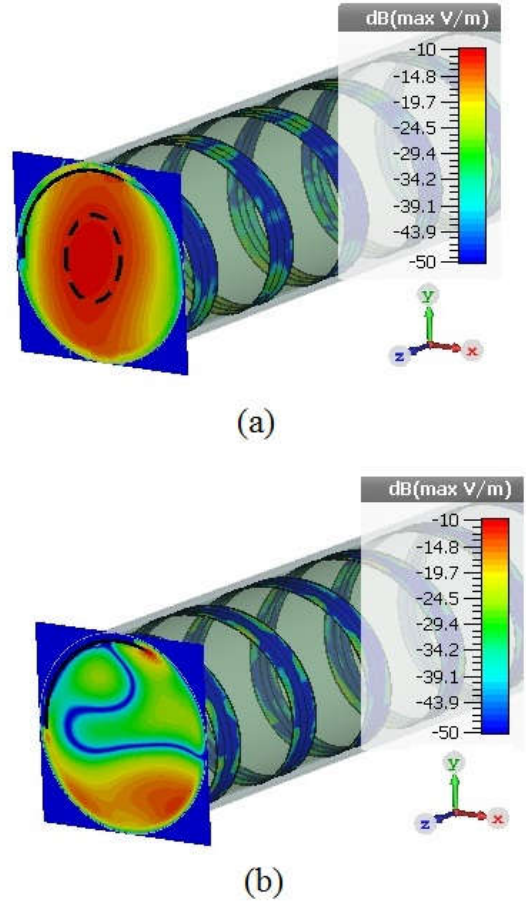

(a)

(b)

**FIG. 4-12.** The electric field distributions of: (a)Ey and (b) Ez on the output aperture of the proposed pulse compressor of length 700μm, at 2.15 THz.

An important conclusion is that the pulse compressor functionalities can be controlled with a feasible gate voltage $Vg$. Figures 4-13 and 4-14 compare the pulse compression performance of the device in the time and frequency domains for both studied optimal lengths. It can be seen that with increasing structure length, a high quality temporal compressed pulse [negligible pedestal and large full-width at half-maximum (FWHM)] can be achieved. Additionally, the compression rate can easily be tuned via the gate voltage, $V_g$.

The mode patterns of the device are shown in the inset of the spectrum plots. These patterns show that the mode confinement can be tuned via two factors, which are the graphene electrostatic bias and the length of the structure.

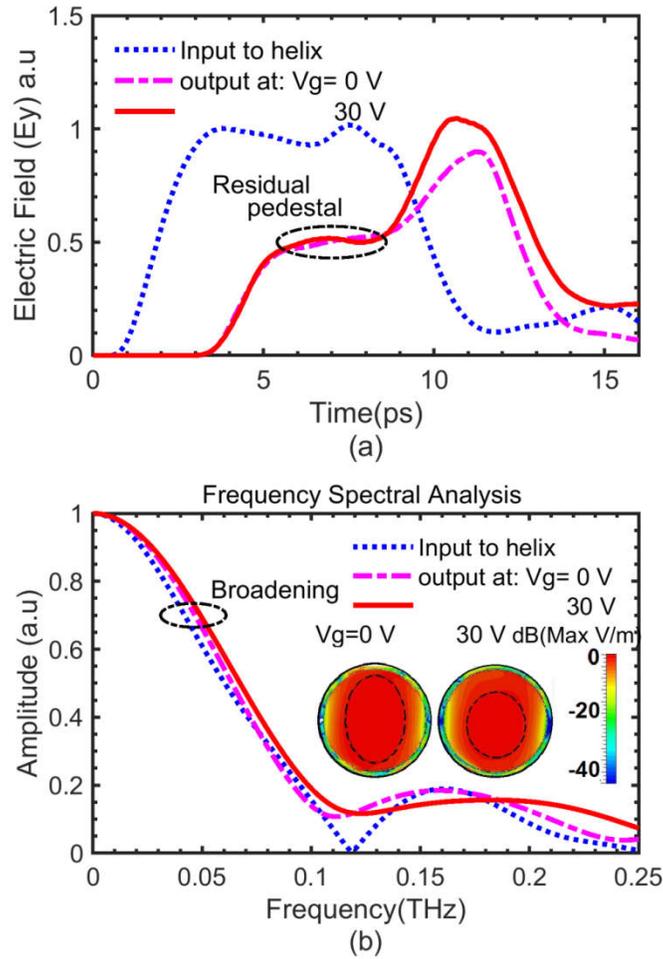

**FIG. 4-13.**(a) Time and (b) spectral descriptions of the pulse compressor output for *Vg* = 0 and 30 V for a length of 720 μm (the mode patterns at the end of the helix are presented in the insets of the figure).

One of the important parameters in the pulse compression applications is the broadening factor (*F*), which is defined as the ratio of FWHM of the compressed pulse to the FWHM of the input pulse. Table 4-2 summarizes the broadening (ΔFWHM) of the two studied model lengths and their tuning percentages, which is defined as the percentage of the broadening change. This parameter is also known as the modulation depth defined by (Δ*F* /*F₀*)× 100%, where Δ*F* is the changes of *F* due to Δ*Vg* = 30 V, and *F₀* is the average of the minimum and maximum obtainable *F* [55]. A theoretical detail of how the pulse broadening is caused by the group velocity dispersion is given in Appendix A.

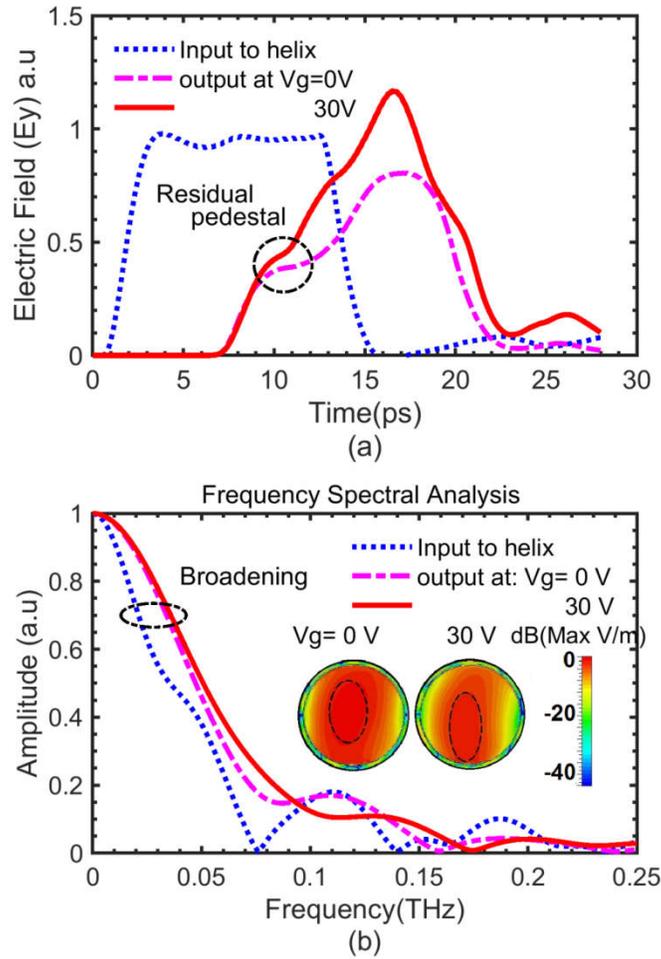

**FIG. 4-14.** (a) Time and (b) spectral descriptions of the pulse compressor output for *Vg* = 0 and 30 V for a length of 1700 μm (the mode patterns at the end of the helix are presented in the insets of the figure).

**Table 4-2**. Tuning Percentage of the Proposed Graphene-Based Pulse Compressor.

| Model Lengths | 720 μm | 1700 μm |
|---|---|---|
| ΔFWHM @ $V_g$ = 0.0 V | 3.96 GHz | 12.20 GHz |
| ΔFWHM @ $V_g$ = 30.0 V | 1.11 GHz | 9.46 GHz |
| Tuning % | 5.9 % | 8.04 % |

To the best of our knowledge, no other studies have examined the combination of graphene and pulse-compression techniques in the THz regime. It is helpful, however, to review the literature to quantitatively justify our work. In Ref. [41], a THz passive pulse compressor shows a 2 ps, 0.5 THz chirped-pulse after transmission through a 10 μm circular aperture in a gold-film GaAs. The pulse was compressed to 1.4 ps with ΔFWHM of 0.07 THz. One of the major drawbacks of this structure is the extremely low output to input peak intensity ratio of $3.7 \times 10^{-5}$. Another example is direct compression of ultra-fast laser pulses using a dispersive and nonlinear thick (550 μm) cholesteric liquid crystal sample [43]. After passing the laser pulses across this material, a broadening of 1.8 THz around the center frequency of 387 THz (777 nm) has been achieved. Comparing the numerical results shown in Table 4-2 with the literature, one can conclude that the proposed graphene-based tunable pulse compressor holds promise for THz applications.

## Sensitivity

In this study, a primary concern is to investigate the sensitivity of the proposed graphene-based pulse shaping device to the crucial device parameters including the geometric dimensions and the refractive index of the dielectric coating. As illustrated in Fig.4-15, the compressed output signal of the proposed zero-biased graphene-based pulse compressor has shown no significant sensitivity to the variation of refractive index of the inner dielectric coating ($n_d$), the dielectric thickness of $t_d$, and the helical graphene-ribbon dimensions: width(w) and pitch size(p).

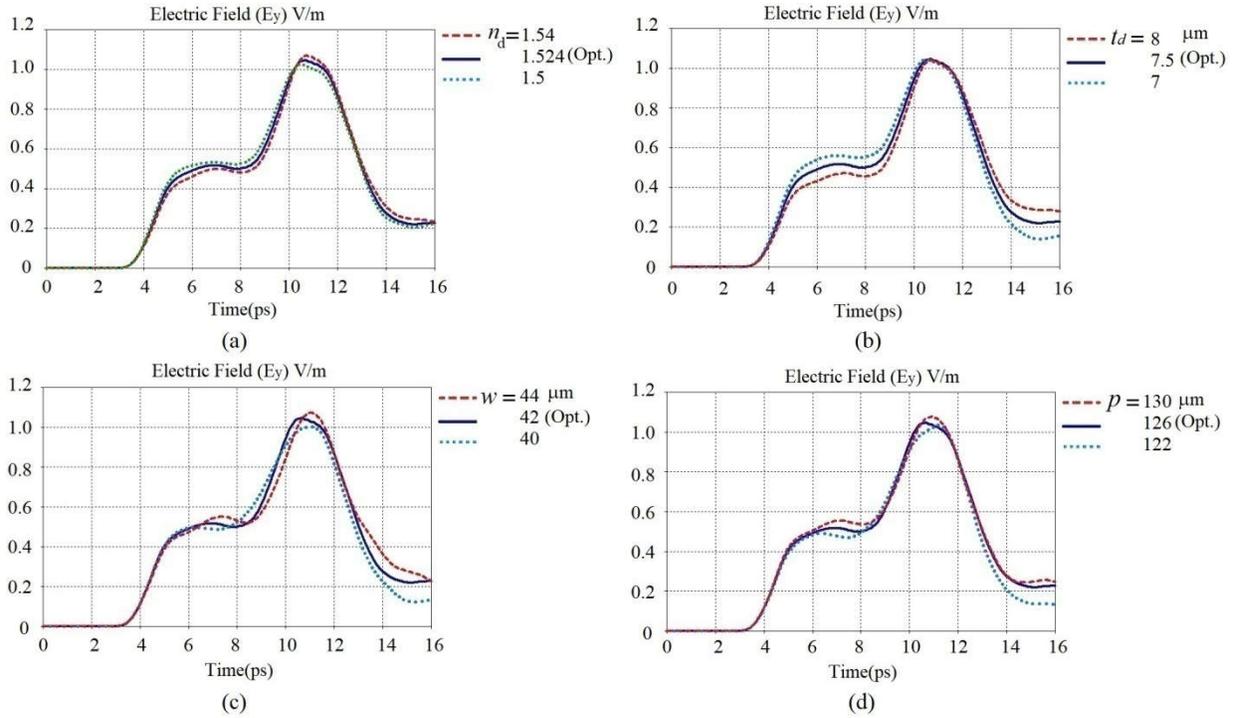

**FIG. 4-15 Parametric study of sensitivity of (a) refractive index of inner dielectric coating layer, and (b) its thickness of t$_d$ , (c) width and (d) pitch size of the helical zero-biased graphene-ribbon of the proposed graphene-based pulse compressor on the compressed output pulse.**

(This page intentionally left blank)

# Chapter 5

# Numerical Analysis Validation based on a Fixed Terahertz Pulse Compressor

Numerical techniques could be a great help in theoretical studies of challenging electromagnetic problems such as complex structures where exact analytical solutions are difficult to obtain.

In this chapter, to confirm the accuracy of the previously achieved results of the proposed tunable graphene-based pulse compressor, first of all, the mesh size sensitivity is analyzed by modeling the of the proposed device with the optimal length of 720 μm, corresponding to an input pulse duration of 8 ps, when controlling DC voltage has been set at 30 V. Then we made a step forward for a passive fixed pulse-shaper by replacing the helical graphene ribbon with a gold ribbon, and compared with our previous work, graphene-based one. Moreover, to prove the validity of the numerical analysis, the proposed device has been examined using the finite element method (FEM) and the finite integral technique (FIT).

The need for system identification in engineering is interesting [88], because of its ability to describe phenomena, which are not to be fully explained by analytic approaches in complicated problems. The use of the system transfer function to analyze the structure is preferable to the full-wave simulation because of saving the execution time. Once the transfer function is determined, one could apply it for the subsequent time-domain analysis of the pulse shaper with various inputs. Finally, the rest of the chapter describes how to estimate the transfer function of the proposed structure from the time domain simulation response.

4
## 5.1    Validation of Numerical Analysis Results

Fast computer-based simulation tools that allow the designers them precisely model their ideas can be responsible for undiscovered model errors. The mesh size sensitivity and convergence test are of importance, so that broadly employed in the various engineering computer-aided designer tools to ensure high accuracy of the final numerical results. Moreover, using different numerical approaches can be helpful to verify the simulation results.

### 5.1.1 Graphene-based Tunable THz Pulse Compression Waveguide

Meshing a complex geometry (see Fig. 3-1), including a dispersive infinitesimal thin layer such as graphene is an essential part of the simulation. Thus, after a convergence test, we have chosen a graphene thickness of $t_{gr}$= 100 nm as the minimum cell size for the helix ribbon, while the material properties of graphene were defined based on its real thickness of 1 nm. Due to computational constraints, a smaller graphene thickness was not viable. This tradeoff approach is commonly found in the literature [89]. As a mesh sensitivity convergence test based on different mesh resolutions of the overall model meshes and the helical ribbons curvatures (graphene and SiO2 spacer) of 20, 50, and 100 segmentations per turn have been performed, Fig. 5-1 shows the dispersion performances of the proposed graphene-based structure. As we can see for smaller mesh resolution, the results do not change significantly.

Execution time is an essential factor in determining the effectiveness of the numerical analysis, which depends on two factors of the minimum machine requirements and mesh cells number of the electromagnetic problems after time-spatial discretization operations. The total mesh cell number, mesh methods, and corresponding computation-time of each mesh resolution are illustrated in Table 5-1.

**TABLE 5-1.** Mesh characteristics and the corresponding computation-time of the proposed graphene-based pulse compressor.

| Mesh setting | | Total mesh cells | Computation time |
|---|---|---|---|
| Mesh cells per wavelength | for curved elements (helical ribbon) segmentation No. | Total mesh cells | Computation time |
| 6 | 20 | 2,290,570 | 0 h, 28 min, 38 s |
| 10 | 50 | 7,223,424 | 1 h, 49 min, 53 s |
| 10 | 100 | 13,727,583 | 3 h, 15 min, 29 s |

For the published article, the default mesh line limit is set at 10 lines per wavelength, resulting in 3,958,080 and 12,533,920 hexahedral mesh units for the 720 and 1700 μm long pulse compressor, respectively.

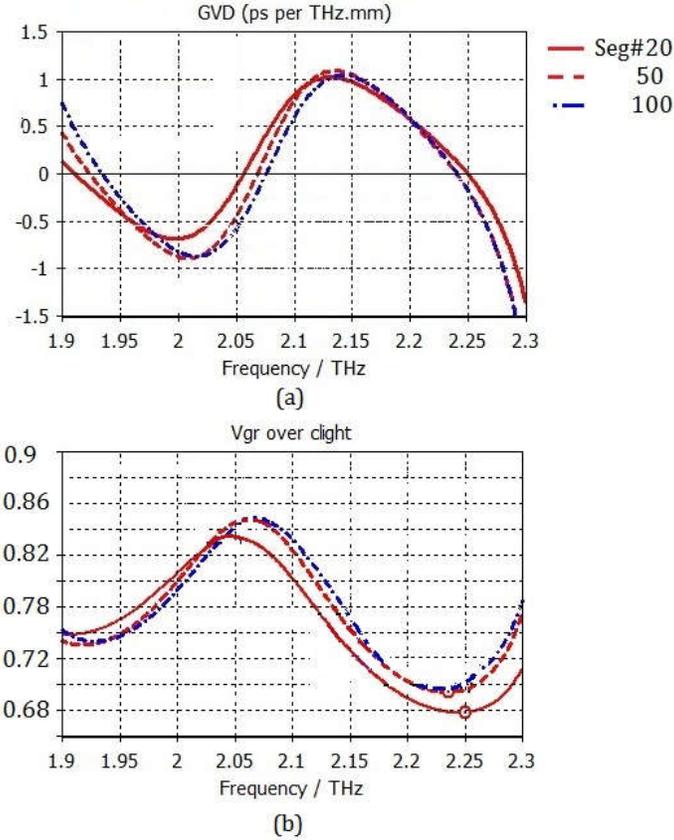

**FIG. 5-1.** (a) GVD and (b) normalized group velocity of the proposed graphene-based tunable pulse compression waveguide in various mesh resolution using by FIT analysis.

## 5.1.2 Gold-based Fixed THz Pulse-shaper Waveguide

In this section in order to prove the validity of the numerical analysis results, another example of our previous work (graphene-based pulse-compressor), a passive fixed pulse-shaper by replacing the helical graphene ribbon with a gold ribbon has been introduced. The idea of using helical-ribbon loaded circular waveguide as a controllable dispersion device can be verified using two different numerical approaches finite element method (FEM) and the finite integral technique (FIT).We see that the results for both methods are in an excellent agreement.

Figure 5-2 sketches the passive waveguide-based terahertz fixed-pulse shaper. A gold ribbon of thickness $t_{gold}$= 1 μm, width $w$ (41 μm), and pitch size $p$ (126 μm) is helically wrapped and inserted inside a metalized dielectric tube with inner radius $R$ of 100 μm and thickness t of 8.5 μm. By carefully choosing the geometric parameters and the dielectric material, an optimum condition for an arbitrary dispersion profile, large mode confinement, minimum mismatch and

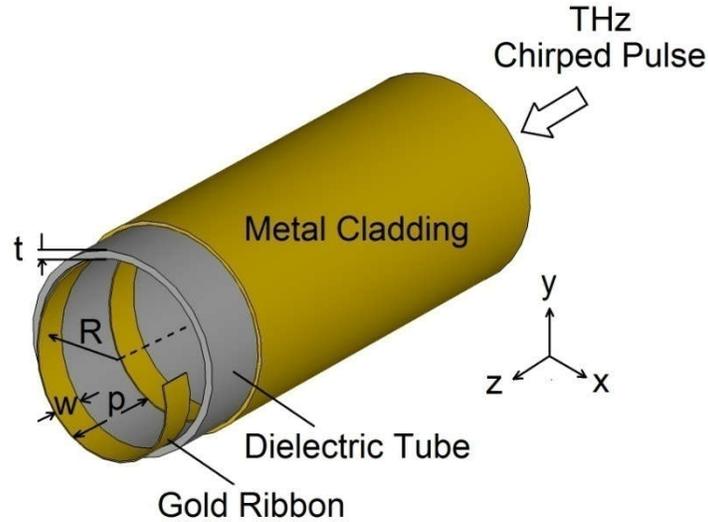

**FIG. 5-2.** 3D geometry and side view of an internally dielectric-coated hollow-core waveguide loaded by a helical gold ribbon

propagation loss with maximum tunability can be achieved. High-density polyethylene (HDPE) is used for the dielectric tube given its good THz transparency ($n = 1.524$, and $\alpha = 0.2$ cm$^{-1}$ at 1 THz). A gold film layer completely covers the dielectric tube as a metal cladding. Gold is modeled with the Drude parameters as stated previously in Chapter 4.

First of all, to prove the accuracy of the numerical analysis, the proposed device has been analyzed using the FEM and the FIT. Fig.5-3a and b, show the transmittance and absorption coefficients of the proposed device across the whole operating bandwidth (2.14 - 2.34 THz) for the dominant $TE_{11}$ mode.

The phase difference between the input and the output signals of the 2-port device is expressed by the phase of the transmission coefficient ($S_{21}$) as depicted in Fig.5-4. The phase of the transmission coefficient, which is calculated numerically using both aforementioned numerical methods demonstrate the dispersion characteristics of the signal transmission. It determines how the input signal delayed when it travels between the input and output ports of the device in the desired frequency spectrum. As shown in Chapter 4, a passive pulse compressor can be designed in a portion of electromagnetic spectra where the normalized group velocity ($V_{gr}/c$) profile has a large negative gradient.

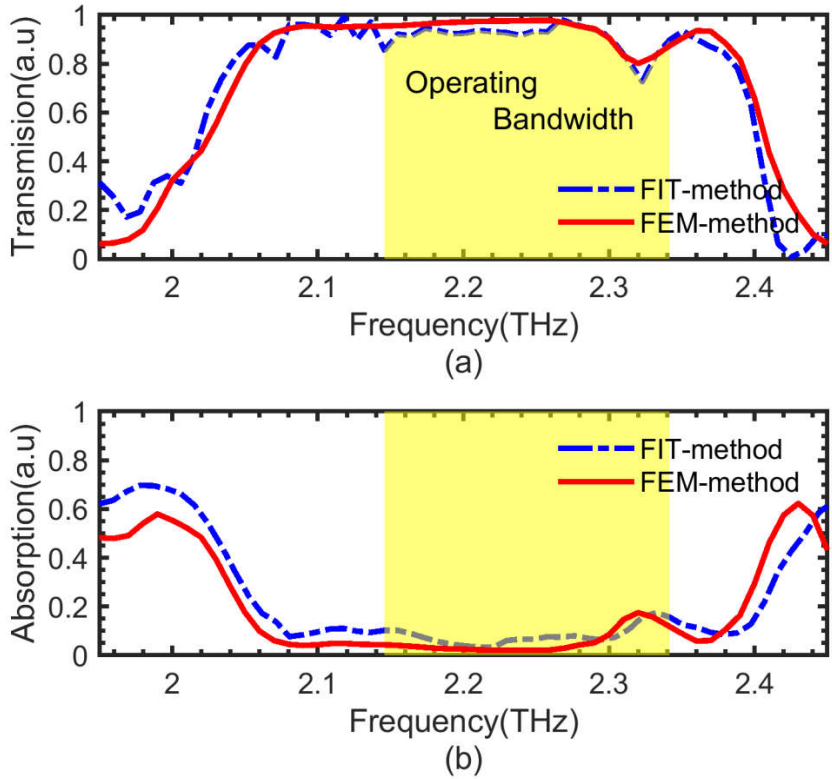

**FIG. 5-3** **(a) Transmittance, and (b) absorption spectra ofthe proposed device using the FIT and the FEM.**

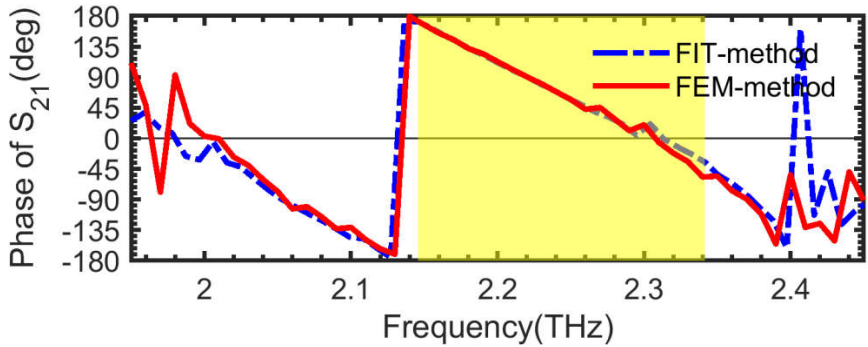

**FIG. 5-4** **The phase of S21 of the proposed device using the FIT and the FEM.**

The optimal bandwidth, which the proposed structure shows the normalized group velocity profile that has a relatively large negative gradient ($f_{\text{Lo}}$ to $f_{\text{Hi}}$), is equal to 2.14-2.34 THz, , as shown in Fig. 5-5. Furthermore, we provide a comprehensive time-domain analysis that verifies the overall pulse-compression performance of the proposed device.

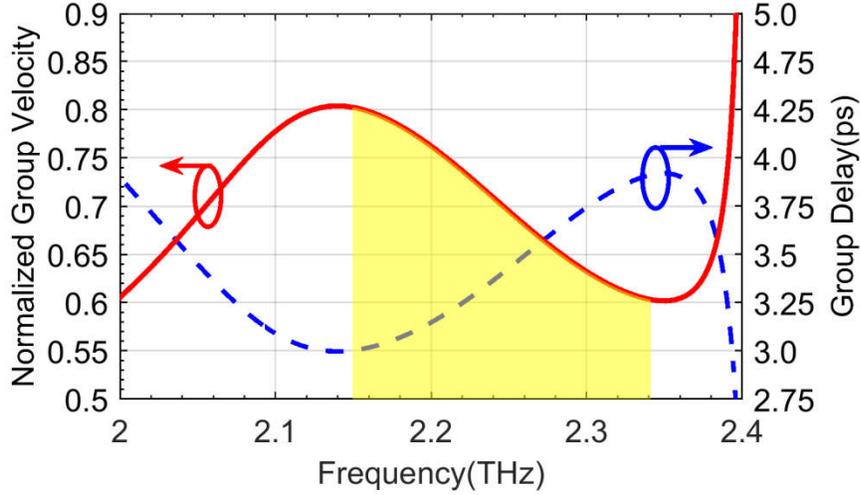

**FIG. 5-5** **Group delay and normalized group velocity extracted from the phase of the transmission coefficient (FIT-method).**

The chirped-input signal is a y-polarized chirped sine function (E$_y$), as stated in Eq. (3-25), where $f_0$ is the upper-band frequency of the negative slope of the GV profile (2.34 THz), $T_0$ is the input pulse width , $E_0$ is the electric field amplitude, and $\mu$ is the chirp factor. The chirp factor for a uniform input pulse spectrum in the negative slope of the GV region from $f_{Lo}$ to $f_{Hi}$, 2.14 -2.34 THz, can be calculated using Eq. (3-26) as 157.079×10$^{21}$ [S$^{-2}$].

As a theoretical background, it is evident that the energy of electromagnetic waves in a dispersive medium travels in a group velocity profile which is dependent on the frequency, as the proposed structure which is shown in Fig.5-5. Therefore, if such dispersive structure is excited by using a linearly down-chirped pulse like E$_y$, it can be expected in a group velocity profile with negative gradient in terms of the frequency spectrum the front pulses with lower group velocity travels with larger delay of T$_{d, Hi}$ than the tail of the input pulses with delay of T$_{d, Lo}$ as described in Eq. (5-1).

$$T_{d,Hi} = \frac{L_0}{V_{gr} \mid_{f_{Hi}}}, T_{d,Lo} = \frac{L_0}{V_{gr} \mid_{f_{Lo}}} \tag{5-1}$$

where L$_0$ is the optimum length of the structure. If the input chirped-pulse duration is chosen as (T$_0$=T$_{d,Hi}$-T$_{d,Lo}$), one can be expected that the tail of the input pulse travel to overtake the front one. Based on the numerically obtained information of group velocity, shown in Fig.5-5, we obtained the optimal traveling length of $L_0$ as 579 μm. As shown in Fig.5-6, the maximum compressed pulse shape is obtained at the output of y-polarized electric field which is located at the distance of 575 μm from the input reference plane of the proposed device. Figure 5-6(a) shows the traveled input pulses just after traveling 50 μm, about 10% of the optimal length while Figs.5-6 (b) and (d) are illustrated the probe outputs located at 525 μm and 625 μm, before and after the optimal position. These figures clearly show that how the probe output signal starts to grow to a peak at the tail position before the optimal position while after passing the optimum length the temporary narrowed pulse peak starts to downgrade.

The 3D demonstration of the spatial temporal electric field distribution along the proposed waveguide device can be used to characterize the field confinement performance for future applications possibilities such as material detection sensors or time domain spectroscopy (TDS). Figure 5-7 gives a comprehensive description of the y-polarized electric field distributions at the optimal probe position in three important moments of the compressed pulse duration, assigned as $t_A$, $t_B$, and $t_C$ those are correspondent to pedestal, peak and ringing positions. As depicted in Figs. 5-7b-d, the field distribution of the peak moment ($t_A$) is significantly more confined than pedestal ($t_B$) and ringing ($t_C$) moments. This ensures the proposed device can be employed as an efficient pulse shaper for various pulse-based THz applications including fast data communications, TDS and imaging. Another important conclusion from the electric field distribution would be the mode purity of the proposed structure at the fundamental propagation linear polarized mode pattern of circular waveguide ($TE_{11}$).

The next Section will describe how to estimate the transfer function of the proposed structure from the time-domain simulation response.

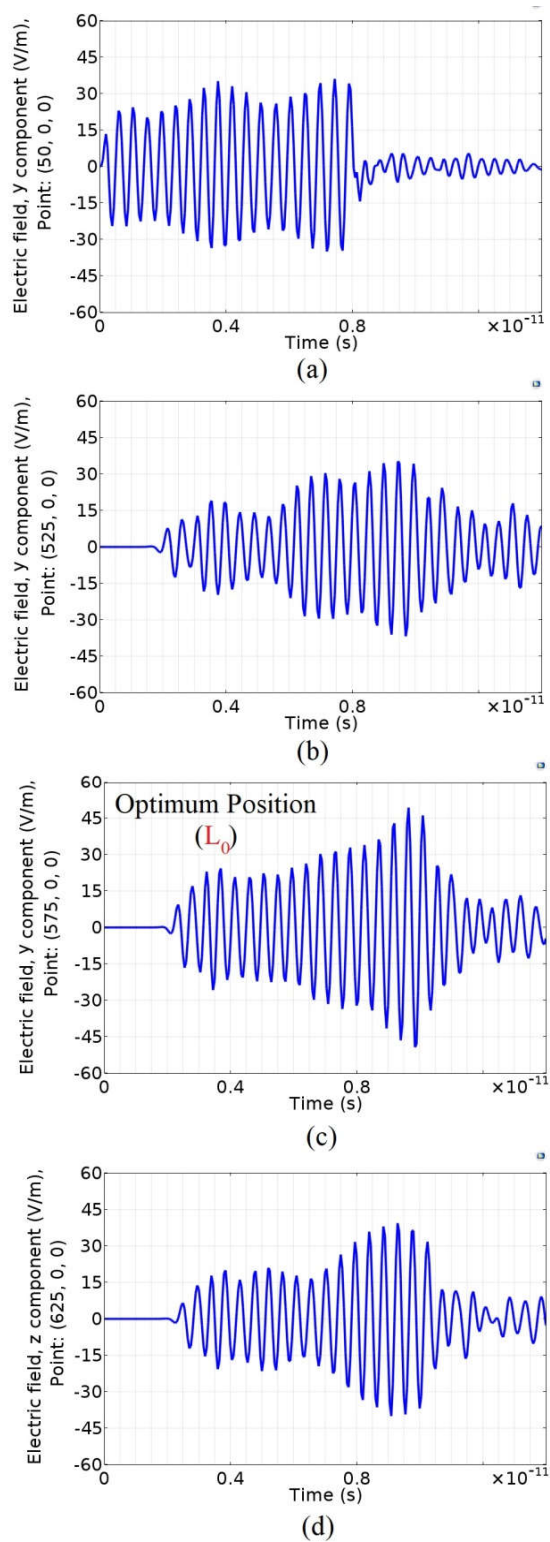

**FIG.5-6** **(a) The output y-polarized E-field probe signals at various positions of the proposed structure (a) front position (z = 50μm), (b) before, and (d) after, and (c) optimal length of $L_0$=575 μm.**

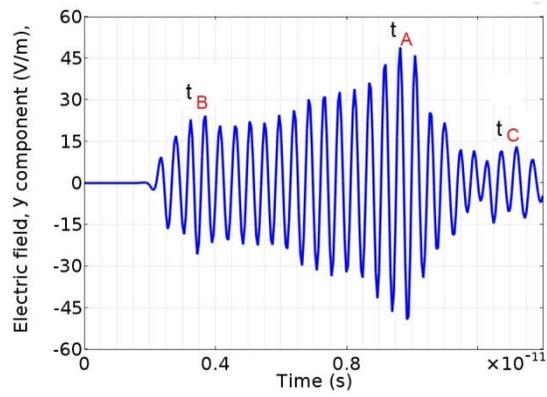

(a)

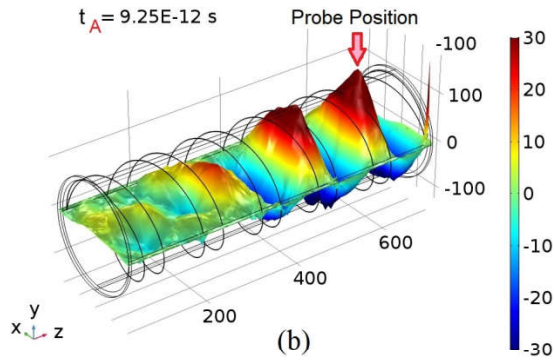

(b)

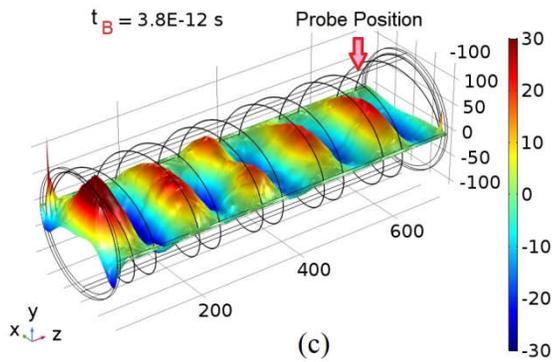

(c)

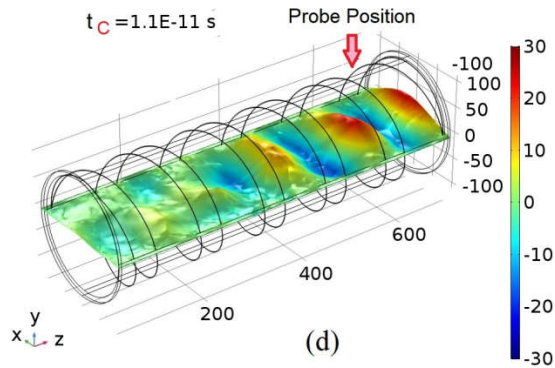

(d)

**FIG 5-7** **The electric field distributions at the optimum positions and the corresponding peak time positions of (b) $t_A$, (c) $t_B$, and (d) $t_C$ as marked on the (a) the probe signal output.**

Figure 5-8 depicts the output envelope waveforms of the proposed device based on a chirped input signal obtained using the FIT and the FEM. As the simulation results exhibit, there is a good agreement between the results of the FIT and the FEM methods over the entire operating bandwidth (2.14- 2.34 THz); moreover, it can be assumed as a time-domain results validation. In Fig. 5-9, we demonstrated the calculated longitudinal electric field distributions (using FEM method) for the fundamental mode at 2.15 THz and 2.25 THz, inside of the proposed structure, the single-mode pattern of the waveguide with good confinement have been maintained. Some of the general modeling consideration and simulation characteristics such as the mesh setting, dispersive material definition in both time and frequency domains, in CST Microwave Studio v.2015 and COMSOL 5.2a besides the postprocessing operations for extracting the principal results are explained in Appendix B.

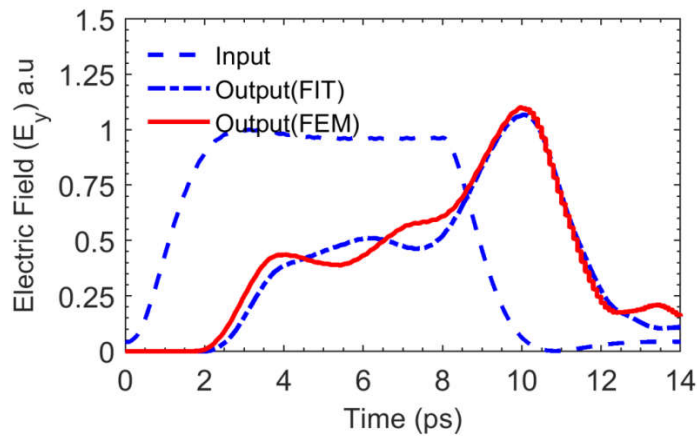

**FIG. 5-8** the output envelope waveforms of the proposed device based on a chirped-input signal using two different numerical approaches of FIT and FEM.

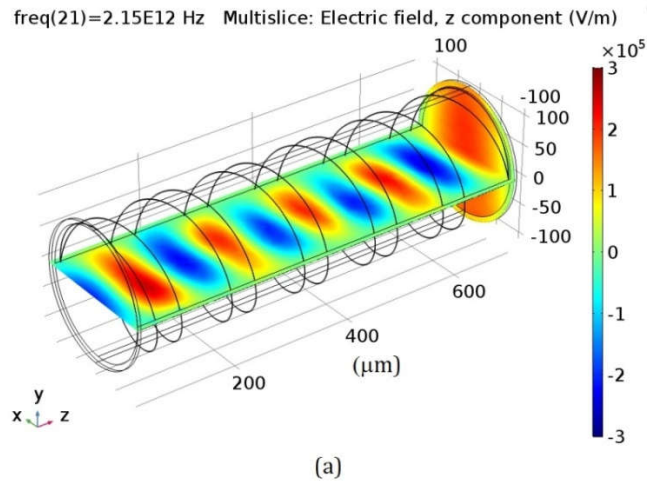

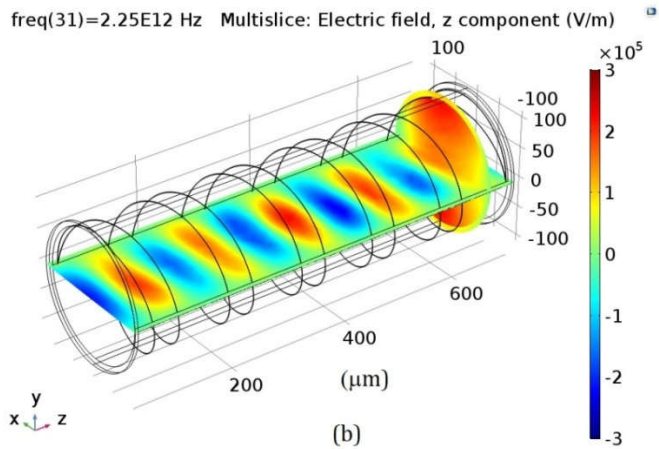

**FIG. 5-9** The fundamental mode pattern(TE$_{11}$) of the proposed pulse shaper structure in xz-planes at (a) 2.15 THz, and (b) 2.25 THz obtained by using FEM.

## 5.2 A System Identification of Au-based Fixed THz Pulse-shaper Waveguide

A transfer function is an important notion in signal processing, which allows one to relate an output signal and an input. System identification, which represents the transfer function allows a great deal of flexibility in building mathematical models by using measurements of the systems input and output signals in the terahertz spectrum [90-93]. Although transfer functions are generally used in the analysis of systems such as single-input single-output filters in the fields of signal processing, control theory, and communication theory, it could also be applied to provide comprehensive system models of electromagnetic problems. This section focuses on the estimation of the general transfer function of the proposed device. In pulse shaping, knowledge of the transfer function makes it possible to achieve the temporal output pulse from any arbitrary input pulses. Here, by incorporating the full-wave time-domain simulations (obtained in the previous section), and the numerical transfer function estimation approach, it becomes possible to extend the application of the system theory to a wider range of optical problems. The proposed waveguide-based structure is assumed as a linear time-invariant (LTI) system, which is stable and can be described by a transfer function of the form Eq.(5-1):

$$H(s) = \frac{b_0 + b_1 s^{\beta_1} + \cdots + b_m s^{\beta_m}}{a_0 + a_1 s^{\alpha_1} + \cdots + a_n s^{\alpha_n}} \tag{5-1}$$

where $b_0$, $b_1$, …, $b_m$, $a_0$, $a_1$, …, $a_n$ are coefficients to be estimated, and $\beta_0$, $\beta_1$, …, $\beta_m$, $\alpha_0$, $\alpha_1$, …, $\alpha_n$ are positive real valued exponents.

The proposed identification approach is based on the transfer function estimation using the input/output temporal data, which are generated in the full-wave transient simulation of the proposed waveguide under a chirped input signal. In this work, the MATLAB signal processing toolbox was employed to estimate the transfer function coefficients given in Table 5-2 based on the obtained time samples of the compressed signal. Figure 5-10 presents the input-chirped signal and the results of the estimated transfer function and those obtained by the FEM simulation. It is noteworthy that the coefficients of the numerators and denominators of H(s) with four zeros and nine poles are determined so that the highest similarity between the FEM output, Fig. 5-10b, and the reconstructed one by using H(s), Fig. 5-10c, has been achieved.

**Table 5-2.** The resulting coefficients of the numerators and denomerators of the estimated transfer function.

| $\beta_0$ | $\beta_1$ | $\beta_2$ | $\beta_3$ | $\beta_4$ | $\beta_5$ | $\beta_6$ | $\beta_7$ | $\beta_8$ | $\beta_9$ |
|---|---|---|---|---|---|---|---|---|---|
| - 7.17e113 | 2.8e100 | - 2.7e88 | 4.9e74 | -2.5e62 | 0 | 0 | 0 | 0 | 0 |
| $\alpha_0$ | $\alpha_1$ | $\alpha_2$ | $\alpha_3$ | $\alpha_4$ | $\alpha_5$ | $\alpha_6$ | $\alpha_7$ | $\alpha_8$ | $\alpha_9$ |
| 1.4e115 | 6.7e102 | 1.3e90 | 5.3e77 | 4.7e64 | 1.5e52 | 7.0e38 | 2.0e26 | 3.8e12 | 1 |

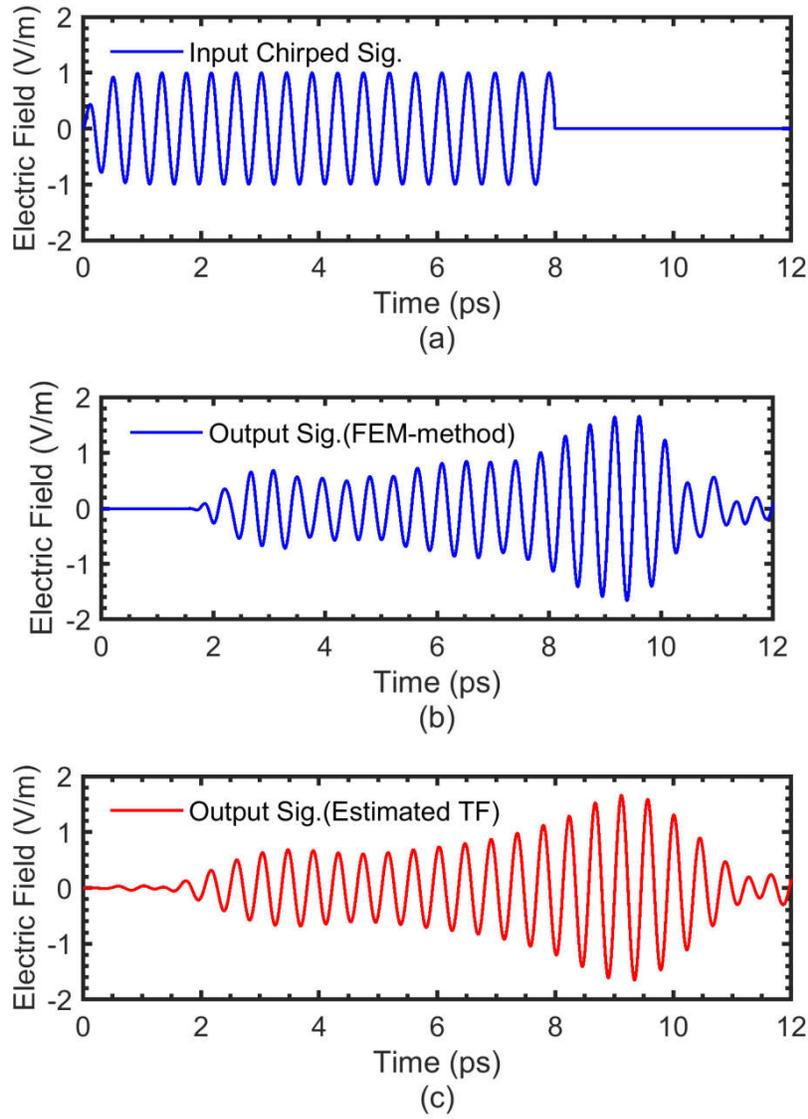

**FIG. 5-10** (a) Chirped-input signal, (b) reference output signal obtained from the FEM-method analysis, and (c) output signal obtained from the estimated transfer function model of the proposed device.

The Bode plot of the estimated transfer function of the proposed pulse-shaper, as shown in Fig. 5-11, resemble an equivalent active bandpass filter as |H(s)|>1 has happened at the passband spectrum. The pole-zero pattern of the H(s) also illustrated in Fig. 5-12, which all poles placed in left side ensure a real estimated system. All the MATLAB codes that were used throughout this thesis, especially including transfer function estimation, have been shared with readers in Appendix C.

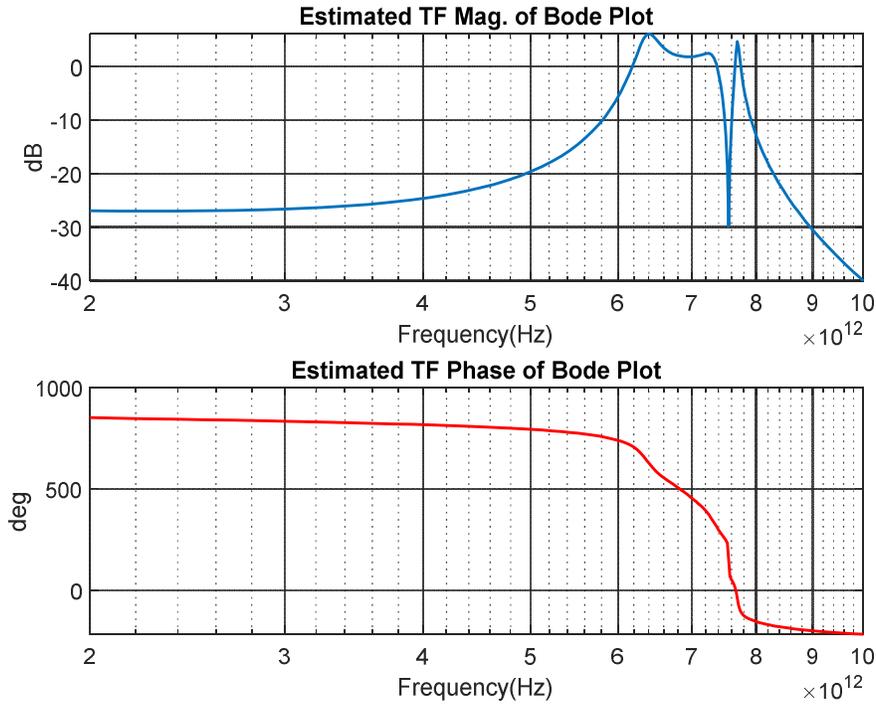

**FIG. 5-11** The magnitude and phase diagram of the Bode plot of the estimated transfer function.

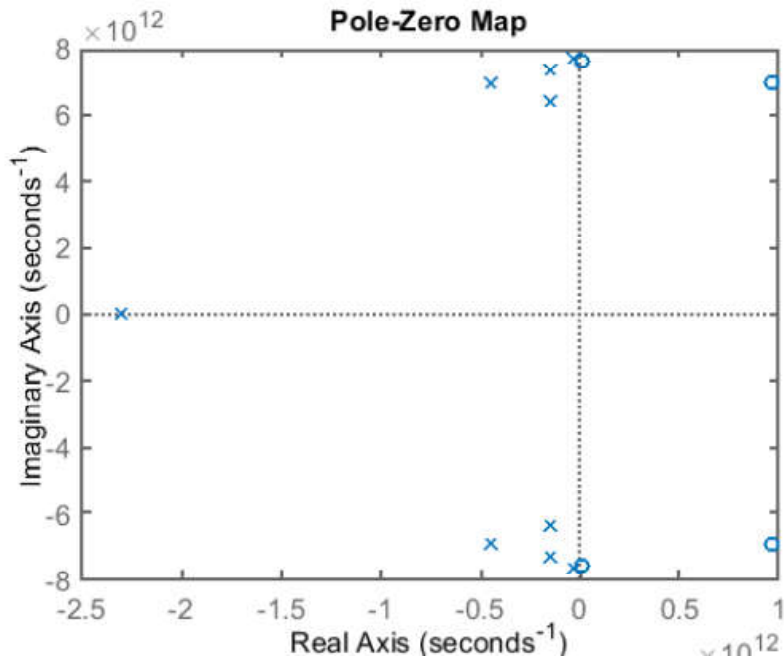

**FIG. 5-12** The pole-zero map of the estimated transfer function.

# Chapter 6

# A Comparative Study on Helical Ribbon Material Dependence

In many optical devices, investigation of the impact of the material of comprised parts is a crucial part of the study. As an example, the helical-ribbon in the proposed pulse compressor, which was introduced and studied in Chapters 4 and 5, has an essential influencer.

5
## 6.1 Dispersion and Matching Performance of PEC, Au, and Graphene-ribbon

A comparative study of material dependence on our proposed terahertz waveguide-based pulse compressor has been presented. Based on dielectric-lined hollow-core waveguide loaded with a helical conductive x-ribbon (where x can be a perfect electric conductor (PEC), Au or graphene-coated Au), pulse compression of a linear chirped sine wave is investigated numerically for different material helical ribbon in cases of (i) PEC, (ii) Au, and (iii) graphene-coated Au with different biases.

Figure 6-1 shows the 3D-geometry and side view of the proposed metal ribbon loaded waveguide: a metal ribbon with the thickness $t_m$, ribbon width $w$, and pitch size $p$ is helically wrapped and inserted inside a metalized dielectric tube with a thickness of $t_d$. The length of the helical x-ribbon loaded waveguide section was introduced with $L_0$. As shown in the side view illustration of the yz-plane, the two ends of the waveguide are tapered sections to ensure the lowest possible reflections when interfacing the THz pulse compressor with a standard smooth circular waveguide. These identical sections are two simple conical circular waveguides with the length of $L_2$ that gradually connect the helix to the ends of the circular waveguides with the length of $L_1$. The dielectric is HDPE with a refractive index of 1.524, and Au is modeled as the Drude model parameter, which stated in Sec. (4.2.3). The optimized dimensions of the proposed helical x-ribbon loaded waveguide are listed in Table 6-1.

**Table 6-1.** The optimized dimensions of the proposed helical x-ribbon loaded waveguide.

| p | W | $t_d$ | $t_m$ | $L_0$ | $L_1$ | $L_2$ |
|---|---|---|---|---|---|---|
| 126 μm | 42.5μm | 8.5μm | 2μm | 665μm | 40μm | 100μm |

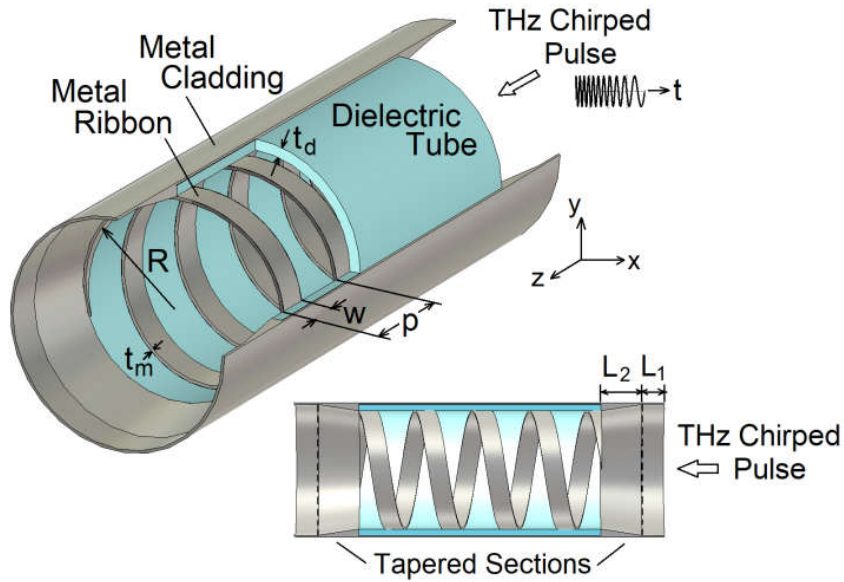

**FIG. 6-1.** Schematic-diagram, 3D-geometry in a, and side-view of an internally dielectric-coated hollow-core waveguide loaded by a helical metal ribbon.

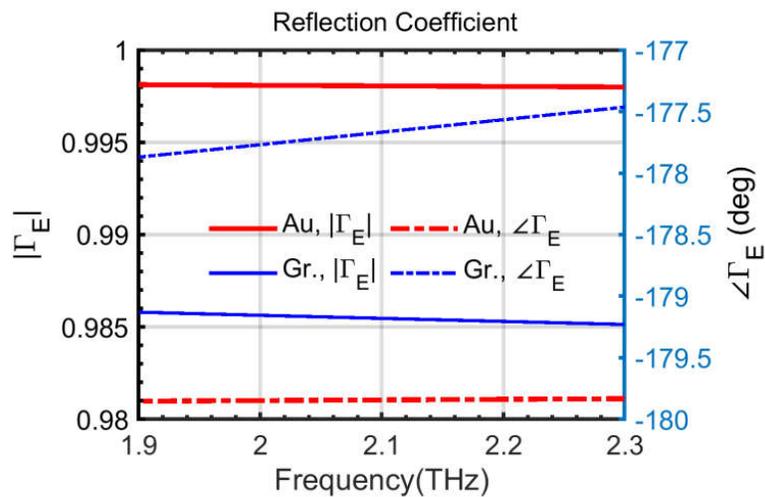

**FIG. 6-2.** The magnitude and phase of reflection coefficients of a gold-air and graphene-air interface at normal incidence of a plane wave.

As successive reflections have formed the overall wavefront inside a metallic hollow waveguide from the inner walls of the waveguide, one can expect the presence of a very high chromatic dispersive structure such as a helical x-ribbon to be able to manipulate the phase of the resultant propagation mode. Hence, the reflection coefficient ($\Gamma$) of a normal incident of each material could be a comprehensive measure that describes how the reflective material can manipulate the transmission phase of the structure; Fig. 6-2 gives a spectral description of the reflection coefficient of Au and graphene films. It can be found that graphene compared to gold-film, as a terahertz dispersive metal, has a more dispersive reflection coefficient in the frequency span of 1.9-2.3 THz that results in equivalently a dispersive transmission coefficient. Of course, PEC with $|\Gamma|$=1 and $\angle\Gamma$=180$^{\circ}$ besides graphene and Au in upcoming qualitative compression performance could be helpful. The S-parameters characteristics of the three cases, as shown in Fig. 6-3, could be better interpreted with the reflection coefficients data, since that the Au and PEC with the better reflectivity than Au-graphene coated ribbon result in the same transmission amplitude ($|S_{21}|$) and phase responses ($\angle S_{21}$).

As illustrated in Fig. 6-4(a), depending on which one of two bands of 2.0 - 2.15THz and 2.15 – 2.25 THz, can be excited in this device, the dispersion can be either normal or anomalous with group velocity dispersion (GVD) values of −0.5 ps/(THz.mm) or 1.2 ps/(THz.mm), respectively, approximately independent of helical-ribbon material. This reveals that the dispersion mechanism is significantly due to helical geometry rather than material dispersion. As we mentioned in Sec.(2.1.1), for passive pulse compression application, the most favorable frequency region is a part of the dispersion characteristic where the operating wave normalized group velocity ($V_{gr}$/c) has a large negative gradient as a function of frequency which one can see in Fig. 6-4(b) all cases have likely the same proper bandwidth, but Au and PEC compared to graphene-coated Au-ribbon have a significantly larger negative gradient of the normalized group velocity.

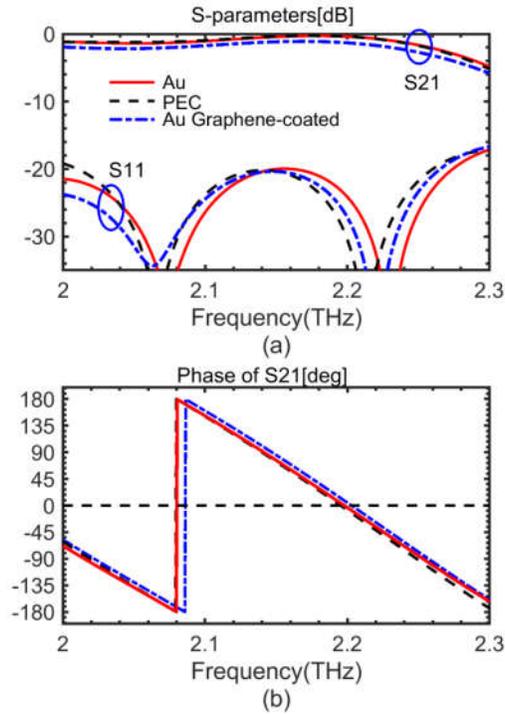

**FIG. 6-3.** A comparative illustration of (a) Magnitude and (b) phase of the S-parameters of the proposed device based on different helical ribbon materials.

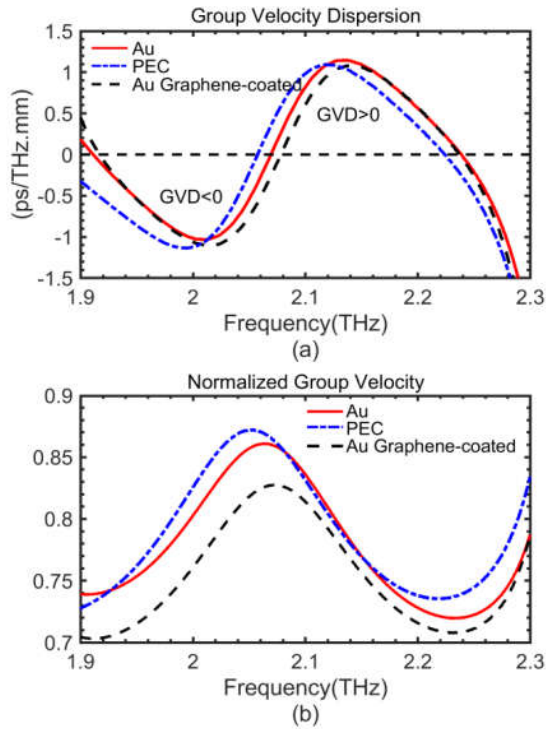

**FIG. 6-4.** The influence of helical ribbon materials on (a) Group velocity dispersion (GVD) and (b) normalized group velocity of the proposed device.

## 6.2   Pulse Compression Analysis

As shown in Fig.6-5, by comparison, we find the pulse compression performance (compression factor and pedestal) in the case of (ii) is further better than that for other ones. High compression factor with low-pedestal is helpful to produce needle-shaped pulses which are required for low ISI data communication. As we have seen earlier in Fig.6-4(a) and (b), the high dispersive behavior of the graphene coating can result in more spectral broadening, as well as more compression ratio, but we should pay the cost of this with more power dissipation. To validate the performance of the device, a unique optimal compression length corresponding to the pulse width (here we choose pulse duration of 8 ps) can be expected.

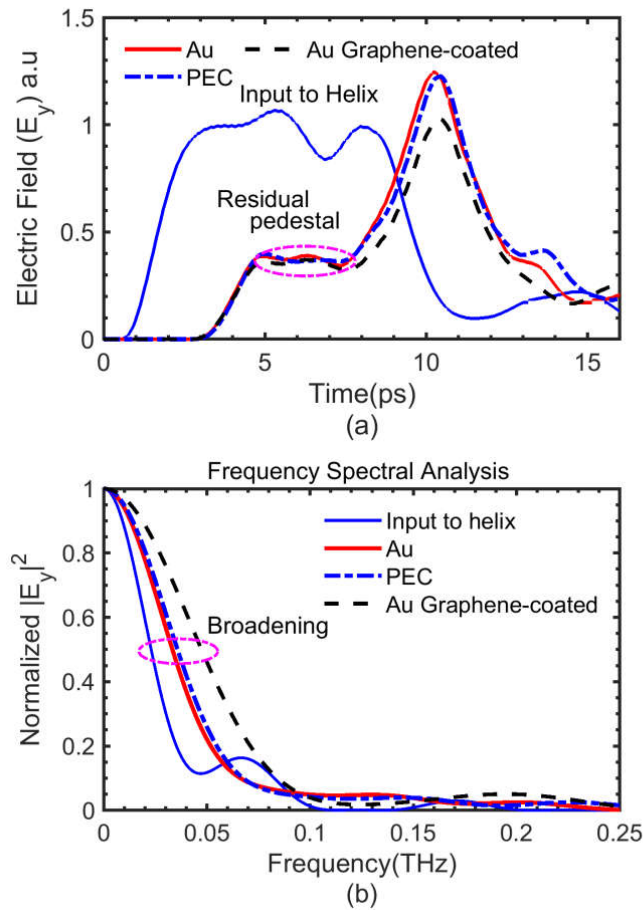

(a)

(b)

**FIG. 6-5.** The impact of helical-ribbon material on the (a) time and (b) frequency responses of the pulse compressor.

The optimization goal in our study is the maximum compression ratio and minimal residual pulse pedestal, which ensures a high-quality compression. Figures 6-6 (a) and (b) show the temporal waveform evolution of the envelope of the proposed device in cases of helical Au- and graphene-coated Au-ribbon, respectively. The corresponding optimal lengths of both cases are around 665 μm distance from the input reference plane. As shown in Fig. 6-6, pulse side lobes of Au-graphene coated compared to Au-ribbon have lower amplitudes and in conclusion higher pulse quality, this could be predicted from higher dispersion that was seen from the phase of the graphene reflectivity coefficient ($\angle\Gamma_E$) which was shown in Fig. 6-2.

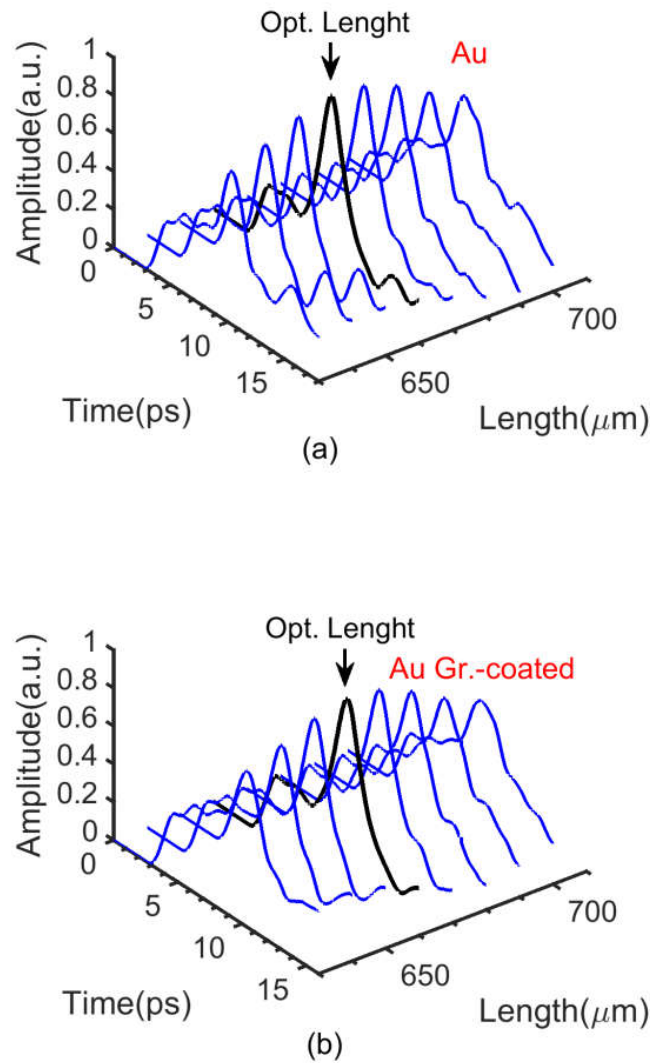

(a)

(b)

**Fig. 6-6.** The waveform evolution of the proposed device μm, corresponding to an input pulse duration of 8 ps in case of (**a**) Au- and (**b**) Au-graphene-coated ribbons.

# Chapter 7

# Conclusion and Future Works

Right from the outset of this thesis, the aim of the work presented was to see how far dispersion tailored pulse shaping devices, in particular DC-controlled graphene-based pulse compression, can go in overcoming some of the challenges in tunable passive THz pulse shaping research. It is hoped that by taking the first steps in applying graphene to a passive hollow-core waveguide, the idea of DC-controlled graphene-based pulse compression could begin to gain traction and help propel tunable graphene-based terahertz devices one step closer to real-world applications where the benefits of THz characterization can be brought out fully.

The first proposed idea was undertaken in this PhD thesis used material and waveguide (chromatic) dispersion in helically wrapped graphene-ribbon inside a dielectric-lined hollow-core waveguide as the passive pulse-shaping mechanism. To the best of our knowledge, no other studies have examined the combination of graphene and pulse-compression techniques in the THz regime.

A spectral pulse broadening factor $\Delta$FWHM of 0.07 THz, the highest for passive THz temporal pulse compression at the time of publication, was achieved from passing a 2 ps, 0.5 THz chirped-pulse through a 10 um circular aperture in a gold-film GaAs [41]. One of the major drawbacks of this structure is the extremely low output to input peak intensity ratio of $3.7 \times 10^{-5}$.

More importantly, the conclusions of this work highlighted certain key factors in THz passive pulse compression:

- High-quality compression
- Large mode confinement
- Excellent tunability of the compression factor
- Minimum mismatch, and low propagation loss, and
- Potential for easy integration into chip-based devices.

For the first time, a proof-of-principle demonstration of THz passive pulse compression was shown based on the portion of electromagnetic spectra where the normalized group velocity ($V_{gr}/c$) profile has a large negative gradient.

The possibilities of terahertz's direct pulse compression in a dielectric-lined circular waveguide with tunability based on a strong both waveguide and material dispersions introduced by helical graphene-loading have been demonstrated. We use helix in our structure as a primary waveguide dispersion controlling element. We show that the proposed structure introduces an excellent tunability of the compression factor via the graphene electrostatic bias (see Chapter 4). The circuit model of the helical waveguide can be expressed by one inductance per unit length and two radial and inter-turn capacitances per unit length (see Fig. 4-5). Thus, a strong dispersion due to helix-shield capacitive coupling voltages and helix equivalent inductive currents can be introduced. This effect mainly depends on the dielectric tube parameters ($t_d$, $\varepsilon_d$) for the former, and the helix characteristics ($\sigma$, $w$, $p$) for the latter, where $\sigma$ is the conductivity of helix-tape that in our proposed structure is an equivalent dispersive conductivity resulted from hybrid graphene-ribbon and DC-bias facilities of gold-ribbon. It is demonstrated that there is an optimal compression waveguide length over which THz chirped pulses reach the maximum compression. It is shown that by applying an electrostatic controlling gate voltage ($V_g$) of 0 and 30 V on the helical graphene ribbon, the temporal input pulses of width 8 and 12 ps, propagating through two different lengths (700 μm and 1700 μm), can be tuned by 5.9% and 8%, respectively, in the frequency range of 2.15–2.28 THz.

In chapter 5, to confirm the accuracy of the previously achieved results of the proposed tunable graphene-based pulse compressor, first of all, the mesh size sensitivity is analyzed,and then we made a step forward for a passive fixed pulse-shaper by replacing the helical graphene ribbon with a gold ribbon and compared to the previous proposed tunable pulse compressor, graphene-based one. Moreover, to prove the validity of the numerical analysis, the proposed device has been examined using the finite element method (FEM) and the finite integral technique (FIT).A good agreement was found between the results of the FIT and the FEM methods.

Another outstanding achievement of this research was to provide a comprehensive system model by incorporating the full-wave time-domain simulations and the numerical transfer function estimation approach. The use of the system transfer function to analyze the structure is preferable to the full-wave simulation because of saving the execution time. Once the transfer function is determined, one could apply it for the subsequent time-domain analysis of the pulse shaper with various inputs.

Based on a comparative study that was performed on the impact of helical-ribbon material on the proposed terahertz waveguide-based pulse compressor performance for different material of helical ribbon in cases of (i) PEC, (ii) Au, and (iii) graphene-coated Au, the material-dispersion of the structure was highlighted. As shown in Fig. 6-6, pulse side lobes of Au-graphene coated compared to Au-ribbon have lower amplitudes and in conclusion higher pulse quality, this could be predicted from higher dispersion that was seen from the phase of the graphene reflectivity coefficient ($\angle\Gamma_E$) which was shown in Fig. 6-2 .

Moving on from this thesis, there are three areas of further development that would be interesting for direct passive pulse compression. Firstly, work can be done with a comparison of different waveguide platforms based on geometry (planar/non-planar). In this thesis, a normal size hollow-core waveguide (HCW) as the main platform was used while usually the over-sized HCWs with the emphasis on the low loss transmission of the $HE_{11}$ mode in the infrared and THz regions should be used. In some applications that an integration development is required, the planar platform is the most popular solution.

Another area in which additional research can be carried out is the THz sensing/spectroscopy experiments have been targeted at particular substances such as toxic materials or biological samples using conventional methods such as transmission spectroscopy. Comparatively, work done on THz materials sensing has been focused on detecting refractive index changes caused by using different materials to come into contact with the materials sensor. Among different types of the optical waveguide, hollow core, and porous core for sensing applications as it allows a greater analyte volume inside its core area are more suitable. Moreover, the guided mode in a hollow-core fiber is strongly confined in the core area that greatly reduces the effect of the background material in the waveguiding properties of the fiber [93]. Therefore, as a possible solution, we can propose a terahertz sensor based on our proposed tunable hollow-core waveguide for chemical or biological sample detection.

In order to realize a compact high-speed optical communications system, nonlinear photonics is seen as a key technology. Optical fiber materials exhibit a nonlinear response to strong electric fields, such those of optical signals confined within the small fiber core [94]. The third research direction would be to incorporate nonlinearity onto the graphene-based pulse compressor for optical waveform generation and pulse shaping and in the applications of advanced pulse profiles in all-optical signal processing. Silicon carbide (SiC), a high-transparency nonlinear material, has very promising properties for THz applications, including extremely low loss, ultrabroad transparency window, very high optical damage threshold, moderate nonlinearity, and birefringence [95]. Of course, these previously proposed ideas are only some suggestions for terahertz pulse shaping on the basis of the development of the waveguide-based structure, there is no doubt that many more ways could be found via the creative use of various materials and geometries.

# Appendix A: A Brief Theory on Dispersion and Pulse Compression

As a temporal pulse travels through a dispersive guided system, a degradation of the shape (broadening, compression, or chirping) could be observed. A simple illustration of how dispersion parameters can be used to determine the pulse shaping is by the evaluation of their effects on a Gaussian pulse of the duration $T$ after passing through the waveguide system of length $L$. For simplicity, we assume only a linearly polarized temporal Gaussian pulse at the reference plane $z=0$ as follows:

$$\boldsymbol{E}(z,t) = \boldsymbol{E_0} \, exp\left(-\frac{1}{2}\left(\frac{t}{T}\right)^2\right) exp(j\omega_0 t) \tag{A.1}$$

Equivalently, the normalized output Gaussian pulse spectrum of a guided system can also be shown as:

$$\boldsymbol{E^{norm}}(L,\omega) = exp\left(-\frac{1}{2}(T)^2(\omega-\omega_0)^2\right) exp(j\beta(\omega)L) \tag{A.2}$$

By replacing $\beta(\omega)$ with its quadratic dispersion approximated expansion (Refer to Chapter 3) and applying inverse Fourier transform, the time domain expression of the output pulse:

$$\boldsymbol{E^n}(z,t)|_{z=L} = exp\left(-\frac{1}{2}\frac{(t-\beta_1 L)^2}{(T^2+(\beta_2 L/T)^2)}\right) exp\left(\frac{i}{2}\cdot(\beta_2 L/T^2)\frac{(t-\beta_1 L)^2}{(T^2+(\beta_2 L/T)^2)}\right) exp(j(\omega_0 t - \beta_0 L)) \tag{A.3}$$

It can be observed that the output pulse is also Gaussian pulse but with a time shift and a new pulse duration for the Gaussian envelope as $\beta_1 L$ and $T^{'}=[T^2+(\beta_2 L/T)^2]^{1/2}$, respectively. For these two meaningful time-related changes, $\beta_1$ and $\beta_2$ are the essential parameters in the time domain analysis of a dispersive propagating system. In the literature [62], [65] the $\beta_1 L$ and $\beta_2$ are so-called group delay (GD) and group velocity dispersion (GVD), respectively. It is clear from the new pulse duration ($T^{'}$) relation that the magnitude of the pulse shaping can be determined by the term of $\Delta\tau$. Therefore, the GVD, as it is stated in the Eq. (A.4) is an important responsible parameter for pulse shaping applications, including pulse broadening, compression, and chirping.

$$\Delta\tau = \frac{\beta_2 L}{T} = GVD \cdot \frac{L}{T} \qquad \text{(A.4)}$$

For a dispersive medium, generally, we can suppose three cases of normal dispersion ($\Delta\tau{>}0$) and anomalous dispersion ($\Delta\tau{<}0$) or even zero-dispersion ($\Delta\tau{=}0$). Another helpful parameter to express the overall dispersion effect of optical or THz fibers is the total induced dispersion coefficient $D$ in *ps/km·nm*, as:

$$D(\omega) = -\frac{\omega^2}{2\pi c} \cdot \beta_2 = -\frac{\omega^2}{2\pi c} \cdot GVD \qquad \text{(A.5)}$$

Chromatic dispersion coefficient $D$ can be zero, negative, or positive over an operating frequency range of interest. The frequency, at which $D$ is zero, is referred to as zero-dispersion frequency. In the case of *D>0*(anomalous dispersion), the higher frequency components travel faster than lower frequency ones, but for the frequency region where the dispersion coefficient $D$ is negative (normal dispersion), the lower frequency components travel faster than higher frequency ones. As an important conclusion, parameter $D$ versus the angular frequency $\omega$ is an essential graph in the dispersion analysis.

# Appendix B

## The General Modeling Consideration and Simulation Characteristics

### B1. Model Designing of a Helix with Rectangular Cross-section

The following section deals with the procedure of creating a helix with the rectangular cross section in CST Microwave Studio and COMSOL environment.

- **CST Microwave Studio v.2015**

Drawing a z-axis aligned helix with the design parameters of: tape helix thickness of tgr=0.1[um], width and pitch of helix {w1=41[um], p1=126[um]}, helix internal radius R=100[um], n_seg=40, and helix final turns nt=6.

1. Define a zero thickness **Brick: x**min=R xmax=R+tgr, ymin=ymax=0, zmin=0 zmax=w1

2. From **Modeling>Picks** select **Pick face**

3. Double click on the **Brick face**

4. From **Modeling>Shapes** select **Rotate**

5. Ask you "**Do you wish to define a rotation axis numerically?**" click **OK**, then click **ESC** button to open "**Enter Edge Numerically**" window, then set **z2=1** and leave all in zero value, click **OK**. The "**Rotate Face**" window will be open.

6. Complete "**Rotate Face**" window as follows:
   **Name:** Helix, **Angle:** nt*360, **Height:** nt*p1, **Radius Ratio:** 1.0 and **Segments per turn:** n_seg.

- **COMSOL 5.2a**

Drawing a helix, x-axis aligned, with rectangular cross section first of all we define the parameters as tape helix thickness of tgr=0.1[um], width and pitch of helix {w1=41[um], p1=126[um]}, helix internal radius R=100[um] and helix final turns nt=6.

*Work Plane 1 (wp1)*

1. On the **Geometry** toolbar, click **Work Plane**.

2. In the **Settings** window for Work Plane, click **Show Work Plane**.

3. In plane Definition, from the **Plane** list, choose **yx-plane**.

*Rectangle 1 (r1)*

**1** On the **Work Plane** toolbar, click **Primitives** and choose **Rectangle**.

**2** In the **Settings** window for Rectangle, locate the **Size and Shape** section.

**3** In the **Width** text field, type tgr.

**4** In the **Height** text field, type w1.

**5** Locate the **Position** section. From the **Base** list, choose **Corner**, with **xw**=R and **yw**=0.

**6** Right-click **Rectangle 1 (r1)** and choose **Build Selected**.

Next add a parametric curve using the helical profile.

*Parametric Curve 1 (pc1)*

**1** On the **Work Plane** toolbar, click **Primitives** and choose **Parametric Curve**.

**2** In the **Settings** window for Parametric Curve, in the **Parameter** section type the parameter **Name** text field s.

**3** In the **Maximum** text field, type 2*pi.

**4** Locate the **Expressions** section. In the **x** text field, type (s*nt*p1)/(2*pi).

**5** In the **y** text field, type cos(nt*s).

**6** In the **z** text field, type sin(nt*s).

**7** Locate the **Position** section. In the **x** text field, type 1, and 0 in the y and z fields.

**8** Locate the **Axis** section. From the **Axis** list, choose **z-axis.**

**9** Build all.

As the final step we should sweep the rectangular cross section *wp1* along the helical parametric curve of *pc1*.

*Sweep 1 (swe1)*

**1** On the **Geometry** toolbar, click **Sweep**.

**2** In the **Settings** window for Sweep, in the **Cross Section** field locate the *wp1* as the Face to sweep.

**3** In the **Spine Curve** field locate the *pc1* as the Edges to follow.

# B2. Transient Solver

- **CST Microwave Studio v.15**

In CST MICROWAVE STUDIO two time domain solvers are available. One is based on the Finite Integration Technique (FIT), just called Transient solve and another is Transmission Line Model (TLM) which work both on hexahedral meshes, we used in this study FIT method. For better postprocessing operations for obtaining dispersive reports such as GD, GV, and GVD its strongly recommended to use a broadband Raised-cosine pulse (generated based on MATLAB codes presented in C2) for limited number of time samples to avoid unnecessary extra computation time and minimum ringing on the tail of the output pulse. The pulse responses of the proposed structure is obtained based on our generated chirped signal using MATLAB codes of C1 instead of default chirped pulse defined in the software.

- **COMSOL 5.2a**

In COMSOL the time samples needed for excitation can be directly defined as follows:

First of all define the required new parameters as E0=30[V/m], T0=8[ps], Tao=0.2[ps], Tf=14[ps], f1=1.95[THz], f2=2.45[THz], Df=f2-f1, and mue=-2*pi*Df/Tf.

Then we should define a Rectangle function with ON and OFF time windows of [0-T0] and [T0-Tf], respectively.

*Rectangle 1 (rect1)*

**1** On the **Home** toolbar, click **Difinition** and choose **More Functions>Rectangular**.

**2** In the **Settings** window for Step function, Locate the **Parameters** section. In the **Lower** and **Upper Limits** text field, type: 0 and T0, respectively.

*Analytic 1 (an1)*

1   On the **Home> Definition** toolbar, click **Local Variables**.

2   In the **Settings** window for Local Variable, type *Echp* in the **variable name** text field.

3   In the **Expression** text field, type:

E0*rect1(t[1/s])*(1-exp(-(t/Tao)))*sin(2*pi*t*f2+0.5*mue*t^2)

For transient analysis of the waveguide problems we define two *Scattering Boundary Conditions* on input and output ports and then the first boundary is chosen as excitation port with the following setting:

*Scattering Boundary Condition 1*

1   On the **Physics** toolbar, click **Boundaries** and choose **Scattering Boundary Condition**.

2   In the **Settings** window for Scattering Boundary Condition, locate the **Boundary Selection** section and choose the input aperture of the waveguide.

3   In **Scattering Boundary Condition,** from the **Incident field** list, choose **Wave given by E field**.

4   Then type the Incident electric field vector as E0: [0, 0, Echp].

## B3. Frequency Domain Solver

▪ **CST Microwave Studio v.2015**

It is recommended that for a proper comparison condition in the *Frequency Domain Solver Parameter*, in *Frequency Samples* Section, set the *type* at "*Equidistance*" with Samples number equal with that will use in COMSOL.

▪ **COMSOL 5.2a**

In electromagnetic problem with more complexity in geometry including dispersive materials, such as our structure the Generalized Minimal Residual Method (GMRES), which is an iterative method for the numerical solution of a nonsymmetric system of linear equations has been chosen for fast convergence.

## B4. Dispersive Material Definition

At optical frequencies and lower (THz), noble metals such as Silver and Gold, behave like "plasmas", this model of the classical gas of free electrons is called the Drude model or Drude–Lorentz model. The optical response of a free electron gas is approximately described by the Drude–Lorentz model as stated in Eq.(3-3) in terms of three fundamental drude parameters of ωp (plasma frequency), $\epsilon_\infty$ (dielectric function at infinite frequency), and τ (relaxation time). Graphene is also can be modeled in linear mode via Drude-like model based on the Kubo formula which was described in Sec (4-2-1) and Eq.(4-5). Here we introduce the instruction of defining Drude-modeled metals such as gold and Drude-like model of graphene in two forms of (i) as a zero-thickness impedance surface of equivalent conductivity and (ii) a dielectric thin-layer with a Drude-like material with complex permittivity.

▪ **CST Microwave Studio v.2015**

*Definition of Gold*

Nobel metals such as Gold in optical regime can be modeled by using Drude model as follows:

1. **Material Properties>Dispersion>Dielectric dispersion**, from **Disp. Model** list choose **Drude**.

2. In the **Epsilon infinity** field type 9.1, and type 1.38e16 (rad/s) in **Plasma frequency** text field, then type in **collision frequency text** field 1.075e14 (1/s).

*Definition of Graphene*

Graphene can be defined as (i) a thin-layer dispersive dielectric with complex permittivity, or (ii) zero thickness surface boundaries with complex surface impedance. First of all, there is a common operation for both two cases, as follows:

1. As a common step go to **Home>Macros>Material>Create Graphene Material for Optical Applications.**

2. Type the required preconditions of the problem in their text fields in "**Define Graphene Materials**" window as the following Table:

| Temperature [K] | Chemical Potential [eV] | Relaxation Time [ps] | Thickness [nm] | Frequency [THz] | | Number Of Points |
|---|---|---|---|---|---|---|
| | | | | Minimum | Maximum | |
| 300 | 0 | 10 | 1 | 1.7 | 3 | 500 |

These items will be appeared in Parameter list table. Then, two subfolders of Graphene and Graphene1, including Dielectric constants and Surface Impedance data, can be used to assign the proper model.

*Graphene as Dispersive Dielectric*

1. Right-click on the **helical ribbon** with thickness of tgr, choose **Change material**.

2. In **Change Material** window, from **Material** list select **Graphene_Eps**.

*Graphene as Complex Surface Impedance*

1. Right-click on the **helical ribbon** with zero thickness, choose **Change material**.

2. In **Change Material** window, from **Material** list select **Graphene**.

▪ **COMSOL 5.2a**

*Definition of Gold*

In COMSOL dispersive materials can be defined by creating a separate wave equation the material domain as follows:

*Wave Equation, Electric 2*

Create a separate wave equation feature for the Gold-ribbon domains.

1. On the **Physics** toolbar, click **Domains** and choose **Wave Equation, Electric**.
2. In the **Settings** window for Wave Equation, Electric, locate the **Domain Selection** section.
3. From the **Domain Selection**, choose **Helical Ribbon**.
4. Locate the **Electric Displacement Field** section. From the **Electric displacement field model** list, choose **Drude-Lorentz dispersion model**.
5. In the ε∞ value field type **9.1**, and in ω**P** text field, type **1.38e16 rad/s**.
6. In the table, enter the following settings:

| Oscillator strength (1) | Resonance frequency (rad/s) | Damping in time (rad/s) |
|:---:|:---:|:---:|
| 1 | 0 | **(1.075e14)/(2*pi)** |

*Definition of Graphene*

Definition of Graphene is similar a dispersive material but the main difference lies in the expressed Plasma frequency (ω$_P$). In the **Model Builder** window, under **Component 1 (comp1)** right-click **Definitions** and choose **Variables**.

*Variables 1*

1. In the **Settings** window for Variables, locate the **Variables** section.
2. In the table, enter the following settings:

| Name | Expression | Unit | Description |
|---|---|---|---|
| temp | 300[K] | [K] | Temperature |
| Ef | 0[eV] | J | Fermi level |
| tgr | 1[nm] | m | Graphene thickness |
| tau | 1e-13[s] | [s] | Relaxation time |
| gamma_1 | 1/tau | [1/s] | Damping in time |
| omega_1 | 0 [rad/s] | [rad/s] | Resonance freq. |
| omega_p | sqrt(log(2*cosh(Ef/(2*k_B_const*temp)))* (2*(e_const^2)*k_B_const*temp)/( pi*(hbar_const^2)* (epsilon0_const*tgr))) | [rad/s] | Plasma frequency for Ef (eV) |

In addition, there is a small difference in the required setting of Drude-Lorentz dispersion model in TD and FD.

## Time Domain Study

*Wave Equation, Electric 1*

1. In the **Model Builder** window, under **Component 1 (comp1)>Electromagnetic Waves, Transient (ewt)** click **Wave Equation, Electric.**

2. In the **Settings** window for Wave Equation, Electric, locate the **Electric Displacement Field** section.

3. From the **Electric displacement field model** list, choose **Drude-Lorentz dispersion model**.

4. From the **Relative permittivity, high frequency** list, choose **User defined**. From the list, choose **Diagonal**.

5. In the **Relative permittivity, high frequency** table, enter **1** for all diagonal elements.

6. In the ωP text field, type **omega_p**.

7. Locate the **Magnetic Field** section. From the $\mu_r$ list, choose **User defined**. Accept the default value 1.

8. Locate the **Conduction Current** section. From the σ list, choose **User defined**. Accept the default value 0.

Next, you add a Drude-Lorentz Polarization feature, as a sub feature to the wave equation. There, more material parameters will be defined for the polarization field.

*Drude-Lorentz Polarization 1*

1. On the **Physics** toolbar, click **Attributes** and choose **Drude-Lorentz Polarization**.

2. In the **Settings** window for Drude-Lorentz Polarization, locate the **Drude-Lorentz Dispersion Model** section.

3. In the *fn* text field, type 1.

4. In the ω*n* text field, type **omega_1**.

5. In the Γ*n* text field, type **gamma_1**.

## Frequency Domain Study

*Wave Equation, Electric 1*

Create a separate wave equation feature for the Gold-ribbon domains.

1. On the **Physics** toolbar, click **Domains** and choose **Wave Equation, Electric**.

2. In the **Settings** window for Wave Equation, Electric, locate the **Domain Selection** section.

3. From the **Domain Selection**, choose **Helical Ribbon**.

4. Locate the **Electric Displacement Field** section. From the **Electric displacement field model** list, choose **Drude-Lorentz dispersion model**.

5. In the $\varepsilon_\infty$ value field type **1**, and in **$\omega_P$** text field, type **omega_p**.

6. In the table, enter the following settings:

| Oscillator strength (1) | Resonance frequency (rad/s) | Damping in time (rad/s) |
|:---:|:---:|:---:|
| **1** | **omega_1** | **gamma_1/(2*pi)** |

## B5. Dispersion Reports Post-processing

▪ **CST Microwave Studio v.2015**

As the important part of our simulation task is the extraction of the required dispersion results such as group delay (GD), normalized group velocity (Vgr/c) and group velocity dispersion (GVD). In CST MWs, in Filter Analysis section of bout Frequency and Transient solvers, in **Post processing>Template Based Post Processing**, the main parameter of GD is available, then from GD-report all required dispersion results can be calculated. Here is a brief about the Post processing operation to produce Vgr/c, GVD and axial propagation phase constant from GD results.

## Vgr/c

1. Choose **General 1D>0D or 1D Results from 1D Results**.
2. In **Specify Action** section choose **Axes: Scale with Function f(x.y)** from the list of **1D(C).**
3. Type in the text field the rescaled relation of Vgr/c and GD based on the current unit set of the CST MWs, here is the proper text field compatible with the unit presets:

$$(V_{gr}/c) = \frac{L(m)}{3e8(m/s)GD(s)} \qquad (B.1)$$

As an example if the CST MWs units that was set on *THz*, *µm* and *Second* the text field can be rescaled as Eq.(B.2)

$$(V_{gr}/c) = \frac{L(um)}{3e8((1e6)um/s)GD(s)} = (1e - 14) * (\frac{L}{3 * y}) \qquad (B.2)$$

4. Choose Apply to: **y-axis.**
5. Set **1D-Results** field from the list on **GDelay_2(1)1(1).**
6. Evaluate.

**β(Axial Propagation Constant)**

1. Choose **General 1D>0D or 1D Results from 1D Results**.

2. In **Specify Action** section choose **Integral** from the list of **1D(C).**

3. Type in the text field the rescaled relation of **β** and **GD** based on the current unit set of the CST MWs, here is the proper text field compatible with the unit presets:

$$V_{gr} = \frac{1}{(\partial\beta/\partial\omega)} \Rightarrow \beta(\frac{rad}{m}) = \frac{1}{L(m)} \int GD(s) \, d\omega (rad \cdot Hz) \qquad (B.3)$$

Once again with units that was set on *THz*, *μm* and *Second* the text field can be rescaled as Eq.(B.4)

$$\beta(\frac{rad}{m}) = \frac{1e-6}{L(um)} \int GD(s) \, d\omega (rad \cdot (1e12)) = (1e6) * (y/L) \qquad (B.4)$$

4. Choose Apply to: **y-axis.**

5. Set **1D-Results** field from the list on **GDelay_2(1)1(1)**

6. Evaluate

7. Renamed this report as **Beta**

## Group Velocity Dispersion (GVD)

Group velocity dispersion is the frequency dependence of the group velocity in a medium, or quantitatively the derivative of the inverse group velocity with respect to angular frequency or equivalently second derivative of axial phase constant (β). Thus, based on the previous calculated parameter of β, simply the GVD can be computed as follows:

1. Choose **General 1D>0D or 1D Results from 1D Results**.

2. In **Specify Action** section choose **Derivative** from the list of **1D(C).**

3. Set **1D-Results** field from the list on **Beta**

4. Evaluate

5. Renamed this report as **Beta1**

Repeat these steps as the second derivation of β to obtain GVD.

    **1**   Choose **General 1D>0D or 1D Results from 1D Results**.

    **2**   In **Specify Action** section choose **Derivative** from the list of **1D(C).**

    **3**   Set **1D-Results** field from the list on **Beta1**

    **4**   Evaluate

    **5**   Renamed this report as **GVD**

    ▪   **COMSOL 5.2a**

In COMSOL the dispersion reports such as Vgr, β and GVD, as previously stated, can be calculated based on GD using $\angle S_{21}$, see Eq. <span style="color:red">(3-16)</span>.

**Phase Angle of S$_{21}$**

*1D Plot Group 1*

    **1**   On the **Home** toolbar, click **Add Plot Group** and choose **1D Plot Group**.

*Global 1*

    **1**   On the **1D Plot Group 1** toolbar, click **Global**.

    **2**   In the **Settings** window for Global, locate the **y-Axis Data** section for **Expression** the following expression: **arg(emw.S21)*180/pi**

    **3**   Renamed this report as **angle of S21(deg)**

This command would be generate $\angle S_{21}$ in degree.

# Appendix C

# MATLAB Codes

## C.1 Linear Chirped Signal Generation

```
%---------------Linear FM
f1=1.18e12;        %Lower Freq of Sweeping Range1THz
f2=2.32e12;        %Upper Freq of Sweeping Range5THz
BW=max(f1,f2); % Bandwidth
m1=15;            %coef. of resolution

Tao=0.2e-12;      %1.4e9
Fs=m1*BW;          % Sampling frequency
T0=12.0e-12; td=0;        % Pulse duration  2e-9
Tf=35.0e-12;      % Time Window for seeing the time signal with a good view!6e-9

DelF=f2-f1;
mue=-2*pi*DelF/Tf;%mue=-2*pi*(f2-f1)/Tf;
%-------------- Create a FM sine wave in Hz.
DT=1/Fs;
N=Tf*Fs;
t=zeros(1,N);
x=zeros(1,N);
for n=1:N
t =n*DT; % Time vector of Tf second
if t<T0   %for sin
%x(n) = exp(-((t-T0-td)/Tao).^2).*sin(2* pi*t*f2+0.5*mue*t.^2);

  x(n) = (1-exp(-(t/Tao))).*sin(2* pi*t*f2+0.5*mue*t.^2);
else      %for sin
   x(n)=0;%for sin
end       %for sin

end
%-------------- Take fft, padding with zeros so that length(X)
nfft = 2^18;  % Length of FFT
X = fft(x,nfft);
%-------------- FFT is symmetric, throw away half
X = X(1:nfft/2);
%-------------- Take the magnitude of fft of x
Fx = abs(X);
AFx=angle(X);
%-------------- Frequency vector
f = (0:nfft/2-1)*Fs/nfft; %
%-------------- Generate the plot, title and labels.

figure(1);
subplot(2,1,1)
n=1:N;
```

```
plot(n*DT,x);
title('Linear Swept FM Signal No chirped  Per14-4.txt ');legend(['Tao: ',num2str(Tao*1e12),'T0:
',num2str(T0*1e12),';Sample No.:',num2str(N),';  Tf=',num2str(Tf),';  f1=',num2str(f1*1e-12),';
f2=',num2str(f2*1e-12),',';  m1=',num2str(m1),'']);
xlabel('Time (s)');
ylabel('x(t) a.u.');ylim([-1.5 1.5]);grid on
data1=[n*DT;x];
data=transpose(data1);
dlmwrite('Per14_4.txt',data)

subplot(2,1,2)
plot(f,Fx);
title('Magnitude Frequency Spectrum of a chirped Signal');
xlabel('Frequency (Hz)');xlim([0 2*BW]);grid on
ylabel('|X(f)|a.u.');
print -djpeg -r300  Per14-4.jpg
```

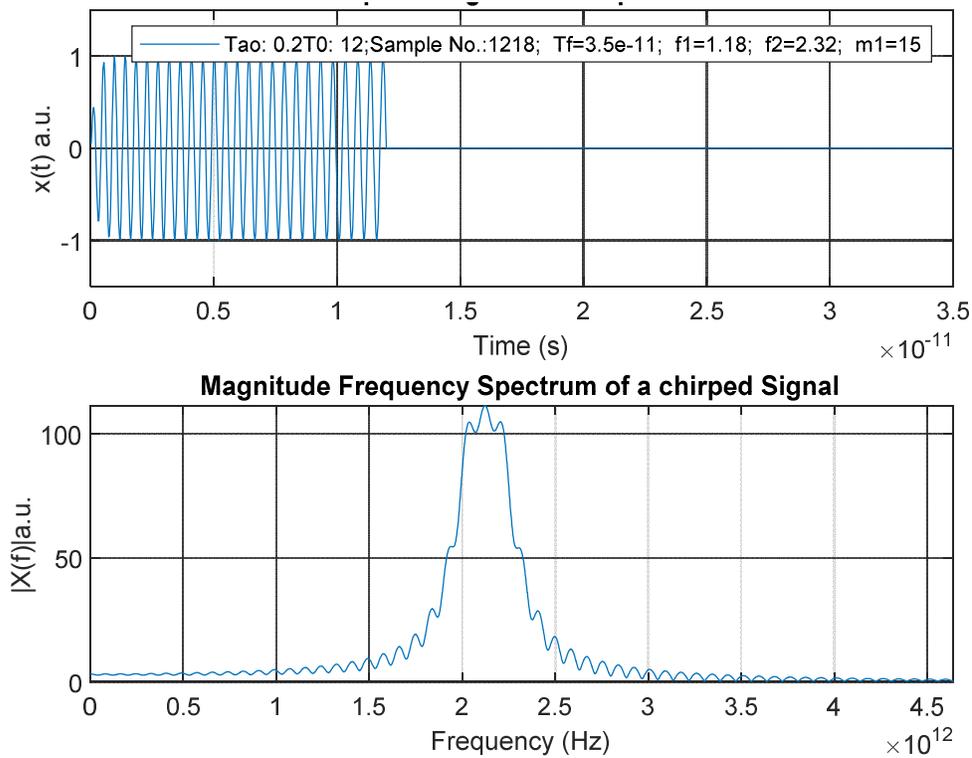

**FIG. C-1 The linear frequency modulated sinusoidal pulse and its frequency spectrum.**

## C.2 Raised Cosine Signal Generation

```
clear all;
close all;
clc;
%-------------- RaisedCosinePulseMod.m --------------%
a1=1.5;          % Roll off Fator
f0=2.0e12;       % Lower Freq of Sweeping Range1THz
BW=max(f0);      % Bandwidth
Ts=1.5e-12;      % Symbol Duration
Fs=20*BW;        % Sampling frequency
T0=5.0e-12;      % Pulse duration  2e-9
Tf=20.0e-12;     % Time
%-------------- Create a Raised Cosine Pulse in Hz.
DT=1/Fs;
N=Tf*Fs;
t=zeros(1,N);
x=zeros(1,N);x1=zeros(1,N);x2=zeros(1,N);
for n=1:N
t =n*DT; % Time vector of Tf second
x1(n)=( sin(((t-T0)/Ts))./(((t-T0)/Ts)));
x2(n)=(cos(2*pi*a1*(t-T0)/Ts)./(1-(16*(a1*(t-T0)/Ts).^2)));
x(n) =x1(n)*x2(n)*cos(2*pi*f0*t);

end
%-------------- Take fft, padding with zeros so that length(X)
nfft = 2^18;  % Length of FFT
X = fft(x,nfft);
%-------------- FFT is symmetric, throw away half
X = X(1:nfft/2);
%-------------- Take the magnitude of fft of x
Fx = abs(X);
AFx=angle(X);
%-------------- Frequency vector
f = (0:nfft/2-1)*Fs/nfft; %
%-------------- Generate the plot, title and labels.
figure(1);
subplot(2,1,1)

n=1:N;
plot(n*DT,x,'r' ,'linewidth', 1.75);set(gca,'linewidth', 1.5,'fontsize',12.5,'fontname','Arial')
title('Moulated Raised Cosine Pulse');%legend(['Ts: ',num2str(Ts*1e12),'T0: ',...
   % num2str(T0*1e12),';Roll-off Fator:',num2str(a1),...
  %  ' Tf=',num2str(Tf),';Sample No.:',num2str(N),';  f0=',num2str(f0*1e-12),'']);
xlabel('Time (s)');set(gca,'XMinorTick','on','YMinorTick','on')
ylabel('x(t) a.u.');grid on %ylim([-1.5 1.5]);
data1=[n*DT;x];
data=transpose(data1);
dlmwrite('RaisedCosinePulseMod1.txt',data)
subplot(2,1,2)
```

```
plot(f,abs(Fx)./max(abs(Fx)),'linewidth', 1.75 );set(gca,'linewidth',
1.5,'fontsize',12.5,'fontname','Arial')
title('Frequency Spectrum');set(gca,'XMinorTick','on','YMinorTick','on')
xlabel('Frequency (Hz)');grid on
xlim([0 2*BW]);
ylabel('|X(f)|a.u.');
print -djpeg -r300   RaisedCosinePulseMod1.jpg
```

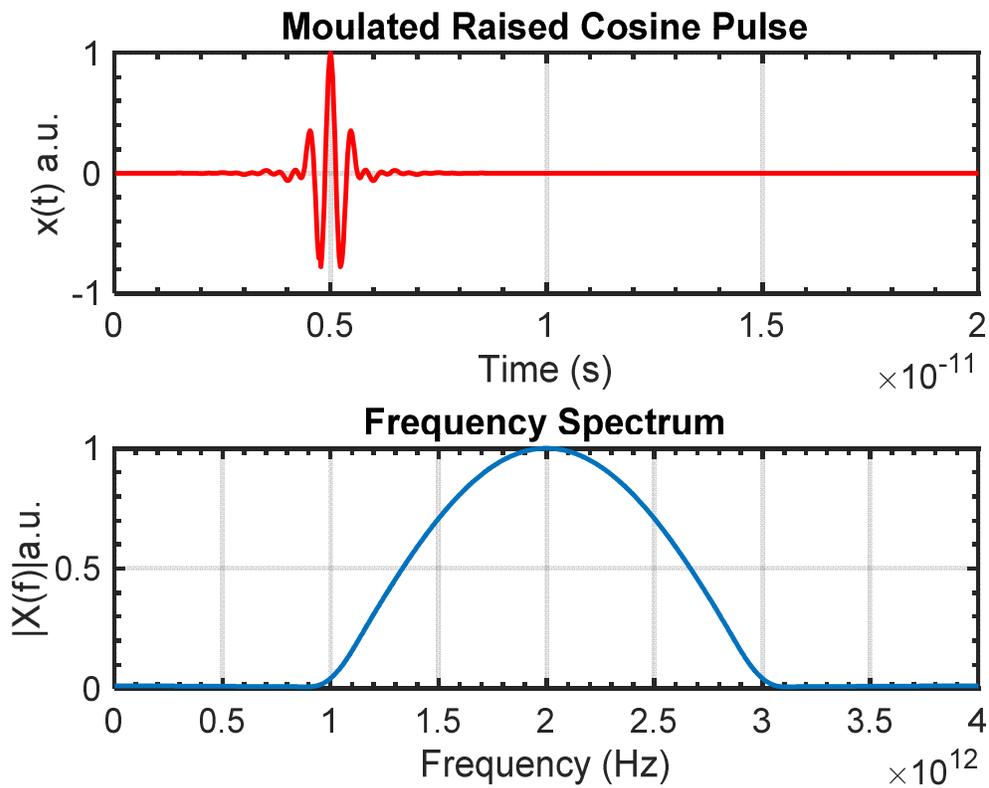

**FIG. C-2** The modulated raised cosine pulse and its frequency spectrum.

## C3. Transfer Function Estimation based on Time Data

```
%%%  Transfer Function Estimation Based on
%%% Time data obtained from simulation
%%% Copyright Line: If you want to use this code
%%% The following notice must appear with the material
%%% (please fill in the citation information):
%%% Reproduced from [FULL CITATION], with the permission of Author."
clc;clear;tic
clear;Ratio_x=3.6;Ratio_y=1.2;
A=xlsread('data011eV.xls');
B = A(:,1)*(1e12);
DT = B(15,1)- B(14,1);
u1=A(:,2);%input
y1=A(:,3);%output
Ts=(B(5,1)-B(4,1))*(1e-12);fs = 1/Ts;
length(A(:,3));
data=iddata(y1,u1,Ts);
%figure(1)
%plot(data);grid on
np=9  ;nz=4;
sys_est  =tfest( data,np,nz)
[num,den] = tfdata(sys_est,'v') ;
%----------------------
Gz=zpk(sys_est) ;
options = bodeoptions;
options.FreqUnits = 'Hz'; % or 'rad/second', 'rpm', etc.
options.FreqScale = 'log';
options.Title.FontSize = 12;
options.XLabel.FontSize = 12;
options.YLabel.FontSize = 12;
options.TickLabel.FontSize = 12;

figure(2)
%bode(sys_est,{5e12 12e12}, options);grid on
%----------------------
[mag,phase,wout] = bode(sys_est,{2e12 1e13}, options);
Mag=20*log10(mag(:));Phase=phase(:);
subplot(2,1,1)
set(gca,'fontsize',13);
semilogx(wout,Mag,'LineWidth',1.25);grid on;
title('Estimated TF Mag. of Bode Plot');ylabel('dB');
xlabel('Frequency(Hz)');
subplot(2,1,2)
set(gca,'fontsize',13);
semilogx(wout,Phase,'r','LineWidth',1.25);grid on;
title('Estimated TF Phase of Bode Plot');ylabel('deg');
xlabel('Frequency(Hz)');
%----------------------
```

```
figure(3)
[y_out, time] = lsim(sys_est,u1,B(:,1)*(1e-12) ); M=length(time);
for m=1:M
time1(m,1)=(m-1)*(14e-12)/M;
end
subplot(3,1,1);
plot(time1*(1e12), u1/max(u1), 'b','linewidth',1.25);
set(gca,'linewidth', 1.5,'fontsize',11.5,'fontname','Arial')
legend('Input Chirped Sig.', 'Box','off','fontsize',11.0,'Location','northwest');
grid off ;xlim([0 12]);ylim([-2 2]);yticks([-2 -1 0 1 2]);
%-------------
set(gca,'XMinorTick','on','YMinorTick','on')
xlabel({'Time (ps)';'(a)'},'fontsize',12.5);
ylabel('Electric Field (V/m)','fontsize',12.0);
%--------------------------
subplot(3,1,3);
plot(time1*(1e12), 1.66*y_out/max(y_out), 'r','linewidth',1.25);
set(gca,'linewidth', 1.5,'fontsize',11.5,'fontname','Arial')
%title('Output based on Estimated TF');
legend('Output Sig.(Estimated TF)', 'Box','off','fontsize',11.0,'Location','northwest');
grid off ;xlim([0 12]);ylim([-2 2]);yticks([-2 -1 0 1 2]);
%-------------
set(gca,'XMinorTick','on','YMinorTick','on')
xlabel({'Time (ps)';'(c)'},'fontsize',12.5);
ylabel('Electric Field (V/m)','fontsize',12.0);
%----
hold on
subplot(3,1,2);
plot(time1*(1E12), 1.66*y1/max(y1),'b','linewidth',1.25);
set(gca,'linewidth', 1.5,'fontsize',11.5,'fontname','Arial')
%title('Output based on CST Simulation ');
legend('Output Sig.(FEM-method)', 'Box','off','fontsize',11.0,'Location','northwest');
grid off ;xlim([0 12]);ylim([-2 2]);yticks([-2 -1 0 1 2]);
%-------------
set(gca,'XMinorTick','on','YMinorTick','on')
xlabel({'Time (ps)';'(b)'},'fontsize',12.5);
ylabel('Electric Field (V/m)','fontsize',12.5);
pbaspect([Ratio_xRatio_y 1])
set(gca,'XMinorTick','on','YMinorTick','on')
%-------------------
set(gcf,'position',[100 100 500 1500])
print -dtiff -r300   Figure3_abc_1
%print('Figure3_abc1','-deps')
%----------------
figure(4)
 [p,z] = pzmap(sys_est)
set(gca,'fontsize',12.5 );
pzmap(sys_est)
toc
```

# Appendix D

## Proposed Fabrication Method

Recent progress in the fabrication of graphene-based devices [96] has shown that our proposed structure, with the dimensions of Table 4-1, could be realized in various ways. The fabrication methods can be broadly categorized as micromechanical exfoliation [97], chemical vapor deposition (CVD) [98, 99], epitaxial growth [100], and the reduction of graphene oxide [101]. Of the various techniques mentioned, CVD has proven to be the most promising approach for the growth of graphene [102].

Cylindrical multi-layered graphene-based structures can be synthesized in two suggested ways: (i) the graphene sheet or ribbon is produced using one of the aforementioned techniques, then rolled up [103, 104] on a dissolvable polymer rod such as polymethyl methacrylate(PMMA)and finally removing the unwanted areas using an effective PMMA solvent like acetone [105]; (ii) CVD growth of graphene on a gold film, which is helically patterned on a PMMA rod by lithography technique.

Then, the hollow-core waveguide will be achieved by removing the PMMA with the help of acetone. The Au ribbon used for applying the external controlling DC voltage could be formed by lithography process. Moreover, HDPE could be realized using a polymer coating solution. Au cladding could be performed as the final step by simple metal coating [106].

## Coupling of Terahertz Wave

The design and implementation of useful guided-wave structures are of increasing interest in transmission of coherent THz radiation at present. This is particularly true for THz radiation in the form of broadband pulses, which are readily available through the use of ultrafast laser systems. In recent years, a number of different waveguide geometries based on dielectric and metal waveguide architectures, have been fabricated and characterized for broadband THz applications. Among these structures, metal waveguides offer a simple means for confining long wavelength electromagnetic radiation.

Based on the theoretical calculations and experimental results it is shown that the linearly polarized incoming THz pulses significantly couple into the TE11, TE12, and TM11 modes of the circular , and TE10 and TM12 modes of rectangular waveguides [107]. Moreover, it is found that using focusing lenses in combination with collimating lenses a high-efficient coupling between terahertz beam and waveguide aperture can be obtained. Figure D-2 shows the schematic a THz-TDS transmission mode setup designed for the optical characterization of the hollow-core THz waveguides with input/output coupling setup [108].

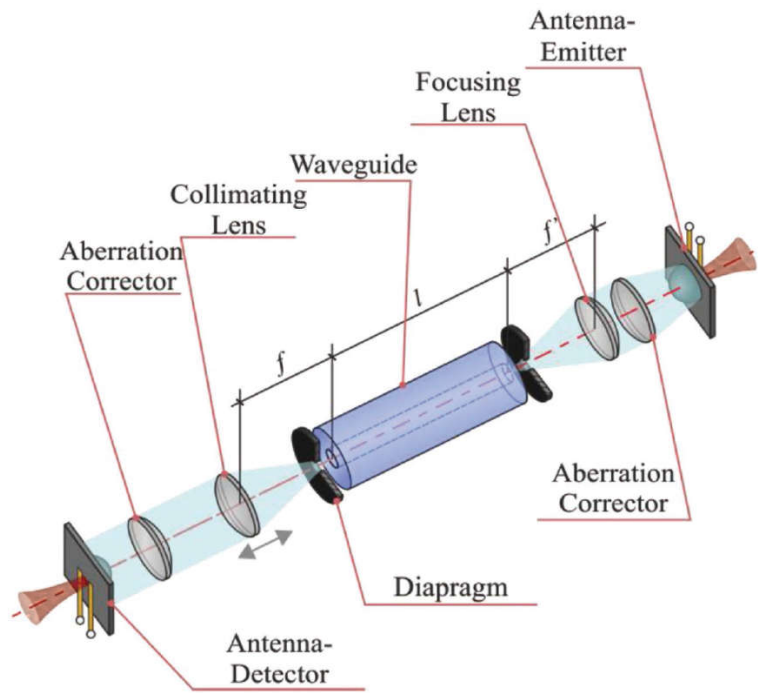

**FIG.D-1** Schematic of the the THz-TDS transmission-mode setup designed for the optical characterization of the hollow-core THz waveguides(Figure 7 with permission from [108]).

## LIST OF PUBLICATIONS

The work presented in this thesis has been published in the following peer-reviewed journal articles:

**Additional Published Work**

In addition to the published works in this dissertation, the author has also published a paper in the research field of terahertz at the following scientific conference:

**چکیده**

مبحث فشرده سازی حوزه زمان پالس های فرکانس بالا در سیستمهای مخابراتی بموجب نیـاز روز افـزون بـه ظرفیت حداکثری انتقال داده از نقش کلیدی برخوردار است. در کاربردهای مدرن مخابراتی باند تراهرتز، استفاده از سیگنالهای مدوله شده بطریقه مدولاسیون فاز یا با چرپ شدگی فرکانس، ضمن بهبود برد انتقال شـرایط لازم برای افزایش پهنای باند انتقال را از طریق افزایش حاصضرب دوره زمانی در پهنای باند پالس فراهم کرده اسـت، ازینرو واحد شکل دهنده پالس کماکان بعنوان یک عنصر کلیدی در سیستمهای مخابراتی مورد توجـه میباشـد. شیوه غالب یک پالس زمانی با دوره زمانی حضور کوتاه بوسیله فشرده سازهای بـا ویژگـی پاشـندگی خطـی یـا غیرخطی که از یک سیستم، خط انتقال یا موجبر بـا پاشـندگی بـالا شـکل گرفتـه، قابـل تحقـق اسـت. چنـین ساختاری میتواند از مکانیزم های متنوع نوری مبتنی بر افزارهای مربوطه نظیر منشورها، سطوح متناوب دندانـه ای، آئینه های چرپ کننده و موجبرهای توخالی پرشده با گازهایی با ضریب شکست غیرخطی تحقق یابند. یک ویژگی بسیار اساسی و کامل کننده برای هر کاربرد مخابراتی قابلیت کنترل پذیری خروجی نهایی میباشد. بهـره گیری از گرافن در تحقق یک سیستم با پاشندگی بالای کنترل پذیربا افـت کـم و بـا قابلیـت کـارکرد در حـوزه فرکانسی تراهرتز، بموجب هم تغییرات مشخصه پاشـندگی آن در ایـن بانـد و هـم تغییرپـذیری قابـل ملاحظـه مشخصه پاشندگی آن با اعمال یک میدان الکتریکی استاتیک بیرونی، راه حلی فن آورانه به حساب می آیـد. در گام نخست انتخاب موجبر دایروی با دیواره تابیده مارپیچی وار بعنوان یک ساختار انتقـال الکترومغناطیسـی بـا قابلیت تامین شرایط پاشندگی بالا که از گذشته برای تولید پالسهای پرقـدرت مـایکروویو و مبتنـی بـر فشـرده سازی خطی پالس ورودی استفاده شده است میتواند الگوی مناسبی برای هدف ما در ابداع یک سامانه موجبری مبتنی بر گرافن باشد. در این تحقیق ما به جای ایجاد تابیدگی مارپیچی در دیـواره مـوجبر اسـتوانه ای، داخـل موجبر استوانه ای یک ترکیب مارپیچی نوار گرافنی بر بستر نوارهای فلزی(طلا و بعنوان الکترودهای اعمال ولتاژ بایاس DC کنترل پاشندگی) را با فاصله مناسب تعبیه کرده ایم. هدف نهایی چنین ساختاری دستیابی به یـک فشرده ساز پالس تراهرتزی کنترل پذیر می باشد که با پذیرش یک پالس چرپ با پهنای باندی منطبق بر ناحیه فرکانسی از ساختار که مشخصه پاشندگی سرعت گروه مثبت میباشد، قادر خواهد بود در خروجی موجبر بارشده با نوار گرافنی، پالسی فشرده با کنترل پذیری الکتریکی و فشردگی قابل قبول بعنوان نتیجـه نهـایی ارایـه کنـد. علاوه بر آن، با بررسی ساختار دیگری که نوار گرافنی با یک نوار فلـزی(طـلا) جـایگزین شـده اسـت، چگـونگی مشارکت پاشندگی ماده بموجب استفاده از گرافن علاوه بر سـاختار هندسـی مـارپیچی نـوار کـه مبـیّن مولفـه پاشندگی موجبری از پاشندگی کروماتیک میباشد، نشان داده شده است. بطور خلاصه مکانیزم عملکرد سـاختار درهر دو سوژه مورد بررسی، اینگونه است که به ازای هر پالس ورودی با دوره زمانی حضور مشـخص تنهـا یـک

طول بهینه ساختار موجبری به بهترین نتیجه پالس خروجی از نظر کیفیت پالس فشرده خروجی( دامنـه ماکزیمم و سطح نسبی دامنه های کناری) منجرخواهد شد. این موضوع در خصوص ساختار پیشنهادی مبتنی بر گرافن در مورد دو پالس ورودی بـا دوره زمـانی حضـور متفـاوت 8ps و 12ps بررسـی و طولهـای 700µm و 1700µm محاسبه گردید. سپس میزان کنترل پذیری ساختار در دو مقدار بایاس 0V و 30V که متنـاظر بـا مقدار تقریبی سطح انرژی فرمی حاملها در گـرافن ($E_F$) بترتیـب 0eV و 0.5eV میباشـند در بـازه فرکانسـی کارکرد ساختار که 2.15THz تـا 2.28 THz میباشـد بترتیـب 5.9% و 8% تعیـین شـد. بعنـوان آخـرین دستاورد این تحقیق راست آزمایی تحلیل عددی نتایج بدست آمده مبتنی بـر تحلیـل عـددی بـروش انتگـرال محدود را با بررسی عددی ساختار مبتنی برنوار فلزی با روش المان محدود انجام و توافق بسیار نزدیک دو روش نشان داده شده است. در آخر با ارایه یک مدل تخمینی سیستمی مبتنـی بـر داده هـای زمـانی پـالس ورودی و خروجی یک مدل تابع انتقال خطی منحصر بفرد برای ساختار پیشنهادی مبتنی بـر نـوار فلـزی ارایـه کنـیم. استفاده ازین مدل راهگشایی برای بررسی از نگاه سیستمی و تحلیل پاسخ گذرای ساختار بـه شـکل مـوج هـای ورودی متنوع با پرهیز از شبیه سازی های وقت گیر در محیط های نرم افزاری تمام موج خواهد بود.

**کلمات کلیدی:** فشرده سازی، پاشندگی، گرافن، تراهرتز، موجبر

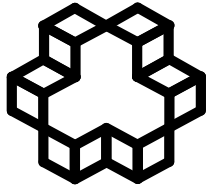

**1307**

دانشگاه صنعتی خواجه نصیرالدین طوسی

دانشکده مهندسی برق

طراحی و تحلیل سامانه فشرده کننده پالس تراهرتزی کنترل پذیر مبتنی بر ساختار نوار گرافنی مارپیچی

رساله برای دریافت درجه دکتری مخابرات–میدان

پدید آورنده:

سید محمدرضا رضوی زاده

استاد راهنما:

دکتر رمضانعلی صادق زاده

استاد مشاور:

دکتر زهرا قطان کاشانی

شهریور ماه 1399